\documentclass[11pt,letterpaper]{article}

\usepackage[utf8]{inputenc}
\DeclareTextCommand{\DJ}{OT1}{\leavevmode\rlap{\raisebox{.15ex}{\kern-.08em-}}D}
\DeclareTextCommand{\dj}{OT1}{\leavevmode d\kern-.1em\raisebox{.55ex}{\rule{.25em}{.3pt}}}
\DeclareTextCommand{\k}{OT1}[1]{#1\kern-.05em\raisebox{-.5ex}{\scriptsize,}}
\usepackage[letterpaper,top=1in,bottom=1in,left=1in,right=1in]{geometry}
\usepackage{amsmath,amssymb}
\usepackage{setspace}
\usepackage[table]{xcolor}
\usepackage{tikz}
\usepackage{placeins}
\usepackage{rotating}
\usepackage{graphicx,adjustbox,booktabs,array,multirow,textcomp,ragged2e,titlesec,longtable,needspace}
\usepackage[super]{cite}
\usepackage[colorlinks=true,citecolor=blue,linkcolor=blue,urlcolor=blue,pdftitle={Single-electron fluorine bonding. II. Benchmarking Minnesota density functional approximations and popular ab initio methods for binding energies and F-bond distances},pdfauthor={Govinda Prasad Khanal, Elangannan Arunan},pdfsubject={Single-electron fluorine bonding; Minnesota functionals; ab initio benchmark},pdfkeywords={fluorine bonding, single-electron bond, SEFB-15, CSFB-14, Minnesota functionals}]{hyperref}
\usepackage[protrusion=true,expansion=false]{microtype}
\newcommand{\orcidicon}[1]{\href{https://orcid.org/#1}{\raisebox{-0.2ex}{\begin{tikzpicture}[baseline=-0.5ex]\fill[color={rgb,255:red,166;green,206;blue,57}] (0,0) circle (0.9ex);\node[white,font=\sffamily\bfseries\fontsize{5}{5}\selectfont,inner sep=0pt] at (0,0) {iD};\end{tikzpicture}}}}
\titleformat{\section}[block]{\normalfont\bfseries\fontsize{17}{20.5}\selectfont}{\thesection.\space}{0pt}{}
\titleformat{\subsection}[block]{\normalfont\bfseries\fontsize{15}{18}\selectfont}{\thesubsection.\space}{0pt}{}
\renewcommand{\thesubsubsection}{\Alph{subsubsection}}
\titleformat{\subsubsection}[block]{\normalfont\bfseries\fontsize{13.5}{16.5}\selectfont}{\thesubsubsection.\space}{0pt}{}
\titlespacing*{\subsubsection}{0pt}{16pt}{6pt}
\titlespacing*{\section}{0pt}{24pt}{9pt}
\titlespacing*{\subsection}{0pt}{20pt}{8pt}
\usepackage{caption}
\renewcommand{\thetable}{\Roman{table}}
\renewcommand{\figurename}{FIG.}
\renewcommand{\tablename}{TABLE}
\DeclareCaptionFormat{gpk}{\setstretch{1}\fontsize{9.5}{11.5}\selectfont\setlength{\parskip}{0pt}\justifying\setlength{\parindent}{0pt}\noindent #1#2#3\par}
\DeclareCaptionFormat{gpkS}{\setstretch{1}\fontsize{10.5}{12.4}\selectfont\setlength{\parskip}{0pt}\justifying\setlength{\parindent}{0pt}\noindent #1#2#3\par}
\usepackage{fancyhdr}
\makeatletter

\renewcommand\@biblabel[1]{\textsuperscript{#1}}
\makeatother
\AtBeginDocument{\renewcommand{\textperiodcentered}{\ensuremath{\cdot}}
\renewcommand{\textdegree}{\ensuremath{^{\circ}}}
\renewcommand{\texttimes}{\ensuremath{\times}}
}

\newsavebox{\rbxb}
\newcommand{\rbx}[1]{\sbox{\rbxb}{#1}\ifdim\wd\rbxb>1.9in\parbox{1.9in}{\centering #1}\else\usebox{\rbxb}\fi}
\newcommand{\rbxs}[1]{\sbox{\rbxb}{#1}\ifdim\wd\rbxb>1.15in\parbox{1.15in}{\centering #1}\else\usebox{\rbxb}\fi}
\begin{document}

\nocite{Mardirossian2016,Cohen2012,Perdew1981,Zhang1998,MoriSanchez2006,Tsuneda2014,Cohen2008,Bryenton2023,Li2018,Tentscher2013,Johnson2013,Zhao2008,Bauza2013,Kozuch2013,OteroDeLaRoza2014,Mller1934,Zhao2006,Zhao2008b,Perdew2001,Perdew2005,Sedlak2013,Riley2012,Karton2015,Bartlett1978,III1982,Pople1987,Frisch2016,Werner2012,Werner2020,Epifanovsky2021,Moller1934,Woon1993,Jr1989,OterodelaRoza2013,Zhao2006b,Peverati2012,Peverati2012b,Yu2016,Zhao2005,Zhao2006c,Haoyu2016,Peverati2011,Peverati2012c,Grimme2010,Perdew1982,Kraisler2013,MoriSanchez2008,Janesko2021,Ranasinghe2017,Jana2018,Ruzsinszky2007,Ruzsinszky2006,Vydrov2007,Keith2012,Mardirossian2017,Brauer2014,Mackie2011,Johnson2013b,Burns2014,Kronik2020,Li2017,Steinmann2012,Anderson2018,Grimme2011,Johnson2017}

\thispagestyle{fancy}
\centering
{\fontsize{17}{21}\selectfont Single-electron fluorine bonding. II. Benchmarking Minnesota density functional approximations and popular \emph{ab initio} methods for binding energies (BEs) and F-bond distances\par}
\vspace{12pt}
Govinda Prasad Khanal\textsuperscript{a}\,\orcidicon{0000-0002-0991-7005}, and Elangannan Arunan\textsuperscript{a}*\,\orcidicon{0000-0001-9669-4307}\par
{\itshape\textsuperscript{a}Inorganic and Physical Chemistry, Indian Institute of Science, Bangalore, India -- 560 012}\par
\textbf{*E-mail Id:} \href{mailto:arunan@iisc.ac.in}{\nolinkurl{arunan@iisc.ac.in}}\par
\vspace{8pt}
\vspace{6pt}{\bfseries\small Abstract}\par\normalfont
\begin{center}\setlength{\fboxsep}{0pt}\setlength{\fboxrule}{0.7pt}\fbox{\includegraphics[width=6.24in]{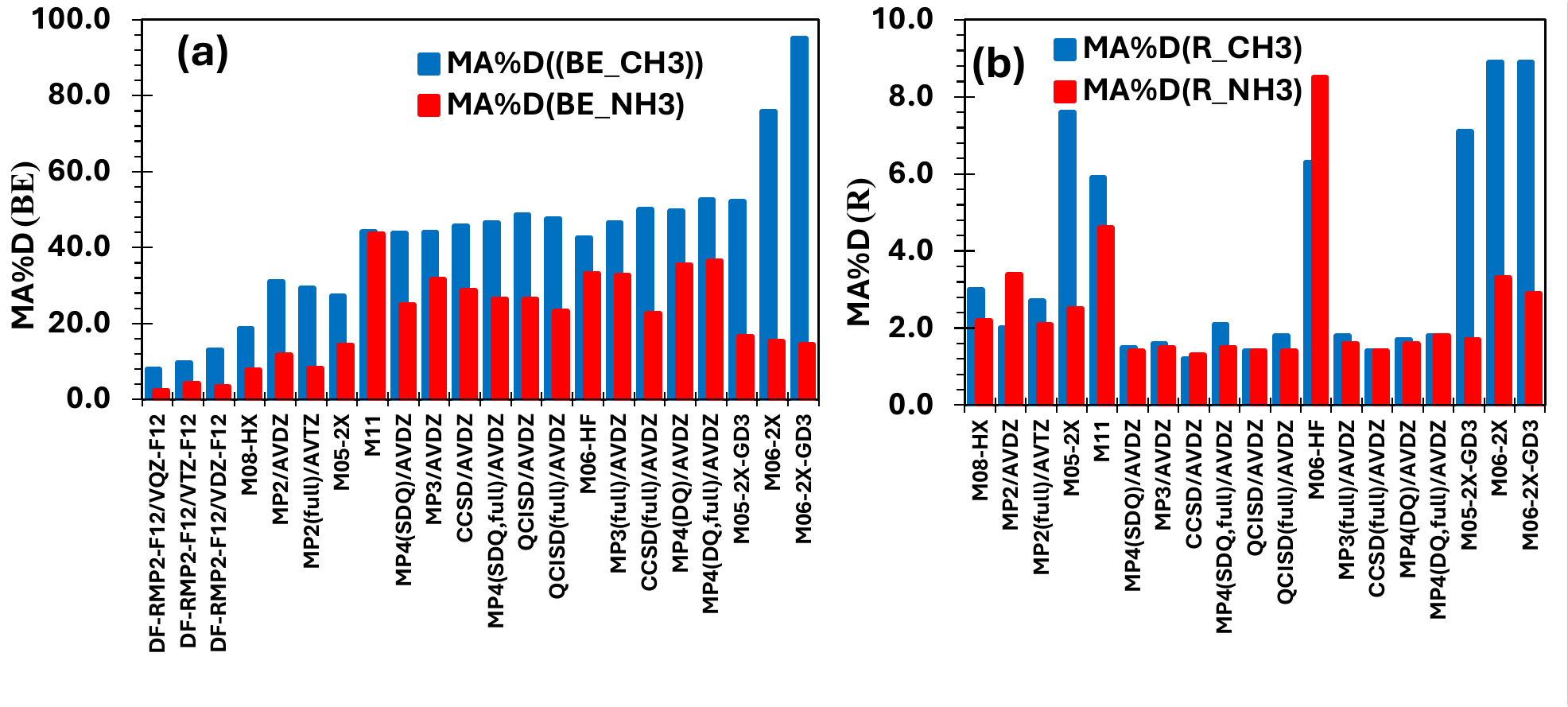}}\end{center}
\vspace{8pt}
{\fontsize{11}{12.9}\selectfont\justifying\setlength{\parindent}{0pt}\leftskip=0.4in\rightskip=0.4in
Companion paper I (Khanal and Arunan, in this issue; \href{https://arxiv.org/abs/2610.16629}{\mbox{arXiv:2610.16629} \mbox{[physics.chem-ph]}}) reported reference BEs and F-bond distances for two new databases SEFB-15 (DF\textperiodcentered{}\textperiodcentered{}\textperiodcentered{}CH\textsubscript{3}) and CSFB-14 (DF\textperiodcentered{}\textperiodcentered{}\textperiodcentered{}NH\textsubscript{3}). Here, we assess performance of popular \emph{ab initio} methods (MP2, MP3, MP4(DQ), MP4(SDQ), CCSD, and QCISD) with valence- and full-electron correlation treatments, and sixteen Minnesota DFAs, including M06-2X already assessed in the companion paper I. CCSD/AVDZ reproduces reference F-bond distances closely, although BEs predicted by it deviate more. Among the 16 Minnesota DFAs, M08-HX reliably reproduces both reference values for DF\textperiodcentered{}\textperiodcentered{}\textperiodcentered{}CH\textsubscript{3} and DF\textperiodcentered{}\textperiodcentered{}\textperiodcentered{}NH\textsubscript{3}, comparable to M05-2X and M06-2X for DF\textperiodcentered{}\textperiodcentered{}\textperiodcentered{}NH\textsubscript{3}. Selected meta-GGAs (M06-L, M11-L, and MN15-L) exhibit spurious fractional charge transfer in cationic complexes even at infinite separation, whereas hybrid Minnesota DFAs do not, except M05 and M06 for NH\textsubscript{3}F\textsuperscript{+}\textperiodcentered{}\textperiodcentered{}\textperiodcentered{}CH\textsubscript{3} and NH\textsubscript{2}F\textsubscript{2}\textsuperscript{+}\textperiodcentered{}\textperiodcentered{}\textperiodcentered{}CH\textsubscript{3}. Although M08-HX shows slightly larger monomer delocalization error than M05-2X and M06-2X, it is more reliable for predicting BEs of SEFB complexes, indicating that monomer delocalization error alone does not govern DFA performance. Bader's intermolecular delocalization indices (DIs) further support M08-HX as more reliable than M05-2X and M06-2X for SEFB, while the larger DIs from M05-2X and M06-2X explain the degraded performance of their dispersion-corrected variants for DF\textperiodcentered{}\textperiodcentered{}\textperiodcentered{}CH\textsubscript{3}, with only a minor impact for DF\textperiodcentered{}\textperiodcentered{}\textperiodcentered{}NH\textsubscript{3}, highlighting the system-dependent nature of functional delocalization error. Present work reveals that M08-HX is more reliable than M05-2X and M06-2X for open-shell DF\textperiodcentered{}\textperiodcentered{}\textperiodcentered{}CH\textsubscript{3} complexes, while all three functionals show comparable reliability for closed-shell DF\textperiodcentered{}\textperiodcentered{}\textperiodcentered{}NH\textsubscript{3} complexes.\par}
\vspace{6pt}
\justifying\setlength{\parindent}{1.5em}
\section{Introduction}\label{sec:1}

Part I of this work, designated as ``Companion paper I'' throughout this article, provides a comprehensive literature survey on the broader significance of open-shell noncovalent bonds (NCBs) and on open-shell single-electron NCBs formed with CH\textsubscript{3} as electron donor across the periodic table. It introduced two new databases SEFB-15 and CSFB-14, for complexes with fluorine bonding (FB), comprising 15 DF\textperiodcentered{}\textperiodcentered{}\textperiodcentered{}CH\textsubscript{3} and 14 DF\textperiodcentered{}\textperiodcentered{}\textperiodcentered{}NH\textsubscript{3} complexes formed with CH\textsubscript{3} and NH\textsubscript{3} as electron donors, respectively. There, we reported benchmark-quality binding energies (BEs) and F-bond distances for both databases, established reference values for both quantities, and compared the open- and closed-shell sets. These databases enable direct assessment of specific DFAs and \emph{ab initio} methods for modelling weakly bound open- and closed-shell FB complexes.

Among several DFAs, Minnesota DFAs are widely used across chemical systems,\cite{Mardirossian2016} yet their performance for SEFB and CSFB complexes had not previously been examined. SEFB complexes (DF\textperiodcentered{}\textperiodcentered{}\textperiodcentered{}CH\textsubscript{3}) are odd-electron systems and are expected to be particularly susceptible to delocalization or many-electron self-interaction error,\cite{Cohen2012,Perdew1981,Zhang1998,MoriSanchez2006,Tsuneda2014,Cohen2008,Bryenton2023,Li2018} consistent with previous reports that radical-molecule complexes generally are prone to such error.\cite{Tentscher2013,Johnson2013} Hybrid DFAs with a substantial fraction of Hartree--Fock exchange can partially reduce this error.\cite{Bryenton2023} M06-2X\cite{Zhao2008} , with 54\% Hartree--Fock exchange, is established as reliable for closed-shell halogen bonding,\cite{Bauza2013,Kozuch2013,OteroDeLaRoza2014} raising the question of whether this reliability extends to the open-shell SEFB and closed-shell CSFB cases. Companion paper I tested this and found that M06-2X fails systematically for open-shell SEFB complexes, regardless of basis set, despite remaining reliable for CSFB complexes. We also assessed MP2\cite{Mller1934}, with and without full electron correlation, and found that it is unreliable for SEFB but reasonably accurate for CSFB. These results raise the questions of whether higher-order perturbation methods, such as MP3 and MP4\cite{Mller1934}, are more reliable than MP2 for SEFB and CSFB complexes and whether the failure of M06-2X is specific to that functional or representative of the Minnesota DFA family as a whole.

Mardirossian and Head-Gordon\cite{Mardirossian2016} found that none of the Minnesota DFAs are equally reliable across various types of NCBs, though M06-2X\cite{Zhao2008} is recommended generally. For open-shell NCBs such as radical-molecule complexes from TA13, Minnesota functionals with larger exact exchange, including M06-2X,\cite{Zhao2008} M06-HF,\cite{Zhao2006} and M08-HX,\cite{Zhao2008b} are recommended, with root mean square deviations (RMSDs) of 5.8, 6.7, and 6.9 kJ/mol, respectively.\cite{Mardirossian2016} M08-HX,\cite{Zhao2008b} despite its improved functional form over M06-2X, nonetheless performs slightly worse on the TA13 database. The authors\cite{Mardirossian2016} found that M11 (7.6 kJ/mol), the range-separated hybrid DFA, performs worse than all three of these functionals. Surprisingly, MN15-L (7.4 kJ/mol), a local functional, also outperforms several hybrid meta-GGAs, such as M05 (12.2 kJ/mol), M05-2X (10.6 kJ/mol), M06 (9.8 kJ/mol), and MN15 (8.9 kJ/mol).\cite{Mardirossian2016} Thus, climbing Jacob\textquotesingle s ladder\cite{Perdew2001,Perdew2005} does not necessarily guarantee improved results. This raises the question of whether Minnesota DFAs show similar behavior for open- and closed-shell SEFB and CSFB complexes.

The questions above motivate a systematic assessment of both \emph{ab initio} methods and Minnesota DFAs for the SEFB-15 and CSFB-14 databases. The present work therefore assesses the reliability of widely used \emph{ab initio} methods, namely MP2-MP4,\cite{Mller1934} MP2.5,\cite{Sedlak2013,Riley2012} MP3.5,\cite{Karton2015} CCSD,\cite{Bartlett1978,III1982} and QCISD\cite{Pople1987}. Both valence-electron and full-electron correlation treatments are considered, and the impact of basis-set superposition error (BSSE) on the predicted BEs of all 29 DF\textperiodcentered{}\textperiodcentered{}\textperiodcentered{}CH\textsubscript{3}/DF\textperiodcentered{}\textperiodcentered{}\textperiodcentered{}NH\textsubscript{3} complexes is examined across the \emph{ab initio} methods used. We further assess 16 Minnesota DFAs, including M06-2X, for BEs and F-bond distances across all complexes, examine the effect of dispersion correction, and compare their performance with that for the TA13 database. We also investigate the presence of delocalization error in the monomers forming the DF\textperiodcentered{}\textperiodcentered{}\textperiodcentered{}CH\textsubscript{3}/DF\textperiodcentered{}\textperiodcentered{}\textperiodcentered{}NH\textsubscript{3} complexes using selected Minnesota DFAs. Finally, we assess the reliability of Bader\textquotesingle s intermolecular delocalization indices (DIs) as descriptors of interaction strength across both open-shell (DF\textperiodcentered{}\textperiodcentered{}\textperiodcentered{}CH\textsubscript{3}) and closed-shell (DF\textperiodcentered{}\textperiodcentered{}\textperiodcentered{}NH\textsubscript{3}) FB complexes, based on previous report for closed-shell halogen bonding.\cite{OteroDeLaRoza2014}

\section{Computational details}\label{sec:2}

All electronic structure calculations in this work were performed using the Gaussian 16 C.01,\cite{Frisch2016} MOLPRO 2015.1,\cite{Werner2012,Werner2020} and Q-Chem 6.0.2.\cite{Epifanovsky2021}

The complete set of F-bonded complexes DF\textperiodcentered{}\textperiodcentered{}\textperiodcentered{}CH\textsubscript{3}/NH\textsubscript{3} (DF = F\textsubscript{2}, SF\textsubscript{6}, NCF, CNF, NF\textsubscript{3}, ClF\textsubscript{3}, OF\textsubscript{2}, NF\textsubscript{5}, FCCF, CF\textsubscript{3}CCF, CF\textsubscript{3}OF, F\textsubscript{3}NO, SF\textsubscript{5}, NH\textsubscript{3}F\textsuperscript{+}, and NH\textsubscript{2}F\textsubscript{2}\textsuperscript{+}), along with reference BEs and F-bond distances (obtained at DF-RMP2-F12/CBS{[}V(T,Q)Z-F12{]}//DF-RMP2-F12/VDZ-F12 and DF-RMP2-F12/VDZ-F12, respectively), were taken from companion paper I. For all DF\textperiodcentered{}\textperiodcentered{}\textperiodcentered{}CH\textsubscript{3}/NH\textsubscript{3} complexes and the corresponding monomers, full geometry optimization was performed using several electronic structure methods employed in this study.

We have used several \emph{ab initio} methods, namely second- through fourth-order Møller--Plesset perturbation theory (MP2-MP4)\cite{Moller1934} composite perturbative schemes {[}MP2.5,\cite{Sedlak2013,Riley2012} and MP3.5\cite{Karton2015}{]}, coupled-cluster theory with single and double excitations {[}CCSD and CCSD(full){]},\cite{Bartlett1978,III1982} and quadratic configuration interaction with single and double excitations {[}QCISD and QCISD(full){]}.\cite{Pople1987} All these methods were considered with both valence-only and all-electron treatments. BEs obtained from these methods were analyzed as uncorrected, BSSE-corrected, and half-half (defined as the arithmetic mean of the former two). All BEs with and without BSSE corrections from DF-RMP2-F12/VnZ-F12 (n = 2-4) level using geometries optimized at DF-RMP2-F12/VDZ-F12 were taken from Supporting material of companion paper I. The corresponding half-half values were computed from these in the present work. For these methods, BSSE calculations were carried out as described in companion paper I. All calculations with conventional \emph{ab initio} methods were carried out using Gaussian 16 program, in conjunction with the augmented correlation-consistent Dunning basis sets, primarily aug-cc-pVDZ (AVDZ).\cite{Woon1993,Jr1989} MP2(full) calculations were performed for all DF\textperiodcentered{}\textperiodcentered{}\textperiodcentered{}CH\textsubscript{3}/NH\textsubscript{3} complexes using the larger aug-cc-pVTZ (AVTZ) basis set. We performed calculations with explicitly correlated method DF-RMP2-F12 with VDZ-F12, VTZ-F12 and VQZ-F12 basis sets using MOLPRO (for detail of such calculations, refer to companion paper I).

To compare the performance of M06-2X functional assessed in paper I with other Minnesota DFAs, we have considered several Minnesota DFAs in conjunction with AVTZ basis set, which is practically close to CBS limit for conventional DFT.\cite{OterodelaRoza2013} BEs from all Minnesota DFAs, including M06-2X, are therefore considered without BSSE correction throughout this work, consistent with companion paper I. These DFAs span multiple rungs of Jacob's ladder, enabling a systematic assessment of the accuracy of M06-2X in predicting both binding energies and F-bond distances relative to other Minnesota DFAs for the complete data set.

The Minnesota DFAs considered are classified as follows:

(i) \textbf{Pure Minnesota functionals}: M06-L,\cite{Zhao2006b} M11-L,\cite{Peverati2012} MN12-L,\cite{Peverati2012b} and MN15-L.\cite{Yu2016}

(ii) \textbf{Global hybrid Minnesota DFAs}: M05,\cite{Zhao2005} M06,\cite{Zhao2008} M05-2X,\cite{Zhao2006c} M06-2X,\cite{Zhao2008} M06-HF,\cite{Zhao2006} M08-HX,\cite{Zhao2008b} and MN15.\cite{Haoyu2016}

(iii) \textbf{Range-separated hybrid Minnesota DFAs}: M11,\cite{Peverati2011} and MN12-SX.\cite{Peverati2012c}

In addition, Grimme's D3 dispersion correction (GD3)\cite{Grimme2010} was applied to selected Minnesota DFAs, namely M06-L, M05-2X , and M06-2X, to examine the influence of GD3 dispersion treatment on the predicted BEs and F-bond distances.

To assess the impact of functional delocalization error on the BEs of DF\textperiodcentered{}\textperiodcentered{}\textperiodcentered{}CH\textsubscript{3}/NH\textsubscript{3} complexes predicted by the Minnesota DFAs, fractional charge transfer between the two monomers (DF and CH\textsubscript{3}/NH\textsubscript{3}) was examined at infinite separation. Fractional charges (q) were imposed on the F-bond donors (DF) and the acceptors (CH\textsubscript{3}/NH\textsubscript{3}), and the corresponding energies, E(q), were computed using Q-Chem 6.0.2.\cite{Epifanovsky2021} The exact density functional requires piecewise linear E(q) behavior between adjacent integer q values.\cite{Cohen2012,Perdew1982} The deviations from this behavior are well-established measures of functional delocalization error, or many-electron self-interaction error (MSIE), inherent in the DFA used.\cite{Cohen2012,Zhang1998,MoriSanchez2006,Tsuneda2014,Cohen2008,Bryenton2023,Kraisler2013,MoriSanchez2008,Janesko2021,Ranasinghe2017,Jana2018,Ruzsinszky2007,Ruzsinszky2006,Vydrov2007} Because the monomers are infinitely separated, these deviations characterize monomer delocalization error exclusively. Exact reference E(q) values for each DFA were constructed following established procedures reported in the literature.\cite{Cohen2012,Bryenton2023,Perdew1982,Kraisler2013,MoriSanchez2008,Ranasinghe2017,Jana2018}

Monomer delocalization error alone, however, may not fully account for failure of a DFA to predict BEs of noncovalent dimers reliably, since the electronic structure of each monomer changes upon dimer formation.\cite{OteroDeLaRoza2014} Therefore, assessing functional delocalization error at the equilibrium dimer geometry can provide a more direct probe.\cite{OteroDeLaRoza2014} Bader's intermolecular delocalization indices (DIs) were computed for all DF\textperiodcentered{}\textperiodcentered{}\textperiodcentered{}CH\textsubscript{3}/NH\textsubscript{3} complexes at their equilibrium geometries using the AIMALL program,\cite{Keith2012} for most of the Minnesota DFAs used, with AVTZ basis set.

\section{Statistical error analysis}\label{sec:3}

We have evaluated the reliability of several Minnesota DFAs and \emph{ab initio} methods used for BEs and F-bond distances of SEFB and CSFB complexes using mean signed error (MSE), mean absolute deviation (MAD), root-mean-square deviation (RMSD), maximum deviation (MAX), and mean absolute percentage deviation (MA\%D), as well as deviation (\ensuremath{\delta}) and percentage deviation (\%\ensuremath{\delta}) for each complex (see companion paper I for definition of these statistical quantities). For all statistical analysis, reference values were taken from companion paper I.

\section{Results and Discussions}\label{sec:4}

\FloatBarrier
\subsection{Performance of several Minnesota DFAs for single electron F-bonded complexes DF\textperiodcentered{}\textperiodcentered{}\textperiodcentered{}CH\textsubscript{3}/NH\textsubscript{3}}\label{sec:4.1}

In this section, we discuss the performance of several Minnesota DFAs for predicting BEs and F-bond distances (R) of DF\textperiodcentered{}\textperiodcentered{}\textperiodcentered{}CH\textsubscript{3}/NH\textsubscript{3} complexes with respect to the reference values described in \textbf{Section \ref*{sec:2}}. We consider individual deviations first.

\subsubsection{Deviations on BEs and F-bond distances predicted by Minnesota functionals}\label{sec:4.1.A}

\begin{table}[!bp]\centering\setlength{\parskip}{0pt}
\caption{Deviations on BSSE-uncorrected BEs (\ensuremath{\delta}\ensuremath{\Delta}\textsubscript{1}) and percentage deviations (\%\ensuremath{\delta}\ensuremath{\Delta}\textsubscript{1}) calculated at M05-2X, M05-2X-GD3, M06-2X, M06-2X-GD3, M06-HF, M08-HX, M11 with AVTZ basis set for all the F-bonded complexes DF\textperiodcentered{}\textperiodcentered{}\textperiodcentered{}CH\textsubscript{3}/NH\textsubscript{3}. BEs are in kJ/\allowbreak{}mol. Reference BEs (\textbf{Ref. BEs}) were calculated at DF-RMP2-F12/\allowbreak{}CBS[V(T,Q)Z-F12]//DF-RMP2-F12/\allowbreak{}VDZ-F12, denoted as DF-RMP2-F12/\allowbreak{}CBS (\mbox{\textbf{TABLE \ref{tab:I}}}\textbf{, }companion paper\textbf{ I}). Least values of \ensuremath{\delta}\ensuremath{\Delta}\textsubscript{1} and \%\ensuremath{\delta}\ensuremath{\Delta}\textsubscript{1} are highlighted in bold and maximum values are italicized.}\label{tab:I}\vspace{7pt}\nopagebreak
\adjustbox{width=1.08\linewidth,max totalheight=0.78\textheight,center}{\fontfamily{ptm}\fontsize{9.3}{10.7}\selectfont\renewcommand{\arraystretch}{1.12}\setlength{\tabcolsep}{1.5pt}
\begin{tabular}{|>{\centering\arraybackslash}m{67.9pt}|>{\centering\arraybackslash}m{23.0pt}|>{\centering\arraybackslash}m{31.2pt}|>{\centering\arraybackslash}m{23.0pt}|>{\centering\arraybackslash}m{31.4pt}|>{\centering\arraybackslash}m{23.2pt}|>{\centering\arraybackslash}m{31.6pt}|>{\centering\arraybackslash}m{23.2pt}|>{\centering\arraybackslash}m{31.6pt}|>{\centering\arraybackslash}m{23.2pt}|>{\centering\arraybackslash}m{31.4pt}|>{\centering\arraybackslash}m{23.2pt}|>{\centering\arraybackslash}m{31.2pt}|>{\centering\arraybackslash}m{23.2pt}|>{\centering\arraybackslash}m{27.9pt}|>{\centering\arraybackslash}m{28.0pt}|}
\hline
\multicolumn{1}{|>{\centering\arraybackslash}m{67.9pt}|}{\multirow{2}{67.9pt}{{\bfseries\boldmath \centering \textbf{DF\textperiodcentered{}\textperiodcentered{}\textperiodcentered{}CH}\textbf{\textsubscript{3}}}}} &\multicolumn{2}{>{\centering\arraybackslash}m{57.2pt}|}{{\bfseries\boldmath M05-2X}} &\multicolumn{2}{>{\centering\arraybackslash}m{57.4pt}|}{{\bfseries\boldmath M05-2X-GD3}} &\multicolumn{2}{>{\centering\arraybackslash}m{57.7pt}|}{{\bfseries\boldmath M06-2X}} &\multicolumn{2}{>{\centering\arraybackslash}m{57.7pt}|}{{\bfseries\boldmath M06-2X-GD3}} &\multicolumn{2}{>{\centering\arraybackslash}m{57.6pt}|}{{\bfseries\boldmath M06-HF}} &\multicolumn{2}{>{\centering\arraybackslash}m{57.4pt}|}{{\bfseries\boldmath M08-HX}} &\multicolumn{2}{>{\centering\arraybackslash}m{54.1pt}|}{{\bfseries\boldmath M11}} & \multirow{2}{28.0pt}{{\bfseries\boldmath \centering Ref. BEs}}   \\
\hline
\end{tabular}}
\par\nopagebreak\vspace{3pt}{\setstretch{1}\setlength{\parskip}{1.5pt}\fontsize{9}{10.5}\selectfont\raggedright\noindent \textsuperscript{b}Convergence error during optimization of respective F-bonded complexes.\par}

\end{table}

The discrepancy in BEs for F\textsubscript{2}\textperiodcentered{}\textperiodcentered{}\textperiodcentered{}CH\textsubscript{3} and F\textsubscript{2}\textperiodcentered{}\textperiodcentered{}\textperiodcentered{}NH\textsubscript{3} predicted by the M06-2X functional, compared with the reference values and those obtained from several other \emph{ab initio} methods and basis sets, including AVTZ, confirms that the observed discrepancy is not an artifact of the AVTZ basis set. Therefore, it is natural to ask whether other Minnesota functionals perform better than M06-2X for predicting BEs and F-bond distances for the complexes in the SEFB-15 data set (15 DF\textperiodcentered{}\textperiodcentered{}\textperiodcentered{}CH\textsubscript{3}). For this purpose, we considered the following Minnesota functionals: M05, M05-2X, M06, M06-2X, M06-HF, M08-HX, M11, MN12-SX, MN15, M06-L, M11-L, MN12-L, and MN15-L. Grimme\textquotesingle s D3 dispersion correction was also considered for three widely used Minnesota DFAs (M05-2X, M06-2X, and M06-L).

\begin{table}[!htbp]\centering\setlength{\parskip}{0pt}
\caption{Deviations on F-bond distances (\ensuremath{\delta}R) and percentage deviations (\%\ensuremath{\delta}R) calculated at M05-2X, M05-2X-GD3, M06-2X, M06-2X-GD3, M06-HF, M08-HX, M11 with AVTZ basis set for all F-bonded complexes considered DF\textperiodcentered{}\textperiodcentered{}\textperiodcentered{}CH\textsubscript{3}/NH\textsubscript{3 }. Units of the bond distances are Å. Reference F-bond distances (Ref. R) were obtained at DF-RMP2-F12/\allowbreak{}VDZ-F12. The \%\ensuremath{\delta}R values around 5.0\% or less than 5.0 \% (in magnitude) are highlighted in red. Least values of \ensuremath{\delta}R and \% \ensuremath{\delta}R are in bold and maximum values are italicized.}\label{tab:II}\vspace{7pt}\nopagebreak
\adjustbox{width=1.08\linewidth,max totalheight=0.78\textheight,center}{\fontfamily{ptm}\fontsize{9.3}{10.7}\selectfont\renewcommand{\arraystretch}{1.12}\setlength{\tabcolsep}{1.5pt}
\begin{tabular}{|>{\centering\arraybackslash}m{67.9pt}|>{\centering\arraybackslash}m{31.1pt}|>{\centering\arraybackslash}m{26.5pt}|>{\centering\arraybackslash}m{31.1pt}|>{\centering\arraybackslash}m{26.5pt}|>{\centering\arraybackslash}m{31.1pt}|>{\centering\arraybackslash}m{26.5pt}|>{\centering\arraybackslash}m{31.1pt}|>{\centering\arraybackslash}m{26.5pt}|>{\centering\arraybackslash}m{31.1pt}|>{\centering\arraybackslash}m{26.5pt}|>{\centering\arraybackslash}m{31.1pt}|>{\centering\arraybackslash}m{25.5pt}|>{\centering\arraybackslash}m{31.0pt}|>{\centering\arraybackslash}m{25.5pt}|>{\centering\arraybackslash}m{30.5pt}|}
\hline
\multicolumn{1}{|>{\centering\arraybackslash}m{67.9pt}|}{\multirow{2}{67.9pt}{{\bfseries\boldmath \centering \textbf{DF\textperiodcentered{}\textperiodcentered{}\textperiodcentered{}CH}\textbf{\textsubscript{3}}}}} &\multicolumn{2}{>{\centering\arraybackslash}m{60.7pt}|}{{\bfseries\boldmath M05-2X}} &\multicolumn{2}{>{\centering\arraybackslash}m{60.7pt}|}{{\bfseries\boldmath M05-2X-GD3}} &\multicolumn{2}{>{\centering\arraybackslash}m{60.7pt}|}{{\bfseries\boldmath M06-2X}} &\multicolumn{2}{>{\centering\arraybackslash}m{60.7pt}|}{{\bfseries\boldmath M06-2X-GD3}} &\multicolumn{2}{>{\centering\arraybackslash}m{60.7pt}|}{{\bfseries\boldmath M06-HF}} &\multicolumn{2}{>{\centering\arraybackslash}m{59.7pt}|}{{\bfseries\boldmath M08-HX}} &\multicolumn{2}{>{\centering\arraybackslash}m{59.6pt}|}{{\bfseries\boldmath M11}} & \multirow{2}{30.5pt}{{\bfseries\boldmath \centering Ref. R}}   \\
\hline
\end{tabular}}
\par\nopagebreak\vspace{3pt}{\setstretch{1}\setlength{\parskip}{1.5pt}\fontsize{9}{10.5}\selectfont\raggedright\noindent \textsuperscript{b}Convergence error during optimization of respective F-bonded complexes.\par}

\end{table}

\begin{table}[!htbp]\captionsetup{type=figure}\centering
\par\vspace{2pt}\noindent\hfil\includegraphics[width=\linewidth]{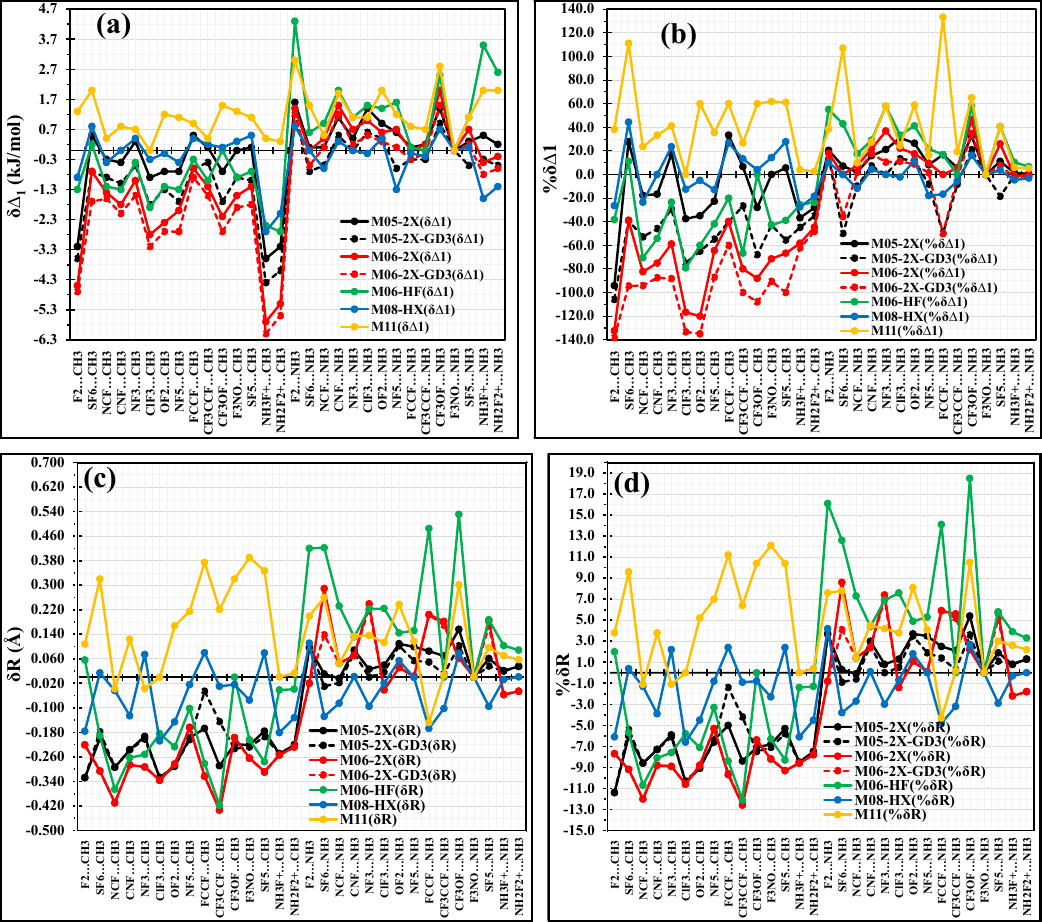}\hfil\null\par\nointerlineskip\vspace{13pt}
\par\vspace{-1pt}\caption{Deviations on BSSE-uncorrected BEs (\ensuremath{\delta}\ensuremath{\Delta}\textsubscript{1}) \textbf{(a)} and percentage deviations (\%\ensuremath{\delta}\ensuremath{\Delta}\textsubscript{1}) \textbf{(b)} calculated at M05-2X, M05-2X-GD3, M06-2X, M06-2X-GD3, M06-HF, M08-HX, M11 with AVTZ basis set for all F-bonded complexes DF\textperiodcentered{}\textperiodcentered{}\textperiodcentered{}CH\textsubscript{3}/NH\textsubscript{3}. Also, deviations on F-bond distances \ensuremath{\delta}R \textbf{(c)} and percentage deviations \%\ensuremath{\delta}R \textbf{(d)} calculated at the same levels of theory are included. Units of BEs and bond distances are kJ/mol and Å, respectively. Reference BEs were calculated at DF-RMP2-F12/CBS{[}V(T,Q)Z-F12{]}//DF-RMP2-F12/VDZ-F12 (denoted as DF-RMP2-F12/CBS level and reference F-bond distances were obtained at DF-RMP2-F12/VDZ-F12 level.}\label{fig:1}
\vspace{10pt}
\end{table}

Out of these 16 Minnesota DFAs used, only 7 (M05-2X, M05-2X-GD3, M06-2X, M06-2X-GD3, M06-HF, M08-HX and M11) predict F\textsubscript{2}\textperiodcentered{}\textperiodcentered{}\textperiodcentered{}CH\textsubscript{3} complex. Deviations (\ensuremath{\delta}\ensuremath{\Delta}\textsubscript{1}) and percentage deviations (\%\ensuremath{\delta}\ensuremath{\Delta}\textsubscript{1}) on BEs and R values calculated from these 7 DFAs in conjunction with AVTZ basis set are summarized in \mbox{\textbf{TABLES \ref{tab:I}}} and \mbox{\textbf{\ref{tab:II}}}, and \mbox{\textbf{FIG. \ref{fig:1}}}(for BEs obtained from all 16 DFAs, refer to \mbox{\textbf{TABLE \ref{tab:S1}}} in supporting information). For the SEFB complexes, the largest deviations on BEs occur for F\textsubscript{2}\textperiodcentered{}\textperiodcentered{}\textperiodcentered{}CH\textsubscript{3}. The percentage deviations for F\textsubscript{2}\textperiodcentered{}\textperiodcentered{}\textperiodcentered{}CH\textsubscript{3} reach -94.1\%, -105.9\%, -132.4\%, and -138.2\% for M05-2X, M05-2X-GD3, M06-2X, and M06-2X-GD3, respectively (\mbox{\textbf{TABLE \ref{tab:I}}} and \mbox{\textbf{FIG. \ref{fig:1}(b)}}). Thus, the large discrepancy associated with F\textsubscript{2}\textperiodcentered{}\textperiodcentered{}\textperiodcentered{}CH\textsubscript{3} is not limited to M06-2X but is also observed for the other popular Minnesota DFAs considered here. In contrast, less popular functional, M08-HX yields a substantially smaller maximum percentage deviation of 44.4\% for the SEFB-15 set. The remaining complexes from SEFB-15 database generally show smaller deviations, although the magnitude and sign of the deviations vary among the DFAs (\mbox{\textbf{TABLE \ref{tab:I}}} and \mbox{\textbf{FIG. \ref{fig:1}}}). For the DF\textperiodcentered{}\textperiodcentered{}\textperiodcentered{}NH\textsubscript{3} complexes, the deviations are generally smaller than those for DF\textperiodcentered{}\textperiodcentered{}\textperiodcentered{}CH\textsubscript{3}, and M08-HX/AVTZ gives the smallest maximum percentage deviation, -17.8\% (see \mbox{\textbf{TABLE \ref{tab:I}}} and \mbox{\textbf{FIG. \ref{fig:1}}}). As noted in the Introduction, M06-2X outperforms both M06-HF and M08-HX for interaction energies of radical-molecule complexes from TA13 benchmark set.\cite{Mardirossian2016,Mardirossian2017} For BEs of the DF\textperiodcentered{}\textperiodcentered{}\textperiodcentered{}CH\textsubscript{3}/NH\textsubscript{3} complexes from SEFB-15 and CSFB-14 databases, however, M08-HX performs better than both M06-2X and M06-HF, indicating that the relative performance observed for TA13 database does not transfer to the FB complexes considered here.

The deviations in F-bond distances for the individual DF\textperiodcentered{}\textperiodcentered{}\textperiodcentered{}CH\textsubscript{3}/NH\textsubscript{3} complexes calculated using the 7 Minnesota DFAs are summarized in \mbox{\textbf{TABLE \ref{tab:II}}} and \mbox{\textbf{FIG. \ref{fig:1}}}.  For the SEFB-15 complexes, M08-HX/AVTZ shows the smallest overall deviations in F-bond distances among the seven DFAs, with a maximum percentage deviation of -6.6\% (\mbox{\textbf{TABLE \ref{tab:II}}}). In contrast, the largest deviations for the SEFB-15 set are obtained for M06-2X and M06-2X-GD3, reaching -12.6\% and -12.5\%, respectively. The deviations predicted by M05-2X, M05-2X-GD3, and M06-HF reach -11.4\%, -11.4\%, and -12.1\%, respectively, whereas M11 gives a maximum deviation of 12.1\%.

For the DF\textperiodcentered{}\textperiodcentered{}\textperiodcentered{}NH\textsubscript{3} complexes, M05-2X, M05-2X-GD3, and M08-HX are all reasonably good for predicting F-bond distances, with maximum percentage deviations of 5.4\%, 3.7\%, and -4.8\%, respectively, all close to the 5.0\% deviation indicated in \mbox{\textbf{TABLE \ref{tab:II}}}. In comparison, the maximum deviations increase to 8.6\% for M06-2X, 7.0\% for M06-2X-GD3, 18.5\% for M06-HF, and 10.5\% for M11. Thus, among the seven Minnesota DFAs, M06-HF gives the largest deviations for F-bond distances for the closed-shell DF\textperiodcentered{}\textperiodcentered{}\textperiodcentered{}NH\textsubscript{3} complexes, whereas M05-2X, M05-2X-GD3, and M08-HX remain close to the reference values.

\begin{table}[!bp]\centering\setlength{\parskip}{0pt}
\caption{MSE, MAD, RMSD, MAX and MA\%D on BEs (kJ/\allowbreak{}mol) without BSSE correction (\ensuremath{\Delta}\textsubscript{1}) for DF\textperiodcentered{}\textperiodcentered{}\textperiodcentered{}CH\textsubscript{3}/NH\textsubscript{3} complexes using different Minnesota functionals with aug-cc-PVTZ basis set. BEs calculated at DF-RMP2-F12/\allowbreak{}CBS[V(T, Q)Z-F12]//DF-RMP2-F12/\allowbreak{}VDZ-F12 level were used as reference values. Out of sixteen Minnesota DFAs used, only the first five methods worked for all fifteen DF\textperiodcentered{}\textperiodcentered{}\textperiodcentered{}CH\textsubscript{3} complexes, whereas seven functionals (M08-HX, M05-2X, M05-2X-GD3, M06-2X, M06-2X-GD3, M11 and M06-HF) produced F\textsubscript{2}\textperiodcentered{}\textperiodcentered{}\textperiodcentered{}CH\textsubscript{3}. These seven functionals are used for overall performance assessment. Number of F-bonded complexes (\textbf{No. of BC}) are also listed. See \mbox{\textbf{Table S}}\textbf{1 }for BEs from all 16 Minnesota DFAs (M05, M05-2X, M05-2X-GD3, M06, M06-2X, M06-2X-GD3, M06-HF, M08-HX, M11, MN12-SX, MN15, M06-L, M06-L-GD3, M11-L, MN12-L and MN15-L) used in this work.\textsuperscript{a}\textsuperscript{ }}\label{tab:III}\vspace{7pt}\nopagebreak
\adjustbox{width=1.08\linewidth,max totalheight=0.78\textheight,center}{\fontfamily{ptm}\fontsize{9.3}{10.7}\selectfont\renewcommand{\arraystretch}{1.12}\setlength{\tabcolsep}{1.5pt}
\begin{tabular}{|>{\centering\arraybackslash}m{60.4pt}|>{\centering\arraybackslash}m{31.0pt}|>{\centering\arraybackslash}m{31.6pt}|>{\centering\arraybackslash}m{25.1pt}|>{\centering\arraybackslash}m{32.2pt}|>{\centering\arraybackslash}m{31.5pt}|>{\centering\arraybackslash}m{33.9pt}|>{\centering\arraybackslash}m{58.4pt}|>{\centering\arraybackslash}m{31.0pt}|>{\centering\arraybackslash}m{31.5pt}|>{\centering\arraybackslash}m{26.5pt}|>{\centering\arraybackslash}m{30.8pt}|>{\centering\arraybackslash}m{31.6pt}|>{\centering\arraybackslash}m{33.9pt}|}
\hline
\multicolumn{7}{|>{\centering\arraybackslash}m{263.8pt}|}{{\bfseries\boldmath \textbf{DF\textperiodcentered{}\textperiodcentered{}\textperiodcentered{}CH}\textbf{\textsubscript{3}}}} &\multicolumn{7}{>{\centering\arraybackslash}m{261.8pt}|}{{\bfseries\boldmath \textbf{DF\textperiodcentered{}\textperiodcentered{}\textperiodcentered{}NH}\textbf{\textsubscript{3}}}}   \\
\hline
\multicolumn{1}{|>{\centering\arraybackslash}m{60.4pt}|}{{\bfseries\boldmath \textbf{DF}\textbf{As}}} & {\bfseries\boldmath No. of BC} & {\bfseries\boldmath MSE} & {\bfseries\boldmath MAD} & {\bfseries\boldmath RMSD} & {\bfseries\boldmath MAX} & {\bfseries\boldmath MA\%D} & {\bfseries\boldmath \textbf{DF}\textbf{As}} & {\bfseries\boldmath No. of BC} & {\bfseries\boldmath MSE} & {\bfseries\boldmath MAD} & {\bfseries\boldmath RMSD} & {\bfseries\boldmath MAX} & {\bfseries\boldmath MA\%D}   \\
\hline
\multicolumn{1}{|>{\centering\arraybackslash}m{60.4pt}|}{\textbf{M08-HX}} & \textbf{15} & \textbf{-0.3} & \textbf{0.6} & \textbf{1.0} & \textbf{-2.7} & \textbf{18.7} & \textbf{M08-HX} & \textbf{14} & \textbf{-0.2} & \textbf{0.5} & \textbf{0.7} & \textbf{-1.6} & \textbf{7.7}  \\
\hline
\multicolumn{1}{|>{\centering\arraybackslash}m{60.4pt}|}{M05-2X} & 15 & -0.8 & 1.0 & 1.6 & -3.6 & 27.2 & M06-2X-GD3 & 14 & 0.2 & 0.5 & 0.7 & 1.5 & 14.4  \\
\hline
\end{tabular}}
\par\nopagebreak\vspace{3pt}{\setstretch{1}\setlength{\parskip}{1.5pt}\fontsize{9}{10.5}\selectfont\raggedright\noindent \textsuperscript{a}DFAs are arranged by increasing MADs on BEs calculated for the DF\textperiodcentered{}\textperiodcentered{}\textperiodcentered{}CH\textsubscript{3} and DF\textperiodcentered{}\textperiodcentered{}\textperiodcentered{}NH\textsubscript{3} complexes, separately by considering number of complexes into account.\par}

\end{table}

\begin{table}[!htbp]\centering\setlength{\parskip}{0pt}
\caption{MSE, MAD, RMSD, MAX and MA\%D on F-bond distances R(F\textperiodcentered{}\textperiodcentered{}\textperiodcentered{}C/\allowbreak{}N) (in Å) for DF\textperiodcentered{}\textperiodcentered{}\textperiodcentered{}CH\textsubscript{3}/NH\textsubscript{3} complexes using different Minnesota functionals with aug-cc-PVTZ basis set. The distances obtained at DF-RMP2-F12/\allowbreak{}VDZ-F12 level of theory were used as reference values. Out of sixteen Minnesota DFAs considered, only seven (M08-HX, M05-2X, M05-2X-GD3, M06-2X, M06-2X-GD3, M11 and M06-HF) will be used for assessing overall performance of the Minnesota DFAs used (For detail, see \mbox{\textbf{TABLE \ref{tab:III}}}). Number of F-bonded complexes (\textbf{No. of BC}) are also listed. F-bond distances (R) obtained from all sixteen Minnesota DFAs are summarized in \mbox{\textbf{Table \ref{tab:S2}}}.\textsuperscript{a}}\label{tab:IV}\vspace{7pt}\nopagebreak
\adjustbox{width=1.08\linewidth,max totalheight=0.78\textheight,center}{\fontfamily{ptm}\fontsize{9.3}{10.7}\selectfont\renewcommand{\arraystretch}{1.12}\setlength{\tabcolsep}{1.5pt}
\begin{tabular}{|>{\centering\arraybackslash}m{67.9pt}|>{\centering\arraybackslash}m{25.4pt}|>{\centering\arraybackslash}m{32.5pt}|>{\centering\arraybackslash}m{32.5pt}|>{\centering\arraybackslash}m{32.4pt}|>{\centering\arraybackslash}m{32.5pt}|>{\centering\arraybackslash}m{25.4pt}|>{\centering\arraybackslash}m{68.2pt}|>{\centering\arraybackslash}m{18.0pt}|>{\centering\arraybackslash}m{32.5pt}|>{\centering\arraybackslash}m{32.4pt}|>{\centering\arraybackslash}m{32.5pt}|>{\centering\arraybackslash}m{32.5pt}|>{\centering\arraybackslash}m{25.4pt}|}
\hline
\multicolumn{7}{|>{\centering\arraybackslash}m{266.4pt}|}{{\bfseries\boldmath \textbf{DF\textperiodcentered{}\textperiodcentered{}\textperiodcentered{}CH}\textbf{\textsubscript{3}}}} &\multicolumn{7}{>{\centering\arraybackslash}m{259.2pt}|}{{\bfseries\boldmath \textbf{DF\textperiodcentered{}\textperiodcentered{}\textperiodcentered{}NH}\textbf{\textsubscript{3}}}}   \\
\hline
\end{tabular}}
\par\nopagebreak\vspace{3pt}{\setstretch{1}\setlength{\parskip}{1.5pt}\fontsize{9}{10.5}\selectfont\raggedright\noindent \textsuperscript{a}DFAs are arranged in the same order as \mbox{\textbf{TABLE \ref{tab:III}}}, based on BEs, not re-ordered for R.\par}

\end{table}

\begin{table}[!htbp]\captionsetup{type=figure}\centering
\par\vspace{2pt}\noindent\hfil\includegraphics[width=\linewidth]{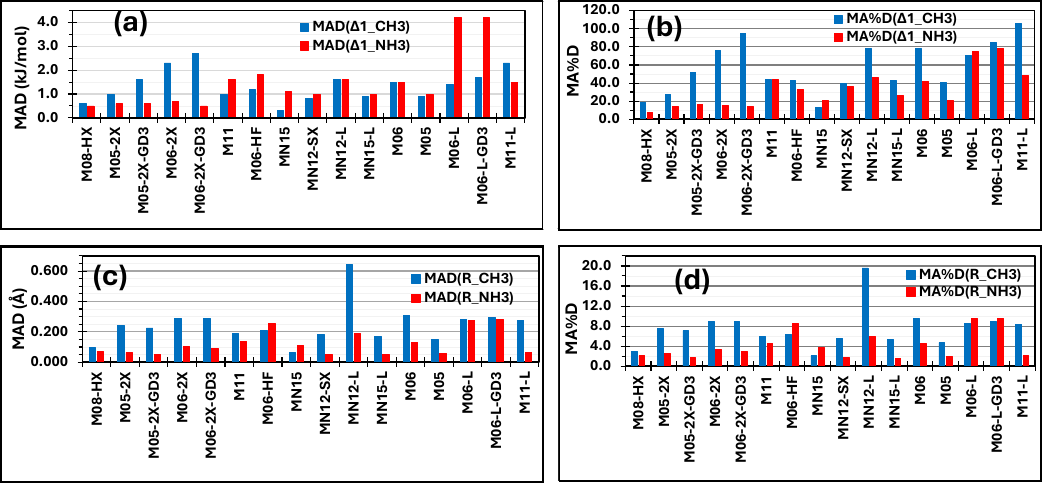}\hfil\null\par\nointerlineskip\vspace{13pt}
\par\vspace{-1pt}\caption{Mean absolute deviations (MADs) \textbf{(a)} and mean absolute percentage deviations (MA\%Ds) \textbf{(b)} on BSSE-uncorrected BEs (\ensuremath{\Delta}\textsubscript{1}) for DF\textperiodcentered{}\textperiodcentered{}\textperiodcentered{}CH\textsubscript{3}/NH\textsubscript{3}, calculated at several Minnesota DFAs using AVTZ basis set. Also, MADs \textbf{(c)} and MA\%Ds \textbf{(d)} on F-bond distances (R) for the same set of complexes, calculated by the same DFAs and basis set are included. Out of sixteen Minnesota DFAs used, only first seven (M08-HX, M05-2X, M05-2X-GD3, M06-2X, M06-2X-GD3, M11 and M06-HF) produced F\textsubscript{2}\textperiodcentered{}\textperiodcentered{}\textperiodcentered{}CH\textsubscript{3} complex. BEs calculated at DF-RMP2-F12/CBS{[}V(T,Q)Z-F12{]}//DF-RMP2-F12/VDZ-F12 level and F-bond distances obtained from DF-RMP2-F12/VDZ-F12 were used as reference values. See \mbox{\textbf{Figure \ref{fig:S1}}} for mean signed errors (MSEs) and root mean square deviations (RMSDs) on BEs and R values for DF\textperiodcentered{}\textperiodcentered{}\textperiodcentered{}CH\textsubscript{3}/NH\textsubscript{3} complexes.}\label{fig:2}
\vspace{10pt}
\end{table}

\subsubsection{Statistical errors on BEs and F-bond distances from Minnesota functionals}\label{sec:4.1.B}

Various statistical measures (MSE, MAD, RMSD, MAX, and MA\%D) for BEs and F-bond distances of the DF\textperiodcentered{}\textperiodcentered{}\textperiodcentered{}CH\textsubscript{3}/NH\textsubscript{3} complexes, calculated using several Minnesota DFAs in conjunction with the AVTZ basis set, are summarized in \mbox{\textbf{TABLES \ref{tab:III}}} and \mbox{\textbf{\ref{tab:IV}}}, respectively.   M08-HX/AVTZ gives MA\%D values of 18.7\% and 7.7\% for BEs of the SEFB and CSFB complexes from SEFB-15 and CSFB-14 databases, respectively. For SEFB-15, the MA\%D values increase from 27.2\% for M05-2X to 52.1\% for M05-2X-GD3 and from 75.8\% for M06-2X to 95.1\% for M06-2X-GD3, showing that the GD3 dispersion correction substantially worsens the BEs for these open-shell complexes (\mbox{\textbf{TABLE \ref{tab:III}}} and \mbox{\textbf{FIG. \ref{fig:2}(b)}}). In contrast, the effect of GD3 is smaller for CSFB-14, where M06-2X-GD3 gives a MA\%D of 14.4\%, compared with 15.3\% for M06-2X (\mbox{\textbf{TABLE \ref{tab:III}}} and \mbox{\textbf{FIG. \ref{fig:2}(b)}}). Among the methods that converge for all 15 SEFB complexes, M08-HX is the only DFA with a MA\%D below 20\% (\mbox{\textbf{TABLE \ref{tab:III}}} and \mbox{\textbf{FIG. \ref{fig:2}(b)}}). There are larger deviations on BEs of the DF\textperiodcentered{}\textperiodcentered{}\textperiodcentered{}CH\textsubscript{3} complexes compared to those for DF\textperiodcentered{}\textperiodcentered{}\textperiodcentered{}NH\textsubscript{3} (see MA\%D in \mbox{\textbf{FIG. \ref{fig:2}(b)}}).

The MA\%D values for F-bond distances (R) for SEFB-15 are lowest for M08-HX (3.0\%), followed by M05-2X-GD3 (7.1\%), M05-2X (7.6\%), and M06-2X/M06-2X-GD3 (8.9\% each) (\mbox{\textbf{TABLE \ref{tab:IV}}}). In contrast, MN12-L shows a substantially larger MA\%D of 19.6\% for DF\textperiodcentered{}\textperiodcentered{}\textperiodcentered{}CH\textsubscript{3} (\mbox{\textbf{TABLE \ref{tab:IV}}}), as also evident in \mbox{\textbf{FIG. \ref{fig:2}(d)}}. The corresponding F-bond distances calculated using all 16 Minnesota DFAs are provided in \mbox{\textbf{TABLE \ref{tab:S2}}}. For CSFB-14, M08-HX again gives the lowest MA\%D (2.2\%); among the 9 Minnesota DFAs that converge for all 14 DF\textperiodcentered{}\textperiodcentered{}\textperiodcentered{}NH\textsubscript{3} complexes, only MN12-L exceeds 5.0\% of MA\%D, with a value of 6.0\% (\mbox{\textbf{TABLE \ref{tab:IV}}} and \mbox{\textbf{FIG. \ref{fig:2}(d)}}).

Thus, while most Minnesota DFAs reproduce F-bond distances for the DF\textperiodcentered{}\textperiodcentered{}\textperiodcentered{}NH\textsubscript{3} complexes reasonably well, reliable prediction of BEs remains more difficult, particularly for the open-shell DF\textperiodcentered{}\textperiodcentered{}\textperiodcentered{}CH\textsubscript{3} complexes. For DF\textperiodcentered{}\textperiodcentered{}\textperiodcentered{}NH\textsubscript{3}, four Minnesota DFAs that converge for all 14 complexes (M08-HX, M06-2X-GD3, M05-2X, and M06-2X) give MA\%D values below 20\%, whereas only M08-HX does so among the DFAs converging for all 15 DF\textperiodcentered{}\textperiodcentered{}\textperiodcentered{}CH\textsubscript{3} complexes (\mbox{\textbf{TABLE \ref{tab:III}}}). Among the Minnesota DFAs that converge for all 15 SEFB complexes, M08-HX gives the lowest MA\%D for both BEs and F-bond distances. The comparison of the Minnesota DFAs that predict the challenging F\textsubscript{2}\textperiodcentered{}\textperiodcentered{}\textperiodcentered{}CH\textsubscript{3} complex with the conventional \emph{ab initio} methods will be discussed in \textbf{Section \ref*{sec:4.3}}.

From \mbox{\textbf{TABLE \ref{tab:III}}} (also refer to \mbox{\textbf{TABLES \ref{tab:S1}-\ref{tab:S3}}} for raw data), only 5 Minnesota DFAs (M08-HX, M05-2X, M05-2X-GD3, M06-2X, and M06-2X-GD3) converge for all 15 SEFB complexes, while two others (M11 and M06-HF) converge for 14, including the challenging F\textsubscript{2}\textperiodcentered{}\textperiodcentered{}\textperiodcentered{}CH\textsubscript{3} complex. The remaining 9 Minnesota DFAs (MN15, MN12-SX, MN12-L, MN15-L, M06, M05, M06-L, M06-L-GD3, and M11-L) fail to converge for F\textsubscript{2}\textperiodcentered{}\textperiodcentered{}\textperiodcentered{}CH\textsubscript{3}. For DF\textperiodcentered{}\textperiodcentered{}\textperiodcentered{}NH\textsubscript{3}, nine Minnesota DFAs (M08-HX, M06-2X-GD3, M05-2X, M06-2X, MN15-L, MN12-SX, M06, M11, and MN12-L) converge for all 14 complexes.

\mbox{\textbf{FIG. \ref{fig:2}}} provides a direct graphical comparison of the MAD and MA\%D for BEs and F-bond distances of the DF\textperiodcentered{}\textperiodcentered{}\textperiodcentered{}CH\textsubscript{3}/NH\textsubscript{3} complexes using these levels of theory, with the reference values given in \mbox{\textbf{TABLES \ref{tab:I}}} and \mbox{\textbf{\ref{tab:II}}} of the companion Paper I.

\FloatBarrier
\subsection{Performance of various \emph{ab initio} methods for BEs and bond distances}\label{sec:4.2}

In \textbf{Section \ref*{sec:4.1}}, we discussed the performance of various Minnesota DFAs for BEs and F-bond distances of open- and closed-shell FB complexes (DF\textperiodcentered{}\textperiodcentered{}\textperiodcentered{}CH\textsubscript{3}/NH\textsubscript{3}). In this section, we discuss the performance of several widely used \emph{ab initio} methods with and without full electron correlation.

\subsubsection{Assessment of BEs from \emph{ab initio} methods}\label{sec:4.2.A}

\begin{table}[!htbp]\centering\setlength{\parskip}{0pt}
\caption{MSE,\textbf{ }MAD, RMSD, MAX and MA\%D on BEs (kJ/\allowbreak{}mol) for DF\textperiodcentered{}\textperiodcentered{}\textperiodcentered{}CH\textsubscript{3}/NH\textsubscript{3} using different popular \emph{ab}\emph{ }\emph{initio} methods with basis sets mentioned. BEs calculated at DF-RMP2-F12/\allowbreak{}CBS[V(T, Q)Z-F12]//DF-RMP2-F12/\allowbreak{}VDZ-F12 were used as reference values. See \mbox{\textbf{Table S}}\textbf{4} for BSSE un-corrected BEs (\ensuremath{\Delta}\textsubscript{1}) and BSSE-corrected BEs (\ensuremath{\Delta}\textsubscript{1+BSSE}) for all the DF\textperiodcentered{}\textperiodcentered{}\textperiodcentered{}CH\textsubscript{3}/NH\textsubscript{3} complexes. Also, \mbox{\textbf{Table S}}\textbf{5} summarizes half-half-corrected BEs for the set of complexes.}\label{tab:V}\vspace{7pt}\nopagebreak
\adjustbox{width=1.08\linewidth,max totalheight=0.78\textheight,center}{\fontfamily{ptm}\fontsize{9.3}{10.7}\selectfont\renewcommand{\arraystretch}{1.0}\setlength{\tabcolsep}{4.0pt}
\begin{tabular}{|>{\centering\arraybackslash}m{109.7pt}|>{\centering\arraybackslash}m{28.3pt}|>{\centering\arraybackslash}m{23.8pt}|>{\centering\arraybackslash}m{19.4pt}|>{\centering\arraybackslash}m{23.9pt}|>{\centering\arraybackslash}m{23.9pt}|>{\centering\arraybackslash}m{19.4pt}|>{\centering\arraybackslash}m{23.8pt}|>{\centering\arraybackslash}m{23.8pt}|>{\centering\arraybackslash}m{23.8pt}|>{\centering\arraybackslash}m{23.9pt}|>{\centering\arraybackslash}m{26.6pt}|>{\centering\arraybackslash}m{19.4pt}|>{\centering\arraybackslash}m{25.5pt}|>{\centering\arraybackslash}m{26.6pt}|>{\centering\arraybackslash}m{24.2pt}|}
\hline
\multicolumn{16}{|>{\centering\arraybackslash}p{510.8pt}|}{{\bfseries\boldmath \textbf{DF\textperiodcentered{}\textperiodcentered{}\textperiodcentered{}CH}\textbf{\textsubscript{3}}}}   \\
\hline
\multicolumn{1}{|>{\centering\arraybackslash}m{109.7pt}|}{\multirow{2}{109.7pt}{{\bfseries\boldmath \centering \emph{ab initio} Methods}}} &\multicolumn{5}{>{\centering\arraybackslash}p{131.2pt}|}{{\bfseries\boldmath \textbf{Without BSSE}}} &\multicolumn{5}{>{\centering\arraybackslash}p{126.6pt}|}{{\bfseries\boldmath \textbf{With BSSE}}} &\multicolumn{5}{>{\centering\arraybackslash}p{134.2pt}|}{{\bfseries\boldmath \textbf{Half-half}}}   \\
\hline
\end{tabular}}

\end{table}

In \mbox{\textbf{TABLE \ref{tab:V}}}, we have listed error statistics (MSE, MAD, RMSD, MAX and MA\%D) on BEs for the DF\textperiodcentered{}\textperiodcentered{}\textperiodcentered{}CH\textsubscript{3}/NH\textsubscript{3} complexes using several \emph{ab initio} methods, including the explicitly correlated method DF-RMP2-F12. The BEs are calculated without BSSE correction, with BSSE correction, and as the average of the two, denoted as half-half. The statistics are over the complexes that converged for each method (see \mbox{\textbf{TABLES}} \mbox{\textbf{\ref{tab:S4}}} and \mbox{\textbf{\ref{tab:S5}}} for BEs obtained from all \emph{ab initio} methods). BEs at the DF-RMP2-F12/CBS{[}V(T,Q)Z-F12{]}//DF-RMP2-F12/VDZ-F12 level were considered as the reference BEs. A negative MSE indicates overbinding.

BEs without BSSE correction obtained from DF-RMP2-F12/VDZ-F12 are closer to the reference values for DF\textperiodcentered{}\textperiodcentered{}\textperiodcentered{}CH\textsubscript{3} (MA\%D = 5.7\%) (\mbox{\textbf{TABLE \ref{tab:V}}}). This agrees with previous findings on the performance of F12 methods with the smaller basis set VDZ-F12 for modeling NCBs.\cite{Brauer2014} This is not surprising, so we focus on the other ab initio methods. Using MP2, BEs for DF\textperiodcentered{}\textperiodcentered{}\textperiodcentered{}CH\textsubscript{3} complexes without BSSE correction are not reliable (MA\%D = 75.5\%) (\mbox{\textbf{TABLE \ref{tab:V}}}). MA\%D values for the BSSE-uncorrected BEs for the same set of complexes, calculated by CCSD, QCISD and MP4(DQ,full) with the AVDZ basis set, are also around 50.0\%. For DF\textperiodcentered{}\textperiodcentered{}\textperiodcentered{}NH\textsubscript{3} complexes, without BSSE correction, MA\%D values for BEs range from 38.6\% (MP4(DQ)/AVDZ) to 69.8\% (MP2/AVDZ) (\mbox{\textbf{TABLE \ref{tab:V}}}). In addition, BSSE-uncorrected BEs from CCSD and QCISD (48.0\% and 51.5\%) are not reliable. \mbox{\textbf{TABLE \ref{tab:V}}} shows that all-electron correlation does not improve BSSE-uncorrected BEs relative to the reference values.

Now, let us discuss error statistics on BSSE-corrected BEs. BSSE-correction does not improve BEs for DF\textperiodcentered{}\textperiodcentered{}\textperiodcentered{}CH\textsubscript{3} complexes calculated from DF-RMP2-F12 method (MA\%D increased from 5.7\% to 13.1\%, 6.5\% to 9.7\% and 6.1\% to 7.9\% with VDZ-F12, VTZ-F12 and VQZ-F12 basis sets, respectively) (\mbox{\textbf{TABLE \ref{tab:V}}}). This is consistent with the error compensation between BSSE and basis-set incompleteness in F12 calculations with small basis sets.\cite{Brauer2014} Contrary to this, BEs for DF\textperiodcentered{}\textperiodcentered{}\textperiodcentered{}NH\textsubscript{3} complexes from DF-RMP2-F12/VDZ-F12 level improve after BSSE correction (11.1\% to 3.4\%) (\mbox{\textbf{TABLE \ref{tab:V}}}). For DF\textperiodcentered{}\textperiodcentered{}\textperiodcentered{}CH\textsubscript{3} complexes, MA\%Ds on BSSE-corrected BEs range from 29.3\% (MP2(full)/AVTZ) to 52.6\% (MP4(DQ,full)/AVDZ), whereas the corresponding range without BSSE-correction is 41.3\%(MP4(DQ)/AVDZ) to 75.5\%(MP2/AVDZ). These data shows that BSSE-correction is useful for BEs of DF\textperiodcentered{}\textperiodcentered{}\textperiodcentered{}CH\textsubscript{3} complexes predicted by conventional \emph{ab initio} methods. For these complexes, in particular, BSSE-correction improves BEs obtained from MP2.5/AVDZ (64.5\% to 38.8\%) and that from from MP4(SDQ,full)/AVDZ (68.6\% to 46.6\%) (for details, see \mbox{\textbf{TABLE \ref{tab:V}}}). For DF\textperiodcentered{}\textperiodcentered{}\textperiodcentered{}NH\textsubscript{3} complexes, MA\%D after BSSE-correction ranges from 8.2\% (MP2(full)/AVTZ) to 36.5\% (MP4(DQ,full)/AVDZ). In general, BSSE-corrected BEs for DF\textperiodcentered{}\textperiodcentered{}\textperiodcentered{}NH\textsubscript{3} complexes are more reliable than for DF\textperiodcentered{}\textperiodcentered{}\textperiodcentered{}CH\textsubscript{3} complexes (\mbox{\textbf{TABLE \ref{tab:V}}}).

One of the important results from \mbox{\textbf{TABLE \ref{tab:V}}} is that BSSE-correction to MP2/AVDZ has improved BEs for both DF\textperiodcentered{}\textperiodcentered{}\textperiodcentered{}CH\textsubscript{3} and DF\textperiodcentered{}\textperiodcentered{}\textperiodcentered{}NH\textsubscript{3} complexes (31.1\% and 11.7\%). Also, BSSE-correction is beneficial for BEs of DF\textperiodcentered{}\textperiodcentered{}\textperiodcentered{}CH\textsubscript{3} complexes with MP2(full)/AVTZ (66.3\% to 29.3\%) although the deviation is still significant. On the other hand, BSSE-corrected BEs of DF\textperiodcentered{}\textperiodcentered{}\textperiodcentered{}NH\textsubscript{3} complexes obtained from MP2(full)/AVTZ level are reliable 8.2\% (\mbox{\textbf{TABLE \ref{tab:V}}} and \mbox{\textbf{FIGs.}} \mbox{\textbf{\ref{fig:3}(a)}} and \mbox{\textbf{\ref{fig:3}(b)}}).

Up to now, we discussed the performance of several \emph{ab initio} methods for BEs (with and without BSSE-correction) of both open- and closed-shell FB complexes (DF\textperiodcentered{}\textperiodcentered{}\textperiodcentered{}CH\textsubscript{3}/NH\textsubscript{3}). Some researchers advocate for using an average of raw (BEs without BSSE-correction) and BSSE-corrected BEs.\cite{Brauer2014,Mackie2011,Johnson2013b,Burns2014} We have denote this average as half-half in \mbox{\textbf{TABLE \ref{tab:V}}} and \mbox{\textbf{FIG. \ref{fig:3}}}. For DF-RMP2-F12/VDZ-F12, half-half BEs for DF\textperiodcentered{}\textperiodcentered{}\textperiodcentered{}CH\textsubscript{3} (8.7\%) are not better than those without BSSE-correction (5.7\%), and error statistics for half-half and uncorrected BEs at VTZ-F12 and VQZ-F12 are the same (\mbox{\textbf{TABLE \ref{tab:V}}}). Half-half BEs for DF\textperiodcentered{}\textperiodcentered{}\textperiodcentered{}CH\textsubscript{3} complexes from DF-RMP2-F12, however, are slightly better than BSSE-corrected ones across all three basis sets used. For DF\textperiodcentered{}\textperiodcentered{}\textperiodcentered{}NH\textsubscript{3} complexes, there is improvement with half-half at VDZ-F12 (5.5\% vs. 11.1\%), with little effect at VTZ-F12 and VQZ-F12 (\mbox{\textbf{TABLE \ref{tab:V}}}). For MP2/AVDZ and MP2(full)/AVTZ, half-half BEs are slightly better than BSSE-corrected ones for DF\textperiodcentered{}\textperiodcentered{}\textperiodcentered{}CH\textsubscript{3} (23.1\% and 20.0\% vs. 31.1\% and 29.3\%), but worse for DF\textperiodcentered{}\textperiodcentered{}\textperiodcentered{}NH\textsubscript{3} (30.5\% and 17.8\% vs. 11.7\% and 8.2\%), though still better than those without BSSE-correction (\mbox{\textbf{TABLE \ref{tab:V}}} and \mbox{\textbf{FIGs. \ref{fig:3}(b)-\ref{fig:3}(d)}}). So, BSSE-corrected BEs for DF\textperiodcentered{}\textperiodcentered{}\textperiodcentered{}NH\textsubscript{3} complexes from MP2/AVDZ and MP2(full)/AVTZ are reliable compared to reference values, as listed in \mbox{\textbf{TABLE \ref{tab:I}}} in companion Paper I. For DF\textperiodcentered{}\textperiodcentered{}\textperiodcentered{}CH\textsubscript{3} complexes, half-half BEs are slightly better than BSSE-corrected ones.

For the remaining Møller--Plesset methods of higher order (MP3, MP4(DQ), MP4(SDQ), MP2.5, MP3.5, and their full-electron analogues), MA\%D on half-half BEs of DF\textperiodcentered{}\textperiodcentered{}\textperiodcentered{}CH\textsubscript{3} complexes ranges from 10.8\% to 16.7\%, much closer to reference values than BSSE-corrected (38.8-52.6\%) or BSSE-uncorrected (41.3-68.6\%). Similarly, for DF\textperiodcentered{}\textperiodcentered{}\textperiodcentered{}NH\textsubscript{3} complexes, half-half BEs (12.8-20.9\%) are closer to reference values than BSSE-corrected (21.0-36.5\%), which are closer than BSSE-uncorrected (38.6-59.0\%) (\mbox{\textbf{TABLE \ref{tab:V}}} and \mbox{\textbf{FIGs. \ref{fig:3}(b)}}, \mbox{\textbf{\ref{fig:3}(d)}}).

\begin{table}[!htbp]\captionsetup{type=figure}\centering
\par\vspace{2pt}\noindent\hfil\includegraphics[width=\linewidth]{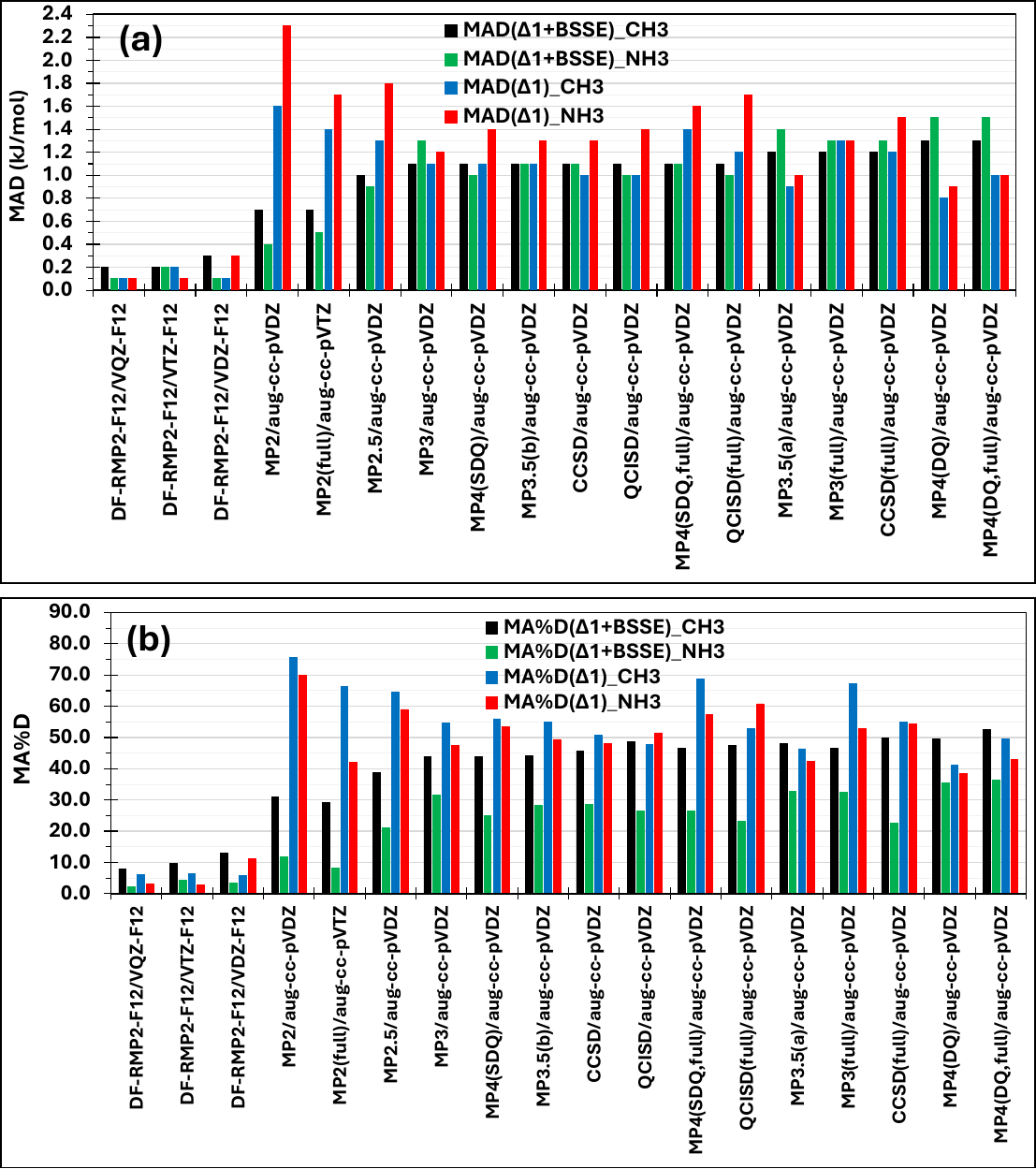}\hfil\null\par\nointerlineskip\vspace{13pt}
\par\vspace{-1pt}\caption{MADs on BSSE-uncorrected (\ensuremath{\Delta}\textsubscript{1}) and BSSE-corrected (\ensuremath{\Delta}\textsubscript{1+BSSE}) BEs (kJ/mol) \textbf{(a)}, MA\%Ds on the same \textbf{(b),} MADs on BSSE-corrected and half-half-corrected BEs \textbf{(c)} and respective MA\%Ds on the same \textbf{(d)} for DF\textperiodcentered{}\textperiodcentered{}\textperiodcentered{}CH\textsubscript{3}/NH\textsubscript{3} complexes using different \emph{ab initio} methods arranged in increasing order of MADs on the BSSE-corrected BEs for DF\textperiodcentered{}\textperiodcentered{}\textperiodcentered{}CH\textsubscript{3} complexes (bar graphs in black, panel \textbf{(a)}). BEs calculated at DF-RMP2-F12/CBS{[}V(T,Q)Z-F12{]} level of theory at DF-RMP2-F12/VDZ-F12 optimized geometries were used as reference values.}\label{fig:3}
\vspace{10pt}
\end{table}

\begin{table}[!htbp]\captionsetup{type=figure}\centering
\par\vspace{2pt}\noindent\hfil\includegraphics[width=\linewidth]{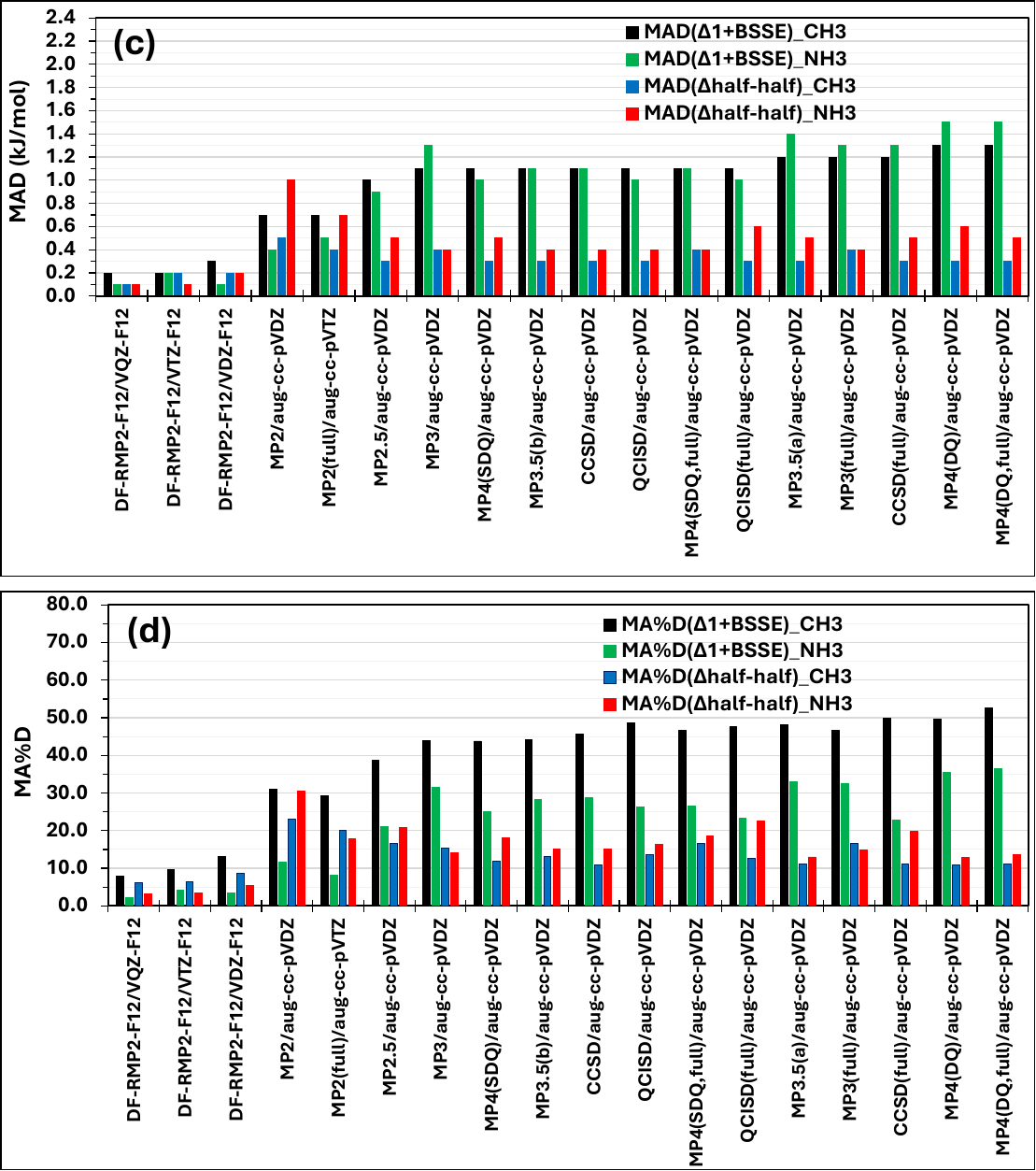}\hfil\null\par\nointerlineskip\vspace{13pt}
\par\vspace{-1pt}\ContinuedFloat\caption[]{(Continued.)}
\vspace{10pt}
\end{table}

Now, let us discuss the performance of CCSD and QCISD as well as CCSD(full) and QCISD(full). For DF\textperiodcentered{}\textperiodcentered{}\textperiodcentered{}CH\textsubscript{3} complexes, MAD values for the half-half BEs obtained from these four methods with the AVDZ basis set are the same, 0.3 kJ/mol. The MA\%D values are also close to each other, ranging from 10.9\% for CCSD/AVDZ to 13.7\% for QCISD/AVDZ (\mbox{\textbf{TABLE \ref{tab:V}}}). Half-half BEs deviate less from the reference BEs than the BSSE-corrected ones (45.7\% to 50.0\%), which are closer than the BSSE-uncorrected BEs (47.9\% to 55.0\%) (with the exception of QCISD/AVDZ). Thus, for the SEFB-15 database, CCSD, QCISD, CCSD(full), and QCISD(full) perform excellently with half-half correction. For DF\textperiodcentered{}\textperiodcentered{}\textperiodcentered{}NH\textsubscript{3} complexes, half-half BEs are similarly improved compared to the BSSE-corrected ones, which are greatly improved compared to the BSSE-uncorrected BEs (48.0\% to 60.8\%) (\mbox{\textbf{TABLE \ref{tab:V}}}).

To sum up, the effect of all-electron correlation on several conventional \emph{ab initio} methods used shows no improvement over valence-electron calculations for reproducing reference BEs of F-bonded complexes DF\textperiodcentered{}\textperiodcentered{}\textperiodcentered{}CH\textsubscript{3}/NH\textsubscript{3}. Half-half results for DF\textperiodcentered{}\textperiodcentered{}\textperiodcentered{}CH\textsubscript{3} complexes are closer to reference values than BSSE-corrected or BSSE-uncorrected BEs, whereas for DF\textperiodcentered{}\textperiodcentered{}\textperiodcentered{}NH\textsubscript{3} complexes, MP2/AVDZ and MP2(full)/AVTZ benefit more from BSSE-correction than from half-half averaging for reliable BEs (\mbox{\textbf{TABLE \ref{tab:V}}}, and \mbox{\textbf{FIGs. \ref{fig:3}}} and \mbox{\textbf{\ref{fig:S2}}}). Leaving few exceptions, half-half BEs are closer to reference BEs than BSSE-corrected ones, which are closer to reference values than BSSE-uncorrected ones (\mbox{\textbf{TABLE \ref{tab:V}}}, and \mbox{\textbf{FIGs. \ref{fig:3}}} and \mbox{\textbf{\ref{fig:S2}}}).  The MA\%D on half-half BEs for DF\textperiodcentered{}\textperiodcentered{}\textperiodcentered{}NH\textsubscript{3} complexes at MP2/AVDZ (30.5\%) is the largest among all the \emph{ab initio} methods used, compared with 11.7\% for BSSE-corrected values at the same level (\mbox{\textbf{TABLE \ref{tab:V}}}, and \mbox{\textbf{FIGs. \ref{fig:3}}} and \mbox{\textbf{\ref{fig:S2}}}). MP2(full)/AVTZ shows the same pattern (17.8\% vs. 8.2\% for half-half vs BSSE-corrected BEs). This larger deviation is due to overestimation of the half-half BE for F\textsubscript{2}\textperiodcentered{}\textperiodcentered{}\textperiodcentered{}NH\textsubscript{3}. The uncorrected BE of F\textsubscript{2}\textperiodcentered{}\textperiodcentered{}\textperiodcentered{}NH\textsubscript{3} from MP2/AVDZ (-11.9 kJ/mol) is overestimated by 4.1 kJ/mol relative to the reference (-7.8 kJ/mol). The BSSE-corrected BE from the same level of theory (-8.0 kJ/mol) (\mbox{\textbf{TABLE \ref{tab:S4}}}) is within 0.2 kJ/mol of it, and the half-half BE (-10.0 kJ/mol) (\mbox{\textbf{TABLE \ref{tab:S5}}}) still deviates by 2.2 kJ/mol. For DF\textperiodcentered{}\textperiodcentered{}\textperiodcentered{}CH\textsubscript{3} complexes, by contrast, half-half BEs for MP2/AVDZ and MP2(full)/AVTZ (23.1\% and 20.0\%) are slightly better than BSSE-corrected ones (31.1\% and 29.3\%). So, we will consider BSSE-corrected BEs for both DF\textperiodcentered{}\textperiodcentered{}\textperiodcentered{}CH\textsubscript{3} and DF\textperiodcentered{}\textperiodcentered{}\textperiodcentered{}NH\textsubscript{3} complexes for further analysis with MP2/AVDZ and MP2(full)/AVTZ. The MAD and MA\%D values for BSSE-corrected, uncorrected and half-half BEs are graphically shown in \mbox{\textbf{FIGs. \ref{fig:3}}}, and \mbox{\textbf{\ref{fig:S2}}}, whereas MSE and RMSD are depicted in \mbox{\textbf{FIG. \ref{fig:S3}}}. For BEs obtained from all \emph{ab initio} methods, refer to \mbox{\textbf{TABLES}} \mbox{\textbf{\ref{tab:S4}}} and \mbox{\textbf{\ref{tab:S5}}} from Supporting Information.

From \mbox{\textbf{FIGs. \ref{fig:3}(a)}} and \mbox{\textbf{\ref{fig:3}(b)}}, BSSE-corrected BEs for both open- and closed-shell FB complexes (DF\textperiodcentered{}\textperiodcentered{}\textperiodcentered{}CH\textsubscript{3}/NH\textsubscript{3}) calculated at MP2/AVDZ and MP2(full)/AVTZ are closer to reference BEs than those obtained from other conventional \emph{ab initio} methods considered. The performance of all the conventional methods used for reproducing reference F-bond distances is discussed next (\textbf{Section \ref*{sec:4.2}.\ref*{sec:4.2.B}}).

\subsubsection{Assessment of F-bond distances from \emph{ab initio} methods}\label{sec:4.2.B}

Now, let us have a brief discussion on the performance of several \emph{ab initio} methods to reproduce the reference F-bond distances (R) for both set of F-bonded complexes (SEFB-15 and CSFB-14). Various error statistics calculated are summarized in \mbox{\textbf{TABLE \ref{tab:VI}}} and \mbox{\textbf{FIG. \ref{fig:4}}} and \mbox{\textbf{\ref{fig:S3}}}.

From \mbox{\textbf{TABLE \ref{tab:VI}}} and \mbox{\textbf{FIG. \ref{fig:4}}}, we can see that CCSD/AVDZ reproduces F-bond distances closer to the reference F-bond distances (\mbox{\textbf{TABLE \ref{tab:II}}} in companion paper I) than any other \emph{ab initio} method used, for both types of complexes. The MA\%D ranges from 1.2\% (CCSD/AVDZ) to 2.7\% (MP2(full)/AVTZ) for DF\textperiodcentered{}\textperiodcentered{}\textperiodcentered{}CH\textsubscript{3} complexes, and from 1.3\% (CCSD/AVDZ) to 3.4\% (MP2/AVDZ) for DF\textperiodcentered{}\textperiodcentered{}\textperiodcentered{}NH\textsubscript{3} complexes (\mbox{\textbf{TABLE \ref{tab:VI}}}). MP2/AVDZ giving the largest MA\%D on F-bond distances of DF\textperiodcentered{}\textperiodcentered{}\textperiodcentered{}NH\textsubscript{3} complexes here is consistent with an independent benchmark  study on closed-shell halogen-bonded complexes, where MP2 intermolecular distances also showed the largest average error (2.3\%) among all methods tested apart from HF (10.5\%), relative to CCSD(T).\cite{Bauza2013} MP2/AVDZ systematically underestimates the F-bond distance for DF\textperiodcentered{}\textperiodcentered{}\textperiodcentered{}NH\textsubscript{3} complexes (MSE = -0.102 Å; \mbox{\textbf{TABLE \ref{tab:VI}}}), most notably for F\textsubscript{2}\textperiodcentered{}\textperiodcentered{}\textperiodcentered{}NH\textsubscript{3} (R(F\textperiodcentered{}\textperiodcentered{}\textperiodcentered{}N) = 2.442 Å vs. the reference 2.616 Å) (see \mbox{\textbf{Table \ref{tab:S6}}} for F-bond distances and \mbox{\textbf{FIGS. \ref{fig:5}a}} and \mbox{\textbf{\ref{fig:5}b}} for deviation and percentage deviations on selected FB complexes), though this single complex accounts for only a small part of the overall distance error (MA\%D drops from 3.4\% to 3.1\% when F\textsubscript{2}\textperiodcentered{}\textperiodcentered{}\textperiodcentered{}NH\textsubscript{3} is excluded). All-electron correlation, instead of valence-electron treatment to the MPn (n = 2, 3, 4), CCSD, or QCISD methods, have minor effect for predicting F-bond distances of either DF\textperiodcentered{}\textperiodcentered{}\textperiodcentered{}CH\textsubscript{3} or DF\textperiodcentered{}\textperiodcentered{}\textperiodcentered{}NH\textsubscript{3} complexes (\mbox{\textbf{TABLE \ref{tab:VI}}} and \mbox{\textbf{FIG. \ref{fig:4}}}). Here, we need to mention that there is apparent except with MP2 and MP2(full) in the case of DF\textperiodcentered{}\textperiodcentered{}\textperiodcentered{}NH\textsubscript{3} complexes (3.4\% vs 2.1\%; \mbox{\textbf{TABLE \ref{tab:VI}}}). In particular, the optimized F-bond distance for F\textsubscript{2}..NH\textsubscript{3} complex calculated at MP2(full)/AVTZ level of theory is 2.561 Å, which is closer to the reference value of 2.616 Å than that calculated at MP2/AVDZ level of theory (R(F\textperiodcentered{}\textperiodcentered{}\textperiodcentered{}N) = 2.442 Å).

\begin{table}[!htbp]\centering\setlength{\parskip}{0pt}
\caption{MSE,\textbf{ }MAD, RMSD, MAX and MA\%D on F-bond distances R(F\textperiodcentered{}\textperiodcentered{}\textperiodcentered{}C/\allowbreak{}N) (Å) for DF\textperiodcentered{}\textperiodcentered{}\textperiodcentered{}CH\textsubscript{3}/NH\textsubscript{3} complexes using different \emph{ab}\emph{ }\emph{initio} methods. F-bond distances obtained from DF-RMP2-F12/\allowbreak{}VDZ-F12 level were used as reference values. See \mbox{\textbf{Table \ref{tab:S6}}} for F-bond distances from all \emph{ab initio} methods used.}\label{tab:VI}\vspace{7pt}\nopagebreak
\adjustbox{width=1.08\linewidth,max totalheight=0.78\textheight,center}{\fontfamily{ptm}\fontsize{9.3}{10.7}\selectfont\renewcommand{\arraystretch}{1.12}\setlength{\tabcolsep}{1.5pt}
\begin{tabular}{|>{\centering\arraybackslash}m{132.0pt}|>{\centering\arraybackslash}m{31.7pt}|>{\centering\arraybackslash}m{29.0pt}|>{\centering\arraybackslash}m{29.0pt}|>{\centering\arraybackslash}m{29.7pt}|>{\centering\arraybackslash}m{17.1pt}|>{\centering\arraybackslash}m{132.0pt}|>{\centering\arraybackslash}m{29.7pt}|>{\centering\arraybackslash}m{29.0pt}|>{\centering\arraybackslash}m{28.0pt}|>{\centering\arraybackslash}m{35.4pt}|>{\centering\arraybackslash}b{22.2pt}|}
\hline
\multicolumn{6}{|>{\centering\arraybackslash}m{283.5pt}|}{{\bfseries\boldmath \textbf{DF\textperiodcentered{}\textperiodcentered{}\textperiodcentered{}CH}\textbf{\textsubscript{3}}}} &\multicolumn{6}{>{\centering\arraybackslash}m{258.5pt}|}{{\bfseries\boldmath \textbf{DF\textperiodcentered{}\textperiodcentered{}\textperiodcentered{}NH}\textbf{\textsubscript{3}}}}   \\
\hline
\multicolumn{1}{|>{\centering\arraybackslash}m{132.0pt}|}{{\bfseries\boldmath \emph{\textbf{ab initio}}\textbf{ Methods}}} & {\bfseries\boldmath \makebox[\linewidth][c]{\rotatebox{90}{\rbxs{MSE}}}} & {\bfseries\boldmath \makebox[\linewidth][c]{\rotatebox{90}{\rbxs{MAD}}}} & {\bfseries\boldmath \makebox[\linewidth][c]{\rotatebox{90}{\rbxs{RMSD}}}} & {\bfseries\boldmath \makebox[\linewidth][c]{\rotatebox{90}{\rbxs{MAX}}}} & {\bfseries\boldmath \makebox[\linewidth][c]{\rotatebox{90}{\rbxs{MA\%D}}}} & {\bfseries\boldmath \emph{\textbf{ab initio}}\textbf{ Methods}} & {\bfseries\boldmath \makebox[\linewidth][c]{\rotatebox{90}{\rbxs{MSE}}}} & {\bfseries\boldmath \makebox[\linewidth][c]{\rotatebox{90}{\rbxs{MAD}}}} & {\bfseries\boldmath \makebox[\linewidth][c]{\rotatebox{90}{\rbxs{RMSD}}}} & {\bfseries\boldmath \makebox[\linewidth][c]{\rotatebox{90}{\rbxs{MAX}}}} &\multicolumn{1}{>{\centering\arraybackslash}m{22.2pt}|}{{\bfseries\boldmath \makebox[\linewidth][c]{\rotatebox{90}{\rbxs{MA\%D}}}}}   \\
\hline
\end{tabular}}

\end{table}

\begin{table}[!htbp]\captionsetup{type=figure}\centering
\par\vspace{2pt}\noindent\hfil\includegraphics[width=\linewidth]{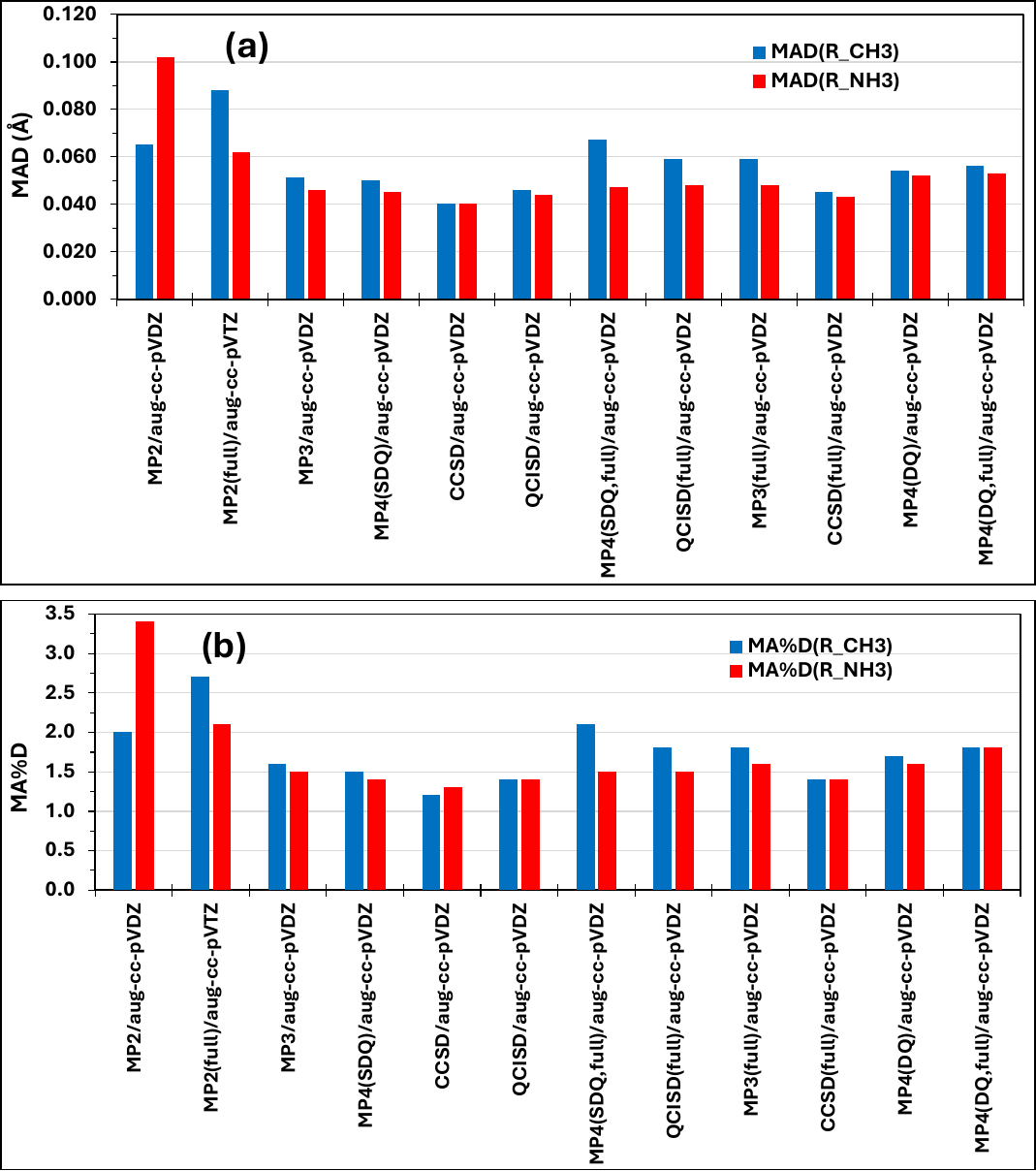}\hfil\null\par\nointerlineskip\vspace{13pt}
\par\vspace{-1pt}\caption{MADs \textbf{(a)} and MA\%Ds \textbf{(b)} on F-bond distance R(F\textperiodcentered{}\textperiodcentered{}\textperiodcentered{}C/N) (Å) for DF\textperiodcentered{}\textperiodcentered{}\textperiodcentered{}CH\textsubscript{3}/NH\textsubscript{3} complexes using different \emph{ab initio} methods, arranged in increasing order of MADs on BSSE-corrected BEs for DF\textperiodcentered{}\textperiodcentered{}\textperiodcentered{}CH\textsubscript{3} complexes. Bond distances obtained at DF-RMP2-F12/VDZ-F12 level were used as reference values. F-bond distances calculated at CCSD/AVDZ level are closer to the reference values for both types of complexes DF\textperiodcentered{}\textperiodcentered{}\textperiodcentered{}CH\textsubscript{3}/NH\textsubscript{3}.}\label{fig:4}
\vspace{10pt}
\end{table}

\FloatBarrier
\subsection{Overall performance of Minnesota DFAs and conventional \emph{ab initio} methods for BEs and F-bond distances: selected systems}\label{sec:4.3}

In \textbf{Section \ref*{sec:4.1}}, we discussed errors on BEs and F-bond distances for individual complexes using seven Minnesota DFAs, along with the various statistical errors. \textbf{Section \ref*{sec:4.2}} assessed several \emph{ab initio} methods using average statistical errors. Here, we compare the same set of Minnesota DFAs and \emph{ab initio} methods for individual complexes, focusing on selected DF\textperiodcentered{}\textperiodcentered{}\textperiodcentered{}CH\textsubscript{3}/NH\textsubscript{3} complexes (DF = F\textsubscript{2}, NCF, CNF, NH\textsubscript{3}F\textsuperscript{+}, and NH\textsubscript{2}F\textsubscript{2}\textsuperscript{+}) (\mbox{\textbf{TABLES \ref{tab:VII}}} and \mbox{\textbf{\ref{tab:VIII}}}, and \mbox{\textbf{FIG. \ref{fig:5}}}). These complexes were also used in companion paper I to validate the reference DF-RMP2-F12/CBS BEs and F-bond distances obtained at DF-RMP2-F12/VDZ-F12.

\subsubsection{Deviation on BEs and F-bond distances for selected complexes}\label{sec:4.3.A}

As already discussed in companion paper I, M06-2X/AVTZ shows an anomaly for several SEFB complexes DF\textperiodcentered{}\textperiodcentered{}\textperiodcentered{}CH\textsubscript{3}, predicting BEs comparable to those for the corresponding DF\textperiodcentered{}\textperiodcentered{}\textperiodcentered{}NH\textsubscript{3} complexes despite their different electronic character. M08-HX/AVTZ resolves this anomaly (see \mbox{\textbf{TABLES \ref{tab:I}}}, \mbox{\textbf{\ref{tab:II}}} and \mbox{\textbf{FIG. \ref{fig:1}}}). DF-RMP2-F12 is not included in analysis, since its good performance here is expected and was independently cross-validated in companion paper I.

No single method is the best for BEs of all five selected DF\textperiodcentered{}\textperiodcentered{}\textperiodcentered{}CH\textsubscript{3} complexes. The least values of \ensuremath{\delta}\textsubscript{BE}(\%\ensuremath{\delta}\textsubscript{BE}) are -0.9 kJ/mol (-26.5\%) {[}M08-HX{]}, -0.3 kJ/mol (-17.6\%) {[}M05-2X, MP2(full){]}, 0.0 kJ/mol (0.0\%) {[}M08-HX{]}, 0.4 kJ/mol (4.1\%) {[}M11{]} for SEFB complexes formed with F\textsubscript{2}, NCF, CNF, NH\textsubscript{3}F\textsuperscript{+}, and NH\textsubscript{2}F\textsubscript{2}\textsuperscript{+}, respectively and the corresponding results in the case CSFB complexes are 0.3 kJ/mol (2.6\%) {[}M11{]}, for F\textsubscript{2}, NCF, CNF, NH\textsubscript{3}F\textsuperscript{+}, and NH\textsubscript{2}F\textsubscript{2}\textsuperscript{+}, respectively. The same holds for DF\textperiodcentered{}\textperiodcentered{}\textperiodcentered{}NH\textsubscript{3}, where the least values are -0.2 kJ/mol (-2.6\%) {[}MP2{]}, 0.1 kJ/mol (2.0\%) {[}M05-2X, M06-2X-GD3{]}, 0.3 kJ/mol (4.3\%) {[}M08-HX{]}, -0.3 kJ/mol (-0.9\%) {[}M05-2X-GD3{]}, and 0.2 kJ/mol (0.5\%) {[}M05-2X, M06-2X{]} (see \mbox{\textbf{TABLE \ref{tab:VII}}}, and \mbox{\textbf{FIG. \ref{fig:5}(a) and FIGs. \ref{fig:5}(b)}}). These results show that the best-performing method varies from one complex to another, even within this small, closely related set of F-bonded complexes.

\begin{table}[!htbp]\centering\setlength{\parskip}{0pt}
\caption{Deviations on BEs (kJ/\allowbreak{}mol), \ensuremath{\delta}\textsubscript{BE} = BE(method) - reference BE, and\textbf{ }percentage deviations on BEs, \%\ensuremath{\delta}\textsubscript{BE} = [100 \texttimes{} \ensuremath{\delta}\textsubscript{BE}]/(- reference BE), for selected DF\textperiodcentered{}\textperiodcentered{}\textperiodcentered{}CH\textsubscript{3}/NH\textsubscript{3} complexes (DF = F\textsubscript{2}, NCF, CNF, NH\textsubscript{3}F\textsuperscript{+} and NH\textsubscript{2}F\textsubscript{2}\textsuperscript{+}), calculated by using seven Minnesota DFAs (M08-HX, M05-2X, M05-2X-GD3, M06-2X, M06-2X-GD3, M11 and M06-HF) that worked for the F\textsubscript{2}\textperiodcentered{}\textperiodcentered{}\textperiodcentered{}CH\textsubscript{3}, and several \emph{ab initio} methods. The reference BEs (BEs) for corresponding complexes were calculated at DF-RMP2-F12/\allowbreak{}CBS[V(T, Q)Z-F12]//DF-RMP2-F12/\allowbreak{}VDZ-F12 level. Least values of \ensuremath{\delta}\textsubscript{BE} and \%\ensuremath{\delta}\textsubscript{BE} are in bold (DF-RMP2-F12 with different basis sets were excluded from this selection as DF-RMP2-F12 is expected to perform well). Also, the maximum values\textsubscript{ }are italicized. See \mbox{\textbf{TABLE \ref{tab:I}}}\textbf{ }and \mbox{\textbf{FIG. \ref{fig:1}}} for \ensuremath{\delta}\textsubscript{BE} and \%\ensuremath{\delta}\textsubscript{BE }for all complexes using Minnesota functionals listed here.}\label{tab:VII}\vspace{7pt}\nopagebreak
\adjustbox{width=1.08\linewidth,max totalheight=0.78\textheight,center}{\fontfamily{ptm}\fontsize{9.3}{10.7}\selectfont\renewcommand{\arraystretch}{1.0}\setlength{\tabcolsep}{4.0pt}
\begin{tabular}{|>{\centering\arraybackslash}m{109.7pt}|>{\centering\arraybackslash}m{41.4pt}|>{\centering\arraybackslash}m{31.0pt}|>{\centering\arraybackslash}m{33.6pt}|>{\centering\arraybackslash}m{30.9pt}|>{\centering\arraybackslash}m{36.2pt}|>{\centering\arraybackslash}m{41.9pt}|>{\centering\arraybackslash}m{27.6pt}|>{\centering\arraybackslash}m{36.3pt}|>{\centering\arraybackslash}m{34.7pt}|>{\centering\arraybackslash}m{35.0pt}|}
\hline
\multicolumn{11}{|>{\centering\arraybackslash}m{488.3pt}|}{{\bfseries\boldmath \textbf{DF\textperiodcentered{}\textperiodcentered{}\textperiodcentered{}CH}\textbf{\textsubscript{3}}}}   \\
\hline
\multicolumn{1}{|>{\centering\arraybackslash}m{109.7pt}|}{\multirow{2}{109.7pt}{{\bfseries\boldmath \centering Methods}}} &\multicolumn{2}{>{\centering\arraybackslash}m{75.3pt}|}{{\bfseries\boldmath F\textsubscript{2}\textperiodcentered{}\textperiodcentered{}\textperiodcentered{}CH\textsubscript{3}}} &\multicolumn{2}{>{\centering\arraybackslash}m{67.5pt}|}{{\bfseries\boldmath NCF\textperiodcentered{}\textperiodcentered{}\textperiodcentered{}CH\textsubscript{3}}} &\multicolumn{2}{>{\centering\arraybackslash}m{81.1pt}|}{{\bfseries\boldmath CNF\textperiodcentered{}\textperiodcentered{}\textperiodcentered{}CH\textsubscript{3}}} &\multicolumn{2}{>{\centering\arraybackslash}m{66.9pt}|}{{\bfseries\boldmath NH\textsubscript{3}F\textsuperscript{+}\textperiodcentered{}\textperiodcentered{}\textperiodcentered{}CH\textsubscript{3}}} &\multicolumn{2}{>{\centering\arraybackslash}m{72.6pt}|}{{\bfseries\boldmath NH\textsubscript{2}F\textsubscript{2}\textsuperscript{+}\textperiodcentered{}\textperiodcentered{}\textperiodcentered{}CH\textsubscript{3}}}   \\
\hline
\end{tabular}}

\end{table}

\begin{table}[!htbp]\captionsetup{type=figure}\centering
\par\vspace{2pt}\noindent\hfil\includegraphics[width=\linewidth]{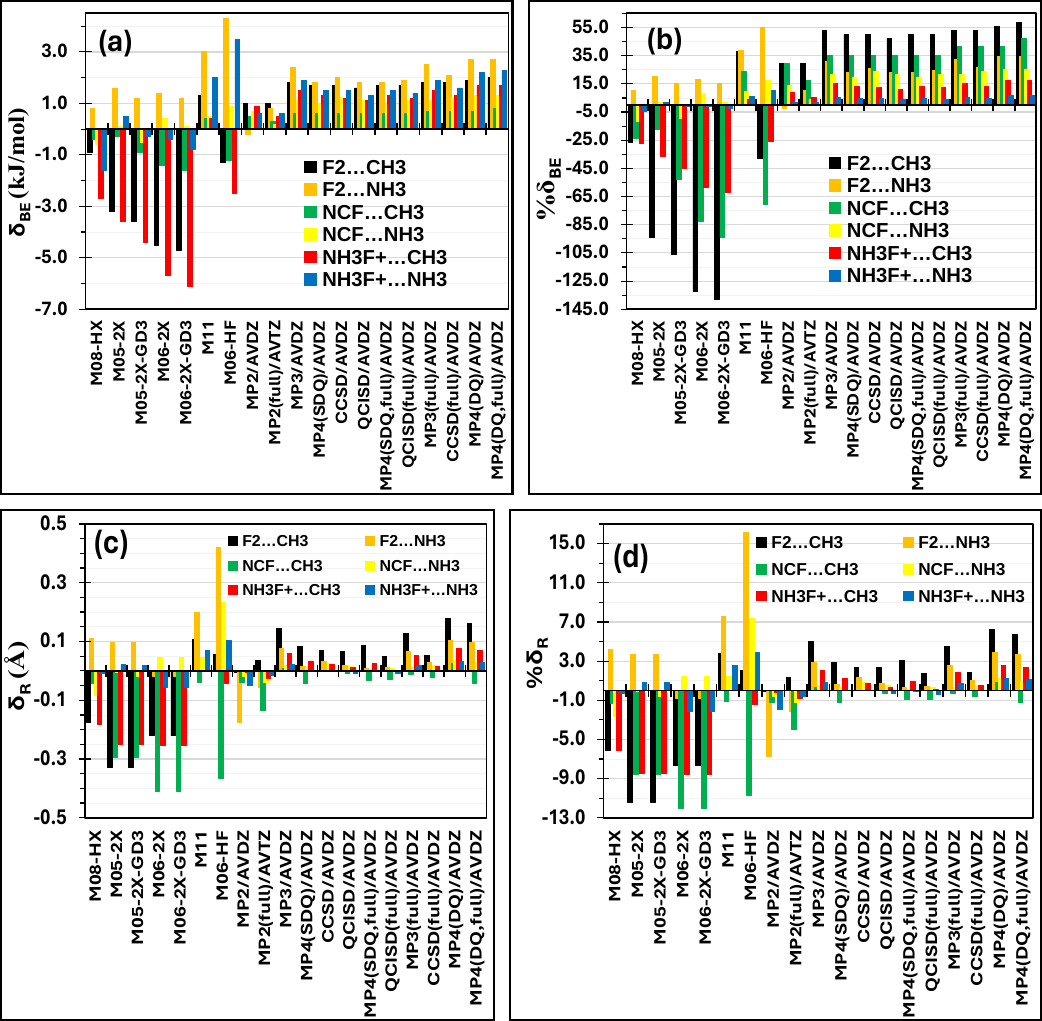}\hfil\null\par\nointerlineskip\vspace{13pt}
\par\vspace{-1pt}\caption{Deviations (\ensuremath{\delta}\textsubscript{BE}) \textbf{(a)} and percentage deviations (\%\ensuremath{\delta}\textsubscript{BE}) \textbf{(b)} on BEs (kJ/mol) and deviations (\ensuremath{\delta}\textsubscript{R}) \textbf{(c)} and percentage deviations (\%\ensuremath{\delta}\textsubscript{R}) \textbf{(d)} on F-bond distances (Å) for selected F-bonded complexes DF\textperiodcentered{}\textperiodcentered{}\textperiodcentered{}CH\textsubscript{3}/NH\textsubscript{3} (DF = F\textsubscript{2}, NCF, NH\textsubscript{3}F\textsuperscript{+}) calculated at seven Minnesota functionals that worked for the F\textsubscript{2}\textperiodcentered{}\textperiodcentered{}\textperiodcentered{}CH\textsubscript{3} (M08-HX, M05-2X, M05-2X-GD3, M06-2X, M06-2X-GD3, M11 and M06-HF) with AVTZ basis set, and several \emph{ab initio} methods. BEs calculated at \emph{ab initio} methods are BSSE-corrected. The reference BEs and reference F-bond distances were calculated at DF-RMP2-F12/CBS{[}V(T, Q)Z-F12{]}//DF-RMP2-F12/VDZ-F12 and DF-RMP2-F12/VDZ-F12 levels of theory, respectively.}\label{fig:5}
\vspace{10pt}
\end{table}

\begin{table}[!htbp]\centering\setlength{\parskip}{0pt}
\caption{Deviations on F-bond distances, \ensuremath{\delta}\textsubscript{R }= R(method) -- reference R (Å) and percentage deviation on F-bond distances, \%\ensuremath{\delta}\textsubscript{R} = [100 \texttimes{} \ensuremath{\delta}\textsubscript{R}]/(reference R) for selected DF\textperiodcentered{}\textperiodcentered{}\textperiodcentered{}CH\textsubscript{3}/NH\textsubscript{3} complexes (DF = F\textsubscript{2}, NCF, CNF, NH\textsubscript{3}F\textsuperscript{+} and NH\textsubscript{2}F\textsubscript{2}\textsuperscript{+}) calculated at several Minnesota functionals that worked for the F\textsubscript{2}\textperiodcentered{}\textperiodcentered{}\textperiodcentered{}CH\textsubscript{3} (M08-HX, M05-2X, M05-2X-GD3, M06-2X, M06-2X-GD3, M11 and M06-HF) and several \emph{ab initio} methods. The reference F-bond distances R(F\textperiodcentered{}\textperiodcentered{}\textperiodcentered{}C/\allowbreak{}N) were calculated at DF-RMP2-F12/\allowbreak{}VDZ-F12 level. The \%\ensuremath{\delta}R values that are above 5.0\% in magnitude are in red. Least values of \ensuremath{\delta}R and \%\ensuremath{\delta}R are in bold. Maximum values are italicized. See \mbox{\textbf{TABLE \ref{tab:II}}} and \mbox{\textbf{FIG. \ref{fig:1}}} for \ensuremath{\delta}\textsubscript{R} and \%\ensuremath{\delta}\textsubscript{R }for all complexes calculated from various Minnesota functionals listed here.}\label{tab:VIII}\vspace{7pt}\nopagebreak
\adjustbox{width=1.08\linewidth,max totalheight=0.78\textheight,center}{\fontfamily{ptm}\fontsize{9.3}{10.7}\selectfont\renewcommand{\arraystretch}{1.0}\setlength{\tabcolsep}{4.0pt}
\begin{tabular}{|>{\centering\arraybackslash}m{111.8pt}|>{\centering\arraybackslash}m{44.5pt}|>{\centering\arraybackslash}m{31.0pt}|>{\centering\arraybackslash}m{35.4pt}|>{\centering\arraybackslash}m{31.0pt}|>{\centering\arraybackslash}m{38.5pt}|>{\centering\arraybackslash}m{38.1pt}|>{\centering\arraybackslash}m{34.2pt}|>{\centering\arraybackslash}m{33.9pt}|>{\centering\arraybackslash}m{35.9pt}|>{\centering\arraybackslash}m{31.9pt}|}
\hline
\multicolumn{11}{|>{\centering\arraybackslash}m{496.2pt}|}{{\bfseries\boldmath \textbf{DF\textperiodcentered{}\textperiodcentered{}\textperiodcentered{}CH}\textbf{\textsubscript{3}}}}   \\
\hline
\multicolumn{1}{|>{\centering\arraybackslash}m{111.8pt}|}{\multirow{2}{111.8pt}{{\bfseries\boldmath \centering Methods}}} &\multicolumn{2}{>{\centering\arraybackslash}m{78.5pt}|}{{\bfseries\boldmath F\textsubscript{2}\textperiodcentered{}\textperiodcentered{}\textperiodcentered{}CH\textsubscript{3}}} &\multicolumn{2}{>{\centering\arraybackslash}m{69.4pt}|}{{\bfseries\boldmath NCF\textperiodcentered{}\textperiodcentered{}\textperiodcentered{}CH\textsubscript{3}}} &\multicolumn{2}{>{\centering\arraybackslash}m{79.7pt}|}{{\bfseries\boldmath CNF\textperiodcentered{}\textperiodcentered{}\textperiodcentered{}CH\textsubscript{3}}} &\multicolumn{2}{>{\centering\arraybackslash}m{71.1pt}|}{{\bfseries\boldmath NH\textsubscript{3}F\textsuperscript{+}\textperiodcentered{}\textperiodcentered{}\textperiodcentered{}CH\textsubscript{3}}} &\multicolumn{2}{>{\centering\arraybackslash}m{70.8pt}|}{{\bfseries\boldmath NH\textsubscript{2}F\textsubscript{2}\textsuperscript{+}\textperiodcentered{}\textperiodcentered{}\textperiodcentered{}CH\textsubscript{3}}}   \\
\hline
\end{tabular}}

\end{table}

M06-2X-GD3/AVTZ gives the largest \ensuremath{\delta}\textsubscript{BE}(\%\ensuremath{\delta}\textsubscript{BE}) for all five selected DF\textperiodcentered{}\textperiodcentered{}\textperiodcentered{}CH\textsubscript{3} complexes, making it unsuitable, even worse than M06-2X itself, for these SEFB complexes (see italicized numbers in \mbox{\textbf{TABLE \ref{tab:VII}}}). M06-HF/AVTZ is the corresponding worst DFA for the selected DF\textperiodcentered{}\textperiodcentered{}\textperiodcentered{}NH\textsubscript{3} complexes (except NCF) (see \mbox{\textbf{TABLE \ref{tab:VII}}}). M05-2X/AVTZ and its GD3 counterpart also give larger errors on BEs and F-bond distances for this selected SEFB complexes than their overall good performance across all 15 DF\textperiodcentered{}\textperiodcentered{}\textperiodcentered{}CH\textsubscript{3} complexes would suggest (\mbox{\textbf{TABLE \ref{tab:III}}}).

F-bond distances show the same lack of a single best method for the selected DF\textperiodcentered{}\textperiodcentered{}\textperiodcentered{}CH\textsubscript{3}/NH\textsubscript{3} complexes. The least values of \ensuremath{\delta}\textsubscript{R}(\%\ensuremath{\delta}\textsubscript{R}) for the selected DF\textperiodcentered{}\textperiodcentered{}\textperiodcentered{}CH\textsubscript{3} complexes are -0.004 Å (-0.1\%) {[}MP2/AVDZ{]}, 0.0 Å (0.0\%) {[}CCSD/AVDZ{]}, 0.009 Å (0.3\%) {[}MP4(DQ)/AVDZ{]}, 0.001 Å (0.0\%) {[}M11/AVTZ{]}, and 0.003 Å (0.1\%) {[}CCSD(full)/AVDZ{]}, for F-bond donors F\textsubscript{2}, NCF, CNF, NH\textsubscript{3}F\textsuperscript{+}, and NH\textsubscript{2}F\textsubscript{2}\textsuperscript{+}, respectively (\mbox{\textbf{TABLE \ref{tab:VIII}}}, and \mbox{\textbf{FIGs. \ref{fig:5}(c)}} and \mbox{\textbf{\ref{fig:5}(d)}}). For the corresponding DF\textperiodcentered{}\textperiodcentered{}\textperiodcentered{}NH\textsubscript{3} complexes, they are 0.009 Å (0.3\%) {[}MP4(SDQ,full)/AVDZ{]}, 0.004 Å (0.1\%) {[}MP4(SDQ,full)/AVDZ{]}, 0.002 Å (0.1\%) {[}M08-HX/AVTZ{]}, -0.001 Å (0.0\%) {[}CCSD/AVDZ, MP4(SDQ)/AVDZ{]}, and 0.001 Å (0.0\%) {[}M08-HX/AVTZ{]}. M05-2X/AVTZ   and M06-2X/AVTZ, with and without GD3, yield \%\ensuremath{\delta}\textsubscript{R} above 5.0\% for the selected DF\textperiodcentered{}\textperiodcentered{}\textperiodcentered{}CH\textsubscript{3} complexes (\mbox{\textbf{TABLE \ref{tab:VIII}}}). Both stay below 4.0\% for the selected DF\textperiodcentered{}\textperiodcentered{}\textperiodcentered{}NH\textsubscript{3} complexes, matching their performance across the whole set of DF\textperiodcentered{}\textperiodcentered{}\textperiodcentered{}NH\textsubscript{3} complexes (\mbox{\textbf{TABLE \ref{tab:IV}}}), indicating that both of these methods are suitable for F-bond distances of these complexes. M08-HX/AVTZ, M05-2X/AVTZ, M05-2X-GD3/AVTZ, M06-2X/AVTZ, and M06-2X-GD3/AVTZ perform very well for F-bond distances for the five selected DF\textperiodcentered{}\textperiodcentered{}\textperiodcentered{}NH\textsubscript{3} complexes (\mbox{\textbf{TABLE \ref{tab:VIII}}}, \mbox{\textbf{FIGs. \ref{fig:5}(c)}} and \mbox{\textbf{\ref{fig:5}(d)}}). M06-HF/AVTZ shows a large difference between the two complex types, with 2.0\% for F\textsubscript{2}\textperiodcentered{}\textperiodcentered{}\textperiodcentered{}CH\textsubscript{3} but 16.1\% for F\textsubscript{2}\textperiodcentered{}\textperiodcentered{}\textperiodcentered{}NH\textsubscript{3}. M08-HX itself performs reasonably well for F-bond distances of both types of complexes, consistent with its reliability for BEs noted above.

\FloatBarrier
\subsection{Best performing Minnesota DFAs and conventional \emph{ab initio} methods}\label{sec:4.4}

Here, we put together all statistical measures (MSE, MAD, RMSD, MAX and MA\%D) for BEs and F-bond distances of the whole SEFB-15 and CSFB-14 databases (15 DF\textperiodcentered{}\textperiodcentered{}\textperiodcentered{}CH\textsubscript{3} and 14 DF\textperiodcentered{}\textperiodcentered{}\textperiodcentered{}NH\textsubscript{3} complexes), obtained from the seven Minnesota DFAs that worked for F\textsubscript{2}\textperiodcentered{}\textperiodcentered{}\textperiodcentered{}CH\textsubscript{3} and several \emph{ab initio} methods (\mbox{\textbf{TABLE \ref{tab:IX}}} and \mbox{\textbf{FIG. \ref{fig:6}}}). The methods are arranged in increasing order of MAD, separately for BEs and F-bond distances and for each database. As in \textbf{Section \ref*{sec:4.2}}, the statistics are over the converged complexes, which are 14 of the 15 DF\textperiodcentered{}\textperiodcentered{}\textperiodcentered{}CH3 complexes for M11 and M06-HF, and 13 of the 14 DF\textperiodcentered{}\textperiodcentered{}\textperiodcentered{}NH3 complexes for M05-2X-GD3 and M06-HF (\mbox{\textbf{TABLES \ref{tab:III}}} and \mbox{\textbf{\ref{tab:IV}}}).

\begin{table}[!htbp]\centering\setlength{\parskip}{0pt}
\caption{Overall performance of seven Minnesota DFAs that worked for F\textsubscript{2}\textperiodcentered{}\textperiodcentered{}\textperiodcentered{}CH\textsubscript{3} (M08-HX, M05-2X, M05-2X-GD3, M06-2X, M06-2X-GD3, M11 and M06-HF) [out of those seven, only five (M08-HX, M05-2X, M05-2X-GD3, M06-2X and M06-2X-GD3) worked for all 15 DF\textperiodcentered{}\textperiodcentered{}\textperiodcentered{}CH\textsubscript{3} complexes], and several \emph{ab initio} methods in terms of MSE, MAD, RMSD, MAX and MA\%D on BEs (kJ/\allowbreak{}mol) and F-bond distances (Å) for whole set of F-bonded complexes (15 DF\textperiodcentered{}\textperiodcentered{}\textperiodcentered{}CH\textsubscript{3} and 14 DF\textperiodcentered{}\textperiodcentered{}\textperiodcentered{}NH\textsubscript{3}). The BEs calculated at \emph{ab initio} methods are BSSE-corrected. The reference BEs and F-bond distances were calculated at DF-RMP2-F12/\allowbreak{}CBS[V(T,Q)Z-F12]//DF-RMP2-F12/\allowbreak{}VDZ-F12 and DF-RMP2-F12/\allowbreak{}VDZ-F12 levels of theory, respectively. See \mbox{\textbf{Tables \ref{tab:I}}} and \mbox{\textbf{\ref{tab:II}}} and \mbox{\textbf{FIG. \ref{fig:1} }}for \ensuremath{\delta}\textsubscript{BE}, \ensuremath{\delta}\textsubscript{R}, \%\ensuremath{\delta}\textsubscript{BE }and\textsubscript{ }\%\ensuremath{\delta}\textsubscript{R} for all the complexes calculated by using Minnesota functionals listed here.\textsuperscript{a}}\label{tab:IX}\vspace{7pt}\nopagebreak
\adjustbox{width=1.08\linewidth,max totalheight=0.78\textheight,center}{\fontfamily{ptm}\fontsize{9.7}{11.2}\selectfont\renewcommand{\arraystretch}{1.0}\setlength{\tabcolsep}{4.0pt}
\begin{tabular}{|>{\centering\arraybackslash}m{114.0pt}|>{\centering\arraybackslash}m{32.5pt}|>{\centering\arraybackslash}m{32.9pt}|>{\centering\arraybackslash}m{30.7pt}|>{\centering\arraybackslash}m{29.6pt}|>{\centering\arraybackslash}m{35.4pt}|>{\centering\arraybackslash}m{114.0pt}|>{\centering\arraybackslash}m{29.5pt}|>{\centering\arraybackslash}m{29.4pt}|>{\centering\arraybackslash}m{30.7pt}|>{\centering\arraybackslash}m{29.6pt}|>{\centering\arraybackslash}m{32.1pt}|}
\hline
\multicolumn{6}{|>{\centering\arraybackslash}m{285.1pt}|}{{\bfseries\boldmath {\fontsize{13}{16}\selectfont\rule{0pt}{17pt}\textbf{Error statistics on }\textbf{BEs}\textbf{ for DF\textperiodcentered{}\textperiodcentered{}\textperiodcentered{}CH}\textbf{\textsubscript{3}}\rule[-7pt]{0pt}{7pt}}}} &\multicolumn{6}{>{\centering\arraybackslash}m{256.9pt}|}{{\bfseries\boldmath {\fontsize{13}{16}\selectfont\rule{0pt}{17pt}\textbf{Error statistics on }\textbf{BEs}\textbf{ for DF\textperiodcentered{}\textperiodcentered{}\textperiodcentered{}NH}\textbf{\textsubscript{3}}\rule[-7pt]{0pt}{7pt}}}}   \\
\hline
\multicolumn{1}{|>{\centering\arraybackslash}m{114.0pt}|}{{\bfseries\boldmath Methods}} & {\bfseries\boldmath MSE} & {\bfseries\boldmath MAD} & {\bfseries\boldmath RMSD} & {\bfseries\boldmath MAX} & {\bfseries\boldmath MA\%D} & {\bfseries\boldmath Methods} & {\bfseries\boldmath MSE} & {\bfseries\boldmath MAD} & {\bfseries\boldmath RMSD} & {\bfseries\boldmath MAX} & {\bfseries\boldmath MA\%D}   \\
\hline
\multicolumn{1}{|>{\centering\arraybackslash}m{114.0pt}|}{DF-RMP2-F12/ VQZ-F12} & 0.2 & 0.2 & 0.2 & 0.3 & 7.9 & DF-RMP2-F12/ VQZ-F12 & 0.1 & 0.1 & 0.1 & 0.3 & 2.3  \\
\hline
\multicolumn{1}{|>{\centering\arraybackslash}m{114.0pt}|}{DF-RMP2-F12/ VTZ-F12} & 0.2 & 0.2 & 0.2 & 0.3 & 9.7 & DF-RMP2-F12/ VDZ-F12 & 0.1 & 0.1 & 0.2 & 0.4 & 3.4  \\
\hline
\multicolumn{1}{|>{\centering\arraybackslash}m{114.0pt}|}{DF-RMP2-F12/ VDZ-F12} & 0.3 & 0.3 & 0.3 & 0.4 & 13.1 & DF-RMP2-F12/ VTZ-F12 & 0.2 & 0.2 & 0.2 & 0.4 & 4.2  \\
\hline
\end{tabular}}
\par\nopagebreak\vspace{3pt}{\setstretch{1}\setlength{\parskip}{1.5pt}\fontsize{9}{10.5}\selectfont\raggedright\noindent \textsuperscript{a}Methods are arranged according to increasing order of MADs on BEs and F-bond distances separately for DF\textperiodcentered{}\textperiodcentered{}\textperiodcentered{}CH\textsubscript{3} and DF\textperiodcentered{}\textperiodcentered{}\textperiodcentered{}NH\textsubscript{3} complexes.\par}

\end{table}

\begin{table}[!htbp]\captionsetup{type=figure}\centering
\par\vspace{2pt}\noindent\hfil\includegraphics[width=\linewidth]{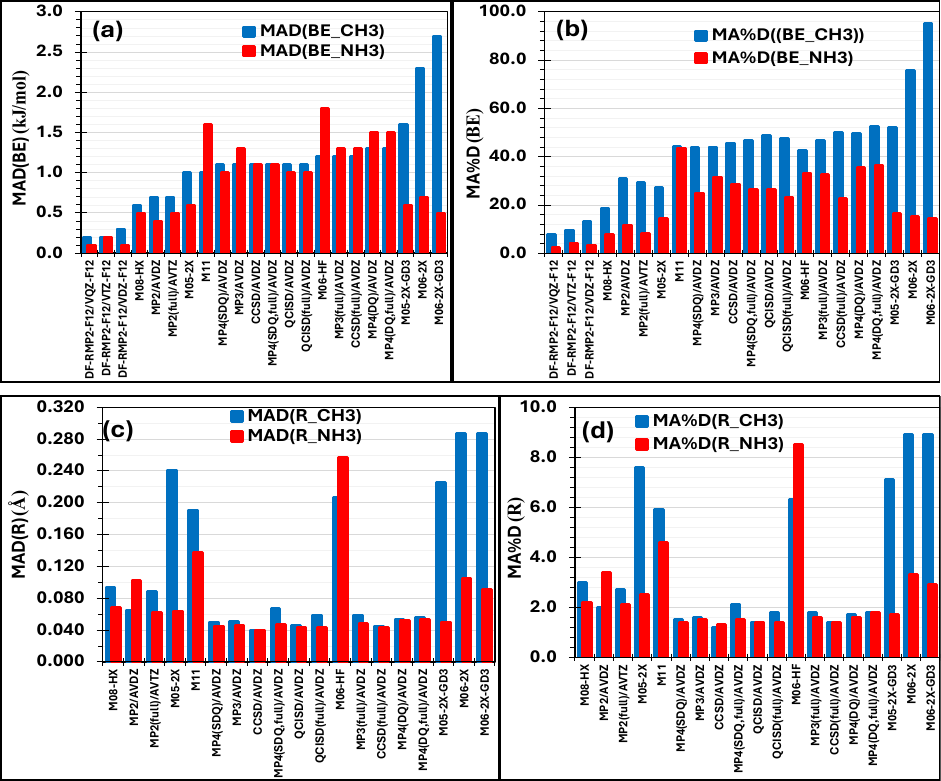}\hfil\null\par\nointerlineskip\vspace{13pt}
\par\vspace{-1pt}\caption{MADs on BEs (kJ/mol) \textbf{(a)}, MA\%Ds for the same \textbf{(b),} MADs on F-bond distances (Å) \textbf{(c)} and MA\%Ds for the same \textbf{(d)}, for the whole set of DF\textperiodcentered{}\textperiodcentered{}\textperiodcentered{}CH\textsubscript{3}/NH\textsubscript{3} complexes (15 SEFB and 14 CSFB complexes) using seven Minnesota DFAs that worked for F\textsubscript{2}\textperiodcentered{}\textperiodcentered{}\textperiodcentered{}CH\textsubscript{3} (M08-HX, M05-2X, M05-2X-GD3, M06-2X, M06-2X-GD3, M06-HF and M11) with AVTZ basis set and several \emph{ab initio} methods. BEs calculated at \emph{ab initio} methods are BSSE-corrected. Methods are arranged according to increasing order of the MADs on BEs for the whole set of the SEFB complexes (DF\textperiodcentered{}\textperiodcentered{}\textperiodcentered{}CH\textsubscript{3}). The reference BEs and F-bond distances were calculated at DF-RMP2-F12/CBS{[}V(T,Q)Z-F12{]}//DF-RMP2-F12/VDZ-F12 and DF-RMP2-F12/VDZ-F12 levels of theory, respectively.}\label{fig:6}
\vspace{10pt}
\end{table}

From \mbox{\textbf{TABLE \ref{tab:IX}}}, the MAD (MA\%D) on BEs for the SEFB complexes ranges from 0.2 kJ/mol (7.9\%) for DF-RMP2-F12/VQZ-F12 to 2.7 kJ/mol (95.1\%) for M06-2X-GD3/AVTZ (\mbox{\textbf{FIGs. \ref{fig:6}(a)}} and \mbox{\textbf{\ref{fig:6}(b)}}). As expected, DF-RMP2-F12/VQZ-F12 is the best among all the methods for BEs of both the SEFB and CSFB complexes, with a MAD (MA\%D) of 0.1 kJ/mol (2.3\%) for the latter.

For the DF\textperiodcentered{}\textperiodcentered{}\textperiodcentered{}CH\textsubscript{3} complexes, M08-HX/AVTZ performs better than MP2/AVDZ and MP2(full)/AVTZ (MA\%D = 18.7\% against 31.1\% and 29.3\%). For the DF\textperiodcentered{}\textperiodcentered{}\textperiodcentered{}NH\textsubscript{3} complexes, it is comparable with them (7.7\% against 11.7\% and 8.2\%). The performance of M05-2X/AVTZ (27.2\%) is comparable to that of the two MP2 methods for the DF\textperiodcentered{}\textperiodcentered{}\textperiodcentered{}CH\textsubscript{3} complexes, and it is good for the DF\textperiodcentered{}\textperiodcentered{}\textperiodcentered{}NH\textsubscript{3} complexes (14.3\%) (\mbox{\textbf{TABLE \ref{tab:IX}}}). M11/AVTZ (44.2\%) is comparable with M06-HF/AVTZ, CCSD/AVDZ and MP3, MP4(SDQ) and QCISD with and without full-electron correlation (42.6-48.7\%) for the DF\textperiodcentered{}\textperiodcentered{}\textperiodcentered{}CH\textsubscript{3} complexes, but it performs badly for the DF\textperiodcentered{}\textperiodcentered{}\textperiodcentered{}NH\textsubscript{3} complexes (43.6\%). M05-2X-GD3/AVTZ (52.1\%) is close to CCSD(full)/AVDZ (50.0\%), MP4(DQ)/AVDZ (49.6\%) and MP4(DQ,full)/AVDZ (52.6\%) for the DF\textperiodcentered{}\textperiodcentered{}\textperiodcentered{}CH3 complexes. For the DF\textperiodcentered{}\textperiodcentered{}\textperiodcentered{}NH\textsubscript{3} complexes, it performs very well (16.6\%), comparable with M05-2X/AVTZ (14.3\%) and M06-2X/AVTZ (15.3\%). M06-2X/AVTZ performs very badly for the DF\textperiodcentered{}\textperiodcentered{}\textperiodcentered{}CH3 complexes (75.8\%) but very well for the DF\textperiodcentered{}\textperiodcentered{}\textperiodcentered{}NH3 complexes (15.3\%), where it is better than M11/AVTZ (43.6\%), M06-HF/AVTZ (33.2\%) and all the ab initio methods other than DF-RMP2-F12 and the two MP2 methods (22.7-36.5\%).

For F-bond distances of the SEFB complexes, all \emph{ab initio} methods (MA\%D = 1.2-2.7\%) perform better than all seven Minnesota DFAs (3.0-8.9\%) (\mbox{\textbf{TABLE \ref{tab:IX}}} and \mbox{\textbf{FIGs. \ref{fig:6}(c)}} and \mbox{\textbf{\ref{fig:6}(d)}}). Among the DFAs, only M08-HX/AVTZ has an average error below 5.0\%, and M06-2X/AVTZ and M06-2X-GD3/AVTZ are the worst of all the methods.

For the DF\textperiodcentered{}\textperiodcentered{}\textperiodcentered{}NH\textsubscript{3} complexes, M05-2X-GD3/AVTZ performs better than MP2(full)/AVTZ and MP2/AVDZ {[}MAD(MA\%D) = 0.050 Å (1.7\%) against 0.062 Å (2.1\%) and 0.102 Å (3.4\%){]}. MP2/AVDZ is also worse than M08-HX/AVTZ, M05-2X/AVTZ and M06-2X-GD3/AVTZ (2.2-2.9\%) and comparable with M06-2X/AVTZ {[}0.105 Å (3.3\%){]}. The worst DFA is M06-HF/AVTZ {[}0.257 Å (8.5\%){]}, followed by M11/AVTZ {[}0.137 Å (4.6\%){]}, and all the other methods reproduce the F-bond distances within 5.0\%.

The poor performance of most Minnesota DFAs for the SEFB complexes agrees with earlier findings for radical-molecule interactions. For TA13 database, none of the methods used reaches chemical accuracy for all types of radical complexes.\cite{Tentscher2013} Conventional DFAs, particularly those with little or no HF exchange, can overbind radical-molecule complexes because of erroneous charge transfer.\cite{Johnson2013} M06-2X, which performs well for closed-shell halogen bonding,\cite{Bauza2013,Kozuch2013,OteroDeLaRoza2014} and is recommended for the TA13 radical-molecule\cite{Tentscher2013} complexes in literature,\cite{Mardirossian2016} is worse than all the conventional \emph{ab initio} methods for BEs of the SEFB complexes (MA\%D = 75.8\% against 29.3-52.6\%) (\mbox{\textbf{TABLE \ref{tab:IX}}}). In contrast, M08-HX is better than MP2 and MP2(full) for the same set of the complexes (18.7\% against 31.1\% and 29.3\%). We discuss the possible reasons for the failure of the popular Minnesota DFAs next.

\FloatBarrier
\subsection{Localization/delocalization errors in F-bond donors and acceptors}\label{sec:4.5}

\subsubsection{Fractional charge-transfer in DF\textperiodcentered{}\textperiodcentered{}\textperiodcentered{}CH\textsubscript{3}/NH\textsubscript{3} complexes at infinite separation}\label{sec:4.5.A}

Otero-de-la-Roza et al.\cite{OteroDeLaRoza2014} attributed the overbinding of closed-shell halogen-bonded complexes by popular DFAs to delocalization error, and showed that dispersion-corrected DFAs with minimal delocalization error, such as XDM-corrected BHandHLYP with 50\% HF exchange (similar to 54\% HF in M06-2X), are reliable. Here, we examine delocalization error in various Minnesota DFAs for\par
\begin{table}[!ht]\captionsetup{type=figure}\centering
\par\vspace{2pt}\noindent\hfil\includegraphics[width=0.88\linewidth]{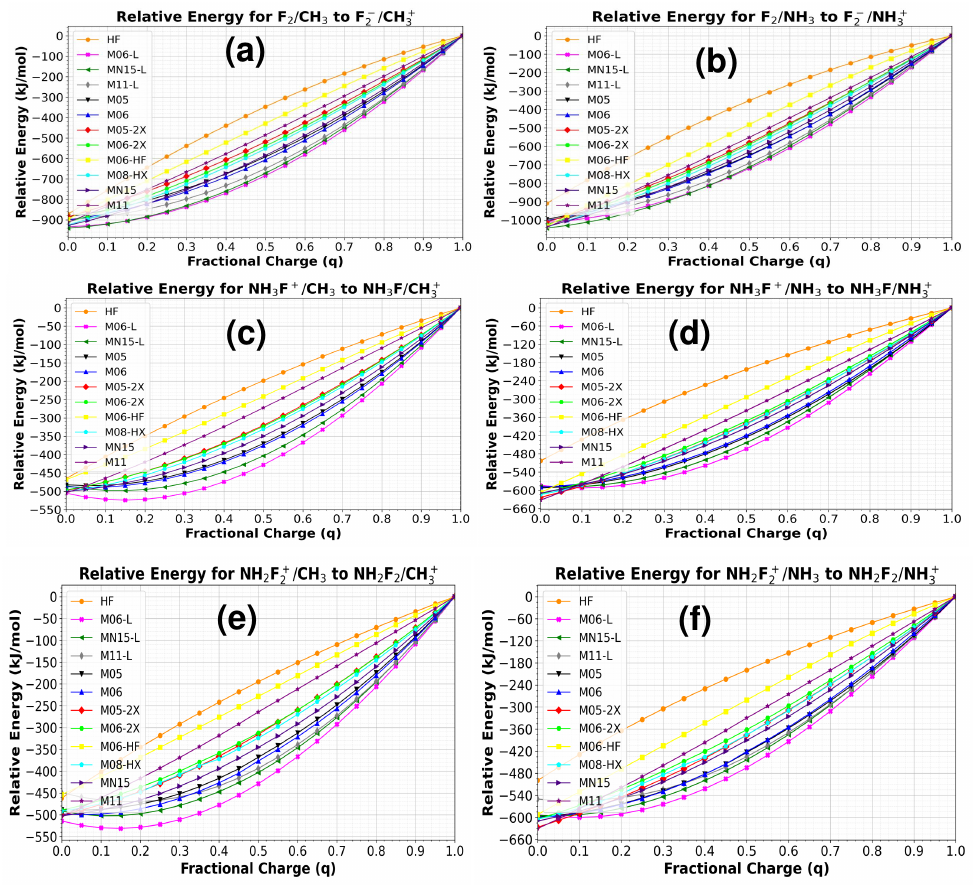}\hfil\null\par\nointerlineskip\vspace{13pt}
\par\vspace{-1pt}\caption{Plots showing relative energy (in kJ/mol) with respect to fractional charge transfer (q) for various F-bonded complexes DF\textperiodcentered{}\textperiodcentered{}\textperiodcentered{}CH\textsubscript{3}/NH\textsubscript{3} (DF = F\textsubscript{2}, NH\textsubscript{3}F\textsuperscript{+} and NH\textsubscript{2}F\textsubscript{2}\textsuperscript{+}) at infinite separation: \textbf{(a)} F\textsubscript{2}/CH\textsubscript{3} \ensuremath{\rightarrow} F\textsubscript{2}\textsuperscript{-}/CH\textsubscript{3}\textsuperscript{+}, \textbf{(b)} F\textsubscript{2}/NH\textsubscript{3} \ensuremath{\rightarrow} F\textsubscript{2}\textsuperscript{-}/NH\textsubscript{3}\textsuperscript{+}, \textbf{(c)} NH\textsubscript{3}F\textsuperscript{+}/CH\textsubscript{3} \ensuremath{\rightarrow} NH\textsubscript{3}F/CH\textsubscript{3}\textsuperscript{+}, \textbf{(d)} NH\textsubscript{3}F\textsuperscript{+}/NH\textsubscript{3} \ensuremath{\rightarrow} NH\textsubscript{3}F/NH\textsubscript{3}\textsuperscript{+}, \textbf{(e)} NH\textsubscript{2}F\textsubscript{2}\textsuperscript{+}/CH\textsubscript{3} \ensuremath{\rightarrow} NH\textsubscript{2}F\textsubscript{2}/CH\textsubscript{3}\textsuperscript{+} and \textbf{(f)} NH\textsubscript{2}F\textsubscript{2}\textsuperscript{+}/NH\textsubscript{3} \ensuremath{\rightarrow} NH\textsubscript{2}F\textsubscript{2}/NH\textsubscript{3}\textsuperscript{+}. Energies of right-hand side complexes are considered as zero to calculate relative energies for each fractional charge-transfer value. Calculations were carried out using Q-Chem 6.0.2 program\cite{Epifanovsky2021} in conjunction with AVTZ basis set.}\label{fig:7}
\vspace{10pt}
\end{table}
\noindent FB complexes from SEFB-15 and CSFB-14 databases. We plot the relative energy (E) against the fractional charge (q) transferred between the infinitely separated monomers, DF and CH\textsubscript{3}/NH\textsubscript{3}, which measures their delocalization error (\mbox{\textbf{FIG. \ref{fig:7}}} and \mbox{\textbf{TABLE \ref{tab:X}}} for DF = F\textsubscript{2}, NH\textsubscript{3}F\textsuperscript{+} and NH\textsubscript{2}F\textsubscript{2}\textsuperscript{+}; \mbox{\textbf{TABLE \ref{tab:S7}}} and \mbox{\textbf{FIGs. \ref{fig:S5}}}, \mbox{\textbf{\ref{fig:S7}}}, \mbox{\textbf{\ref{fig:S9}}} and \mbox{\textbf{\ref{fig:S11}}} for all systems). As already mentioned in \textbf{Section \ref*{sec:2}}, for an exact functional, the E(q) should be linear between two consecutive integer electron numbers (q).\cite{Perdew1982,Kraisler2013,Kronik2020} For conventional DFAs, this piecewise linearity condition is generally violated.

Relative to piece-wise linear E(q) behavior for an exact functional (\textbf{Section \ref*{sec:2}}), convex curves indicate delocalization error and concave curves indicate localization error.\cite{Cohen2012,Cohen2008,MoriSanchez2008,Li2017} Delocalization error overstabilizes systems with fractional charge, whereas localization error underestimates the BEs. HF gives concave curves, whereas pure (semi-local) DFAs give convex curves, so adding HF exchange reduces delocalization error through error cancellation. Thus, the convexity decreases as HF exchange increases and the curves turn concave at high HF exchange.\cite{Johnson2013,OteroDeLaRoza2014}

\begin{table}[!htbp]\centering\setlength{\parskip}{0pt}
\caption{Minimum relative energy (kJ/\allowbreak{}mol) and corresponding fractional charge-transfer (q) for selected infinitely separated F-bonded complexes DF\textperiodcentered{}\textperiodcentered{}\textperiodcentered{}CH\textsubscript{3}/NH\textsubscript{3 }(DF = F\textsubscript{2}, NH\textsubscript{3}F\textsuperscript{+} and NH\textsubscript{2}F\textsubscript{2}\textsuperscript{+}) from SEFB-15 and CSFB-14 data sets, predicted by several Minnesota DFAs with AVTZ basis set. See \mbox{\textbf{Table \ref{tab:S7} }}from supporting information for all systems considered in this work.}\label{tab:X}\vspace{7pt}\nopagebreak
\adjustbox{width=0.85\linewidth,max totalheight=0.78\textheight,center}{\fontfamily{ptm}\fontsize{11.2}{12.9}\selectfont\renewcommand{\arraystretch}{1.05}\setlength{\tabcolsep}{1.5pt}
\begin{tabular}{|>{\centering\arraybackslash}p{53.9pt}|>{\centering\arraybackslash}p{46.7pt}|>{\centering\arraybackslash}p{60.8pt}|>{\centering\arraybackslash}p{67.8pt}|>{\centering\arraybackslash}p{75.2pt}|>{\centering\arraybackslash}p{63.8pt}|>{\centering\arraybackslash}p{78.9pt}|}
\hline
\multicolumn{7}{|>{\centering\arraybackslash}p{465.1pt}|}{{\bfseries\boldmath \textbf{SEFB-15 data set: DF}\textperiodcentered{}\textperiodcentered{}\textperiodcentered{}\textbf{CH}\textbf{\textsubscript{3}}\textbf{ complexes}}}   \\
\hline
\multicolumn{1}{|>{\centering\arraybackslash}p{53.9pt}|}{\multirow{2}{53.9pt}{{\bfseries\boldmath \centering \textbf{Methods}}}} &\multicolumn{2}{>{\centering\arraybackslash}p{110.5pt}|}{{\bfseries\boldmath \textbf{F}\textbf{\textsubscript{2}}\textbf{/CH}\textbf{\textsubscript{3}}\textbf{ \ensuremath{\rightarrow} F}\textbf{\textsubscript{2}}\textbf{\textsuperscript{-}}\textbf{/CH}\textbf{\textsubscript{3}}\textbf{\textsuperscript{+}}\textbf{ }}} &\multicolumn{2}{>{\centering\arraybackslash}p{146.0pt}|}{{\bfseries\boldmath \textbf{NH}\textbf{\textsubscript{3}}\textbf{F}\textbf{\textsuperscript{+}}\textbf{/CH}\textbf{\textsubscript{3}}\textbf{ \ensuremath{\rightarrow} NH}\textbf{\textsubscript{3}}\textbf{F/\allowbreak{}CH}\textbf{\textsubscript{3}}\textbf{\textsuperscript{+}}\textbf{ }}} &\multicolumn{2}{>{\centering\arraybackslash}p{145.8pt}|}{{\bfseries\boldmath \textbf{NH}\textbf{\textsubscript{2}}\textbf{F}\textbf{\textsubscript{2}}\textbf{\textsuperscript{+}}\textbf{/CH}\textbf{\textsubscript{3}}\textbf{ \ensuremath{\rightarrow}NH}\textbf{\textsubscript{2}}\textbf{F}\textbf{\textsubscript{2}}\textbf{/CH}\textbf{\textsubscript{3}}\textbf{\textsuperscript{+}}\textbf{ }}}   \\
\cline{2-7}
\multicolumn{1}{|>{\centering\arraybackslash}p{53.9pt}|}{} & {\bfseries\boldmath \textbf{q}} & {\bfseries\boldmath \textbf{E }} & {\bfseries\boldmath \textbf{q}} & {\bfseries\boldmath \textbf{E }} & {\bfseries\boldmath \textbf{q}} & {\bfseries\boldmath \textbf{E }}   \\
\hline
\end{tabular}}

\end{table}

In principle, no electron should transfer between two monomers at infinite separation. Finite charge transfer nevertheless occurs only for the cationic complexes NH\textsubscript{3}F\textsuperscript{+}\textperiodcentered{}\textperiodcentered{}\textperiodcentered{}CH\textsubscript{3}/NH\textsubscript{3} and NH\textsubscript{2}F\textsubscript{2}\textsuperscript{+}\textperiodcentered{}\textperiodcentered{}\textperiodcentered{}CH\textsubscript{3}/NH\textsubscript{3} at infinite separation (\mbox{\textbf{TABLE \ref{tab:X}}} and \mbox{\textbf{FIG. \ref{fig:7}}}). The meta-GGAs (M06-L, MN15-L and M11-L) and hybrids (M05, and M06) predict q = 0.05-0.20 for these complexes.

The largest value is 0.20 for M11-L (NH\textsubscript{2}F\textsubscript{2}\textsuperscript{+}\textperiodcentered{}\textperiodcentered{}\textperiodcentered{}CH\textsubscript{3}), and M06-L gives 0.15 for both cationic DF\textperiodcentered{}\textperiodcentered{}\textperiodcentered{}CH\textsubscript{3} complexes. Among the hybrids, only M05 and M06 predict charge transfer (q = 0.05), and only for the two cationic DF\textperiodcentered{}\textperiodcentered{}\textperiodcentered{}CH\textsubscript{3} complexes. These two functionals have the lowest HF exchange among the Minnesota hybrids used (28\% and 27\%). M05-2X, M06-2X, M08-HX, M06-HF, MN15 and M11, with 42.8-100\% HF exchange, predict none, and no DFA predicts any fractional charge-transfer for the neutral complexes (\mbox{\textbf{TABLE \ref{tab:S7},}} and \mbox{\textbf{FIGs. \ref{fig:S5}}}, \mbox{\textbf{\ref{fig:S7}}}, \mbox{\textbf{\ref{fig:S9}}} and \mbox{\textbf{\ref{fig:S11}}}), as for the neutral dimers in literature.\cite{OteroDeLaRoza2014} This spurious sharing of fractional electron numbers between the monomers at infinite separation indicates delocalization error within them, and it is at least as strong for M06-L as for MN15-L. M11-L/AVTZ did not converge errors when calculating single-point energies across the full range of fractional charges (q = 0-1) for certain cases, including NH\textsubscript{3}F\textsuperscript{+}\textperiodcentered{}\textperiodcentered{}\textperiodcentered{}CH\textsubscript{3}/NH\textsubscript{3} at infinite separation (see \mbox{\textbf{TABLES \ref{tab:X}, \ref{tab:S7}}} and \mbox{\textbf{\ref{tab:S8}}}).

\FloatBarrier
\subsubsection{Deviations of total energies from linearity for DF\textperiodcentered{}\textperiodcentered{}\textperiodcentered{}CH\textsubscript{3}/NH\textsubscript{3} complexes at infinite separation}\label{sec:4.5.B}

The absence of charge transfer predicted by a DFA for the two monomers of a dimer does not necessarily imply that the DFA is exact. Since the exact functional should give no deviations of the total energies from linearity, we have analyzed these deviations (\mbox{\textbf{TABLE \ref{tab:XI}}} and \mbox{\textbf{FIG. \ref{fig:8}}} for the selected systems; \mbox{\textbf{TABLE \ref{tab:S8}}} and \mbox{\textbf{FIGs. \ref{fig:S6}}}, \mbox{\textbf{\ref{fig:S8}}}, \mbox{\textbf{\ref{fig:S10}}} and \mbox{\textbf{\ref{fig:S12}}} for all systems). In \mbox{\textbf{FIG. \ref{fig:8}}} and \mbox{\textbf{TABLE \ref{tab:XI},}} positive deviations from linearity indicate localization error and negative deviations indicate delocalization error. HF shows localization error for all the complexes (\mbox{\textbf{TABLE \ref{tab:XI}}}). Also, M06-HF shows localization error for F\textsubscript{2}\textperiodcentered{}\textperiodcentered{}\textperiodcentered{}CH\textsubscript{3} and all the DF\textperiodcentered{}\textperiodcentered{}\textperiodcentered{}NH\textsubscript{3} complexes, and delocalization error for NH\textsubscript{3}F\textsuperscript{+}\textperiodcentered{}\textperiodcentered{}\textperiodcentered{}CH\textsubscript{3} (-10.2 kJ/mol) and NH\textsubscript{2}F\textsubscript{2}\textsuperscript{+}\textperiodcentered{}\textperiodcentered{}\textperiodcentered{}CH\textsubscript{3} (-5.5 kJ/mol) (\mbox{\textbf{TABLE \ref{tab:XI}}} and \mbox{\textbf{FIG. \ref{fig:8}}}). For the other DF\textperiodcentered{}\textperiodcentered{}\textperiodcentered{}CH\textsubscript{3} complexes, it shows delocalization error for four (NF\textsubscript{3}\textperiodcentered{}\textperiodcentered{}\textperiodcentered{}CH\textsubscript{3}, FCCF\textperiodcentered{}\textperiodcentered{}\textperiodcentered{}CH\textsubscript{3}, CF\textsubscript{3}CCF\textperiodcentered{}\textperiodcentered{}\textperiodcentered{}CH\textsubscript{3}, and SF\textsubscript{5}\textperiodcentered{}\textperiodcentered{}\textperiodcentered{}CH\textsubscript{3} ; \mbox{\textbf{TABLE \ref{tab:S8}}} and \mbox{\textbf{FIGs. \ref{fig:8}, \ref{fig:S8}, \ref{fig:S10},}} and \mbox{\textbf{\ref{fig:S12}}}) and localization error for six (SF\textsubscript{6}\textperiodcentered{}\textperiodcentered{}\textperiodcentered{}CH\textsubscript{3}, ClF\textsubscript{3}\textperiodcentered{}\textperiodcentered{}\textperiodcentered{}CH\textsubscript{3}, OF\textsubscript{2}\textperiodcentered{}\textperiodcentered{}\textperiodcentered{}CH\textsubscript{3}, NF\textsubscript{5}\textperiodcentered{}\textperiodcentered{}\textperiodcentered{}CH\textsubscript{3}, CF\textsubscript{3}OF\textperiodcentered{}\textperiodcentered{}\textperiodcentered{}CH\textsubscript{3}, and F3NO\textperiodcentered{}\textperiodcentered{}\textperiodcentered{}CH\textsubscript{3}, as depicted in \mbox{\textbf{TABLE \ref{tab:S8}}}, and \mbox{\textbf{FIGs. \ref{fig:8}, \ref{fig:S6}, \ref{fig:S8}, \ref{fig:S10}}} and \mbox{\textbf{\ref{fig:S12}}}), and two did not converge (\mbox{\textbf{TABLE \ref{tab:S8}}}). All other Minnesota DFAs show delocalization error for all the complexes. Thus, the type of error is system dependent, and conclusions from closed-shell systems may not hold for open-shell F-bonded radical-molecule complexes.

\begin{table}[!htbp]\captionsetup{type=figure}\centering
\par\vspace{2pt}\noindent\hfil\includegraphics[width=\linewidth]{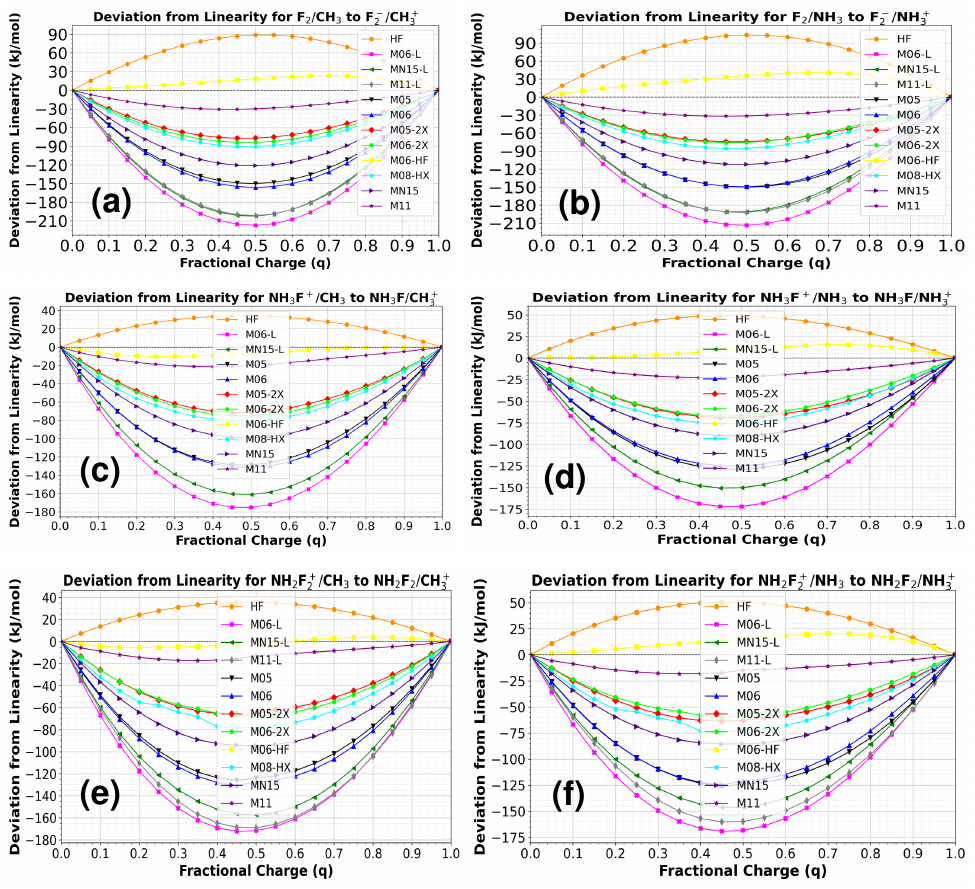}\hfil\null\par\nointerlineskip\vspace{13pt}
\par\vspace{-1pt}\caption{Plots showing deviations from linearity on total energies (in kJ/mol) with respect to fractional charge transfer (q) for various F-bonded complexes DF\textperiodcentered{}\textperiodcentered{}\textperiodcentered{}CH\textsubscript{3} /NH\textsubscript{3} (DF = F\textsubscript{2}, NH\textsubscript{3}F\textsuperscript{+} and NH\textsubscript{2}F\textsubscript{2}\textsuperscript{+}), considered at infinite separation : \textbf{(a)} F\textsubscript{2}/CH\textsubscript{3} \ensuremath{\rightarrow} F\textsubscript{2}\textsuperscript{-}/CH\textsubscript{3}\textsuperscript{+}, \textbf{(b)} F\textsubscript{2}/NH\textsubscript{3} \ensuremath{\rightarrow} F\textsubscript{2}\textsuperscript{-}/NH\textsubscript{3}\textsuperscript{+}, \textbf{(c)} NH\textsubscript{3}F\textsuperscript{+}/CH\textsubscript{3} \ensuremath{\rightarrow} NH\textsubscript{3}F/CH\textsubscript{3}\textsuperscript{+}, \textbf{(d)} NH\textsubscript{3}F\textsuperscript{+}/NH\textsubscript{3} \ensuremath{\rightarrow} NH\textsubscript{3}F/NH\textsubscript{3}\textsuperscript{+}, \textbf{(e)} NH\textsubscript{2}F\textsubscript{2}\textsuperscript{+}/CH\textsubscript{3} \ensuremath{\rightarrow} NH\textsubscript{2}F\textsubscript{2}/CH\textsubscript{3}\textsuperscript{+} and \textbf{(f)} NH\textsubscript{2}F\textsubscript{2}\textsuperscript{+}/NH\textsubscript{3} \ensuremath{\rightarrow} NH\textsubscript{2}F\textsubscript{2}/NH\textsubscript{3}\textsuperscript{+}. Q-Chem 6.0.2 program\cite{Epifanovsky2021} in conjunction with AVTZ basis set has been used for all calculations.}\label{fig:8}
\vspace{10pt}
\end{table}

\begin{table}[!htbp]\centering\setlength{\parskip}{0pt}
\caption{Maximum deviations on electronic energies from linearity (kJ/\allowbreak{}mol), and corresponding fractional charge-transfer (q), for selected infinitely separated complexes DF\textperiodcentered{}\textperiodcentered{}\textperiodcentered{}CH\textsubscript{3}/NH\textsubscript{3 }(DF = F\textsubscript{2}, NH\textsubscript{3}F\textsuperscript{+} and NH\textsubscript{2}F\textsubscript{2}\textsuperscript{+}) from SEFB-15 and CSFB-14 data sets, predicted by several Minnesota DFAs with AVTZ basis set. Positive deviations represent inherent localization error and negative deviations signify delocalization error in the corresponding functionals. See \mbox{\textbf{Table \ref{tab:S8}}}\textbf{ }for all F-bonded systems at infinite separation.}\label{tab:XI}\vspace{7pt}\nopagebreak
\adjustbox{width=1.0\linewidth,max totalheight=0.78\textheight,center}{\fontfamily{ptm}\fontsize{11.2}{12.9}\selectfont\renewcommand{\arraystretch}{1.1}\setlength{\tabcolsep}{1.5pt}
\begin{tabular}{|>{\centering\arraybackslash}m{60.4pt}|>{\centering\arraybackslash}m{52.6pt}|>{\centering\arraybackslash}m{52.8pt}|>{\centering\arraybackslash}m{66.3pt}|>{\centering\arraybackslash}m{73.4pt}|>{\centering\arraybackslash}m{98.3pt}|>{\centering\arraybackslash}m{8.0pt}|>{\centering\arraybackslash}m{39.1pt}|}
\hline
\multicolumn{8}{|>{\centering\arraybackslash}m{471.9pt}|}{{\bfseries\boldmath \textbf{Selected infinitely separated DF}\textbf{\textperiodcentered{}\textperiodcentered{}\textperiodcentered{}}\textbf{CH}\textbf{\textsubscript{3}}\textbf{ complexes}}}   \\
\hline
\multicolumn{1}{|>{\centering\arraybackslash}m{60.4pt}|}{\multirow{2}{60.4pt}{{\bfseries\boldmath \centering \textbf{Methods}}}} &\multicolumn{2}{>{\centering\arraybackslash}m{108.4pt}|}{{\bfseries\boldmath \textbf{F}\textbf{\textsubscript{2}}\textbf{/CH}\textbf{\textsubscript{3}}\textbf{ \ensuremath{\rightarrow} F}\textbf{\textsubscript{2}}\textbf{\textsuperscript{-}}\textbf{/CH}\textbf{\textsubscript{3}}\textbf{\textsuperscript{+}}\textbf{ }}} &\multicolumn{2}{>{\centering\arraybackslash}m{142.7pt}|}{{\bfseries\boldmath \textbf{NH}\textbf{\textsubscript{3}}\textbf{F}\textbf{\textsuperscript{+}}\textbf{/CH}\textbf{\textsubscript{3}}\textbf{ \ensuremath{\rightarrow} NH}\textbf{\textsubscript{3}}\textbf{F/\allowbreak{}CH}\textbf{\textsubscript{3}}\textbf{\textsuperscript{+}}\textbf{ }}} &\multicolumn{3}{>{\centering\arraybackslash}m{151.4pt}|}{{\bfseries\boldmath \textbf{NH}\textbf{\textsubscript{2}}\textbf{F}\textbf{\textsubscript{2}}\textbf{\textsuperscript{+}}\textbf{/CH}\textbf{\textsubscript{3}}\textbf{ \ensuremath{\rightarrow} NH}\textbf{\textsubscript{2}}\textbf{F}\textbf{\textsubscript{2}}\textbf{/CH}\textbf{\textsubscript{3}}\textbf{\textsuperscript{+}}\textbf{ }}}   \\
\hline
\end{tabular}}
\par\nopagebreak\vspace{3pt}{\setstretch{1}\setlength{\parskip}{1.5pt}\fontsize{9}{10.5}\selectfont\raggedright\noindent \textsuperscript{b}Convergence issues with M11-L functional.\par}

\end{table}

M06-L shows the strongest delocalization error in every complex where it converged (\mbox{\textbf{TABLE \ref{tab:X}, \ref{tab:S7}}}, and \mbox{\textbf{\ref{tab:S8}}}, and \mbox{\textbf{FIGs.}} \mbox{\textbf{\ref{fig:7}, \ref{fig:8}, \ref{fig:S6}, \ref{fig:S8}, \ref{fig:S10}}} and \mbox{\textbf{\ref{fig:S12}}}), larger than MN15-L and M11-L, and the three meta-GGAs generally show larger errors than the hybrid DFAs (\mbox{\textbf{TABLES \ref{tab:XI}}} and \mbox{\textbf{\ref{tab:S8}}}). M06-L did not converge for the NCF, CNF and CF\textsubscript{3}CCF complexes (\mbox{\textbf{TABLE \ref{tab:S7}}}). Together with their large BE errors for both databases and convergence for only 8-10 of the 15 DF\textperiodcentered{}\textperiodcentered{}\textperiodcentered{}CH\textsubscript{3} complexes (\mbox{\textbf{TABLE \ref{tab:III}}}). This suggests that these meta-GGAs are not suitable for predicting BEs of F-bonded radical-molecule as well as closed-shell complexes. Significant deviations from linearity also occur for the neutral complexes. So, the absence of charge transfer does not mean a DFA is free from delocalization or localization error.

In every complex where all three (M05-2X, M06-2X and M08-HX) converged, M08-HX shows larger delocalization error than M05-2X and M06-2X, (-90.6 against -77.1 and -83.8 kJ/mol for F\textsubscript{2}\textperiodcentered{}\textperiodcentered{}\textperiodcentered{}CH\textsubscript{3}; \mbox{\textbf{TABLES \ref{tab:XI}}} and \mbox{\textbf{\ref{tab:S8}}}), consistent with its slightly smaller HF exchange (52.23\% against 56\% and 54\%.\cite{Mardirossian2016} Yet M08-HX is more reliable than both for BEs and F-bond distances at equilibrium, for the DF\textperiodcentered{}\textperiodcentered{}\textperiodcentered{}CH\textsubscript{3} as well as the DF\textperiodcentered{}\textperiodcentered{}\textperiodcentered{}NH\textsubscript{3} complexes (\mbox{\textbf{TABLES \ref{tab:I}}} and \mbox{\textbf{\ref{tab:IX}}}, and \mbox{\textbf{FIG. \ref{fig:1}}}). Thus, monomer delocalization error separates the meta-GGAs from the hybrids, but it does not rank M08-HX, M05-2X and M06-2X, and so it does not by itself govern DFA performance. This agrees with previous report on closed-shell halogen bonding,\cite{OteroDeLaRoza2014} where the failure of semi-local DFAs for neutral halogen-bonded complexes was not attributed to self-interaction error in the monomers, since the monomers change on dimer formation and fractional charges occur only at the dimer geometry. It is contrary to the suggestion by Steinmann et al.\cite{Steinmann2012} The overbinding of the SEFB complexes by M05-2X and M06-2X may rather arise from delocalization error in the dimers, which we probe next with Bader\textquotesingle s intermolecular delocalization indices (DIs), the average number of electrons delocalized between two atoms.\cite{OteroDeLaRoza2014}

\FloatBarrier
\subsection{Bader's intermolecular delocalization indices (DIs): Is M08-HX functional free from Delocalization error?}\label{sec:4.6}

\textbf{Section \ref*{sec:4.5}} showed that monomer delocalization error is significant for meta-GGA functionals but does not rank the performance of M08-HX, M05-2X and M06-2X. Overbinding of the SEFB complexes by the latter two, and even worse results after GD3 dispersion correction (\textbf{Section \ref*{sec:4.1}}), may rather arise from delocalization error in the dimers. Following Otero-de-la-Roza et al.,\cite{OteroDeLaRoza2014} we therefore probed the dimers at their equilibrium geometries through Bader's intermolecular DIs. These were computed with AIMAll program\cite{Keith2012} using geometries optimized at M05-2X, M06-2X, M06-HF, M08-HX and M11 (and the GD3-corrected variants of M05-2X and M06-2X) in conjunction with AVTZ basis set, and at CCSD/AVDZ level (\mbox{\textbf{FIG. \ref{fig:9}}} and \mbox{\textbf{TABLE \ref{tab:S10}}}). DIs and BEs using geometries optimized at DF-RMP2-F12/VDZ-F12 are given in \mbox{\textbf{FIG. \ref{fig:S13}}} and \mbox{\textbf{TABLE \ref{tab:S11}}}.

\begin{table}[!htbp]\captionsetup{type=figure}\centering
\par\vspace{2pt}\noindent\hfil\includegraphics[width=\linewidth]{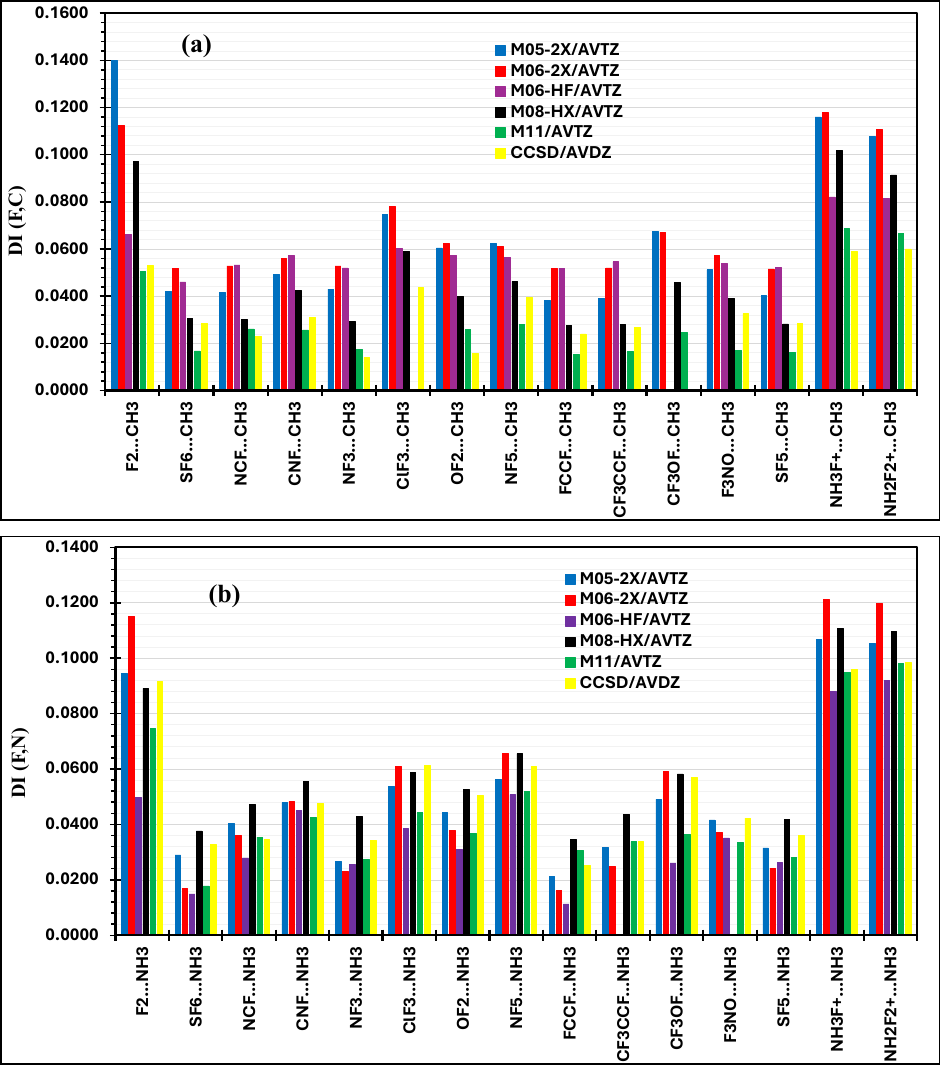}\hfil\null\par\nointerlineskip\vspace{13pt}
\par\vspace{-1pt}\caption{Bader's intermolecular delocalization indices {[}DI(F, C/N){]} for F-bonded complexes \textbf{(a)} DF\textperiodcentered{}\textperiodcentered{}\textperiodcentered{}CH\textsubscript{3} and \textbf{(b)} DF\textperiodcentered{}\textperiodcentered{}\textperiodcentered{}NH\textsubscript{3} obtained from five selected Minnesota DFAs (M05-2X, M06-2X, M06-HF, M08-HX and M11) at dimer geometries optimized at the respective levels of theory using AVTZ basis set for DFAs and AVDZ for CCSD. See \mbox{\textbf{Table \ref{tab:S10}}} for DI and BE values obtained from these five Minnesota DFAs and from M05-2X-GD3/AVTZ and M06-2X-GD3/AVTZ levels. Also, see \mbox{\textbf{Figure \ref{fig:S13}}} and \mbox{\textbf{Table \ref{tab:S11}}} for the corresponding results obtained at geometries optimized at DF-RMP2-F12/VDZ-F12 level.}\label{fig:9}
\vspace{10pt}
\end{table}

\begin{table}[!htbp]\centering\setlength{\parskip}{0pt}
\caption{R\textsuperscript{2} values for the linear correlations between DI values and BEs, and DI values and reference BEs, along with MA\%D and MAD on BEs (kJ/\allowbreak{}mol), for the DF\textperiodcentered{}\textperiodcentered{}\textperiodcentered{}CH\textsubscript{3}/NH\textsubscript{3} complexes predicted by seven selected Minnesota DFAs with AVTZ basis set. Results at CCSD/\allowbreak{}AVDZ level are included for comparison. All BEs and DIs are calculated using fully optimized geometries at the corresponding levels of theory, and separately using optimized geometries obtained from DF-RMP2-F12/\allowbreak{}VDZ-F12 level. BSSE corrections are not applied to BEs. The necessary source data are in \mbox{\textbf{Tables \ref{tab:S10} }}and \textbf{S11}. \mbox{\textbf{Table \ref{tab:S9}}} reports the R\textsuperscript{2} values for DI(DFAs) versus DI(CCSD), and for BE(DFAs) versus reference BEs. DI and BE values obtained for all Minnesota functionals, using geometries optimized at DF-RMP2-F12/\allowbreak{}VDZ-F12 level are summarized in \mbox{\textbf{Tables \ref{tab:S12} }}and \textbf{S13}, respectively.}\label{tab:XII}\vspace{7pt}\nopagebreak
\adjustbox{width=1.0\linewidth,max totalheight=0.78\textheight,center}{\fontfamily{ptm}\fontsize{11.03}{12.7}\selectfont\renewcommand{\arraystretch}{1.15}\setlength{\tabcolsep}{1.5pt}
\begin{tabular}{|>{\centering\arraybackslash}m{85.2pt}|>{\centering\arraybackslash}m{47.1pt}|>{\centering\arraybackslash}m{47.1pt}|>{\centering\arraybackslash}m{47.1pt}|>{\centering\arraybackslash}m{47.1pt}|>{\centering\arraybackslash}m{47.1pt}|>{\centering\arraybackslash}m{47.1pt}|>{\centering\arraybackslash}m{47.1pt}|>{\centering\arraybackslash}m{47.4pt}|}
\hline
\multicolumn{9}{|>{\centering\arraybackslash}m{486.5pt}|}{{\bfseries\boldmath \textbf{R}\textbf{\textsuperscript{2 }}\textbf{values for correlation of DI with BEs, ref. BEs, MAD and MA\%D on BEs for DF\textperiodcentered{}\textperiodcentered{}\textperiodcentered{}CH}\textbf{\textsubscript{3}}\textbf{/NH}\textbf{\textsubscript{3}}\textbf{ complexes using corresponding optimized geometries}}}   \\
\hline
\multicolumn{1}{|>{\centering\arraybackslash}m{85.2pt}|}{\multirow{2}{85.2pt}{{\bfseries\boldmath \centering \textbf{Methods}}}} &\multicolumn{2}{>{\centering\arraybackslash}m{97.3pt}|}{{\bfseries\boldmath \textbf{DI vs BEs}}} &\multicolumn{2}{>{\centering\arraybackslash}m{97.3pt}|}{{\bfseries\boldmath \textbf{DI vs ref. BEs}}} &\multicolumn{2}{>{\centering\arraybackslash}m{97.3pt}|}{{\bfseries\boldmath \textbf{MAD (kJ/\allowbreak{}mol)}}} &\multicolumn{2}{>{\centering\arraybackslash}m{97.5pt}|}{{\bfseries\boldmath \textbf{MA\%}\textbf{D(}\textbf{\%)}}}   \\
\hline
\end{tabular}}

\end{table}

Irrespective of the method, F\textsubscript{2}\textperiodcentered{}\textperiodcentered{}\textperiodcentered{}CH\textsubscript{3}/NH\textsubscript{3}, NH\textsubscript{3}F\textsuperscript{+}\textperiodcentered{}\textperiodcentered{}\textperiodcentered{}CH\textsubscript{3}/NH\textsubscript{3} and NH\textsubscript{2}F\textsubscript{2}\textsuperscript{+}\textperiodcentered{}\textperiodcentered{}\textperiodcentered{}CH\textsubscript{3}/NH\textsubscript{3} generally show larger DIs (F, C/N) than the other complexes (\mbox{\textbf{FIG. \ref{fig:9}}}). For all 15 DF\textperiodcentered{}\textperiodcentered{}\textperiodcentered{}CH\textsubscript{3} complexes, M08-HX gives lower DI(F,C) than M05-2X and M06-2X, and lower than M06-HF (M06-HF failed for CF\textsubscript{3}OF\textperiodcentered{}\textperiodcentered{}\textperiodcentered{}CH\textsubscript{3}), except for these three complexes (F\textsubscript{2}\textperiodcentered{}\textperiodcentered{}\textperiodcentered{}CH\textsubscript{3}, NH\textsubscript{3}F\textsuperscript{+}\textperiodcentered{}\textperiodcentered{}\textperiodcentered{}CH\textsubscript{3} and NH\textsubscript{2}F\textsubscript{2}\textsuperscript{+}\textperiodcentered{}\textperiodcentered{}\textperiodcentered{}CH\textsubscript{3}) (\mbox{\textbf{FIG. \ref{fig:9}(a)}}). Among the DFAs used, M11 gives the lowest DIs for 14 of them (M11 failed for ClF3\textperiodcentered{}\textperiodcentered{}\textperiodcentered{}CH3). Since larger DIs signal larger delocalization error in the dimer,\cite{OteroDeLaRoza2014} the lower DIs of M08-HX are consistent with its better BEs for the SEFB complexes than those from M05-2X and M06-2X (\mbox{\textbf{TABLE \ref{tab:IX}}}). Conversely, the larger DIs of M05-2X and M06-2X point to a larger delocalization error already in the base functionals, and may explain why GD3 further overbinds these complexes (\mbox{\textbf{TABLE \ref{tab:III}}} and \mbox{\textbf{FIG. \ref{fig:1}}}). For DF\textperiodcentered{}\textperiodcentered{}\textperiodcentered{}NH3 complexes, M05-2X and M06-2X give DIs lower than or comparable to those of M08-HX, in line with the comparable reliability of the three and the minor effect of GD3 (\mbox{\textbf{FIG. \ref{fig:9}(b)}}, \mbox{\textbf{TABLES \ref{tab:III}}} and \mbox{\textbf{\ref{tab:IX}}}). A low DI from a DFA is nevertheless not sufficient for its reliability. In particular, M11 (a range-separated DFA) yields the smallest DIs for DF\textperiodcentered{}\textperiodcentered{}\textperiodcentered{}CH\textsubscript{3} complexes (\mbox{\textbf{FIG. \ref{fig:9}(a)}}) but larger errors on BEs and F-bond distances than M08-HX (a hybrid DFA) (\mbox{\textbf{TABLES \ref{tab:III}}}, \mbox{\textbf{\ref{tab:IV}}} and \mbox{\textbf{\ref{tab:IX}}}), while DIs predicted by it for most DF\textperiodcentered{}\textperiodcentered{}\textperiodcentered{}NH\textsubscript{3} complexes are not the smallest (\mbox{\textbf{FIG. \ref{fig:9}(b)}}). Thus, the poor performance of the M11 functional indicates that climbing Jacob's ladder of DFAs may not necessarily be fruitful for reliable BEs of NCBs, consistent with results in the literature.\cite{Mardirossian2016} These trends refer to the DFA-optimized geometries. Furthermore, in the case of geometries optimized at the DF-RMP2-F12/VDZ-F12 level (\mbox{\textbf{FIG. \ref{fig:S13}}} and \mbox{\textbf{TABLES \ref{tab:S11}, \ref{tab:S12}}} and \mbox{\textbf{\ref{tab:S13}}}), the DIs obtained from M08-HX are slightly higher than (or comparable) to those from M05-2X and M06-2X for most DF\textperiodcentered{}\textperiodcentered{}\textperiodcentered{}CH\textsubscript{3}/NH\textsubscript{3} complexes (\mbox{\textbf{FIG. \ref{fig:S13}}}). Therefore, the lower DIs of M08-HX are geometry-dependent.

These observations raise the question of whether the DIs correlate with BEs. \mbox{\textbf{TABLE \ref{tab:XII}}} lists the linear correlation coefficients (R²) for DIs vs BEs and DIs vs reference BEs, together with the MADs and MA\%Ds on BEs (for source data, refer to \mbox{\textbf{TABLES \ref{tab:S10}-\ref{tab:S13}}}). No functional shows a consistent linear correlation. The range of R² for DIs vs reference BEs is 0.470-0.891 at DFA-optimized geometries and 0.591-0.653 at DF-RMP2-F12/VDZ-F12 geometries (\mbox{\textbf{TABLE \ref{tab:XII}}}). The highest R² values do not coincide with the best BEs. M06-HF (0.931/0.851 for DI vs BEs of DF\textperiodcentered{}\textperiodcentered{}\textperiodcentered{}CH\textsubscript{3}/NH\textsubscript{3}) and M11 (0.835/0.810) have MA\%D of 33.2-44.2\%, whereas M08-HX, with moderate R² (0.716/0.803), has the smallest errors among the DFAs (MA\%D =  18.7\%/7.7\%). CCSD/AVDZ at its own geometries likewise shows weak R² (0.590-0.665) but small MA\%Ds (10.9\%/15.2\%). In addition, the DF-RMP2-F12/VDZ-F12 geometries do not change this picture; R² for DI vs BEs obtained from the DFAs used remains moderate (0.566-0.812 for DF\textperiodcentered{}\textperiodcentered{}\textperiodcentered{}CH\textsubscript{3} and 0.469-0.580 for DF\textperiodcentered{}\textperiodcentered{}\textperiodcentered{}NH\textsubscript{3}, and 0.494/0.616 from CCSD). DIs and BEs at these geometries for the extended set of DFAs (including HF, M06-L, MN15-L, M11-L, M05, M06 and MN15) are provided in \mbox{\textbf{TABLES \ref{tab:S12}}} and \mbox{\textbf{\ref{tab:S13}}}.

The results discussed above contrast with those for closed-shell halogen bonds,\cite{OteroDeLaRoza2014} where DIs were found to correlate well with reference BEs (nearly linearly for NCH and slightly concave for OCH\textsubscript{2} electron donors; mostly independent of the donor and acceptor identity, with few outliers). Also, the authors\cite{OteroDeLaRoza2014} found that the delocalization error was larger for pure DFAs than for hybrid DFAs with sufficient HF exchange, such as BHandHLYP and LC-\ensuremath{\omega}PBE. Our results reveal that there is no consistent correlation between DIs and BEs (or reference BEs), even for the closed-shell systems in the CSFB-14 set. So, open-shell character alone cannot be the reason for the lack of correlation. DIs therefore should not be taken as a universal quantitative predictor of BEs. For the correlation of DIs@DFAs vs DIs@CCSD, and that of BEs@DFAs vs reference BEs, refer to \mbox{\textbf{TABLE \ref{tab:S9}}} in Supporting Information.

Coming back to the question in the title, M08-HX is not free from delocalization error. Its monomer delocalization error is larger than that of M05-2X and M06-2X (-90.6 against -77.1 and -83.8 kJ/mol for F\textsubscript{2}\textperiodcentered{}\textperiodcentered{}\textperiodcentered{}CH\textsubscript{3}; \mbox{\textbf{TABLES \ref{tab:XI}}} and \mbox{\textbf{\ref{tab:S8}}}), and its DIs are non-zero (\mbox{\textbf{FIG. \ref{fig:9}}}). Intermolecular DIs for the DF\textperiodcentered{}\textperiodcentered{}\textperiodcentered{}CH\textsubscript{3} complexes obtained from M08-HX/AVTZ level at its own geometries are, however, lower than those of M05-2X and M06-2X, consistent with its better BEs (\mbox{\textbf{TABLE \ref{tab:IX}}}). They do not correlate with BEs better than those of the other DFAs (\mbox{\textbf{TABLE \ref{tab:XII}}}).

\FloatBarrier
\section{Conclusions and outlook}\label{sec:5}

In companion paper I, we introduced new databases SEFB-15 and CSFB-14, consisting of 15 DF\textperiodcentered{}\textperiodcentered{}\textperiodcentered{}CH\textsubscript{3} and 14 DF\textperiodcentered{}\textperiodcentered{}\textperiodcentered{}NH\textsubscript{3} complexes, respectively. M06-2X was found to fail systematically for binding energies (BEs) and F-bond distances of open-shell SEFB complexes, despite being established as reliable for closed-shell halogen bonding.\cite{Bauza2013,Kozuch2013,OteroDeLaRoza2014} The present work extends this to 16 Minnesota DFAs and broad range of popular \emph{ab initio} methods.

Out of the 16 Minnesota DFAs, only 7 (M05-2X, M05-2X-GD3, M06-2X, M06-2X-GD3, M06-HF, M08-HX and M11) with AVTZ basis set predict F\textsubscript{2}\textperiodcentered{}\textperiodcentered{}\textperiodcentered{}CH\textsubscript{3}, and most of these remain unreliable for open-shell SEFB complexes despite most of them reliable for CSFB. GD3 dispersion correction to M05-2X/AVTZ and M06-2X/AVTZ worsens BEs of SEFB complexes, with only minor effect for CSFB complexes. Thus, dispersion correction is not always beneficial for BEs of open-shell SEFB complexes, as already pointed out even for closed-shell halogen bonding.\cite{Kozuch2013} Our results caution against generalizing the importance of dispersion correction for halogen bonding, as concluded by Anderson et al.\cite{Anderson2018} from the correction limited to LC-wPBE functional. Surprisingly, M08-HX/AVTZ resolves the anomaly from companion paper I, where M06-2X/AVTZ predicted comparable BEs of F\textsubscript{2}\textperiodcentered{}\textperiodcentered{}\textperiodcentered{}CH\textsubscript{3} and F\textsubscript{2}\textperiodcentered{}\textperiodcentered{}\textperiodcentered{}NH\textsubscript{3}. It yields the lowest mean absolute percentage deviation (MA\%D) on BEs among all DFAs used for both databases (18.7\% for DF\textperiodcentered{}\textperiodcentered{}\textperiodcentered{}CH\textsubscript{3}, and 7.7\% for DF\textperiodcentered{}\textperiodcentered{}\textperiodcentered{}NH\textsubscript{3}).

Among conventional \emph{ab initio} methods, CCSD/AVDZ yields F-bond distances closest to reference values for both databases (MA\%D of 1.2\% and 1.3\%). Among the same set of methods, half-half BEs outperform BSSE-corrected or uncorrected results for both databases, with few exceptions. MP2/AVDZ and MP2(full)/AVTZ are the exceptions for BEs of CSFB complexes, where BSSE correction performs (11.7\% and 8.2\%) better than those without BSSE correction (69.8\% and 42.1\%) and half-half treatment (30.5\% and 17.8\%). All-electron correlation to several \emph{ab initio} methods generally does not improve BEs or F-bond distances over valence-electron treatment at the same basis set.

Fractional charge-transfer in cationic F-bonded complexes even at infinite separation obtained from meta-GGAs, including M06-L, indicates delocalization error within the monomers involved in the F-bonded dimers. This is consistent with similar conclusions for closed-shell halogen bonding using other semi-local DFAs.\cite{OteroDeLaRoza2014} This delocalization error at monomer level is not responsible for unreliable BEs at equilibrium geometries, contrary to Steinmann et al.\cite{Steinmann2012} M08-HX functional shows this directly. It shows slightly larger monomer delocalization error than M05-2X and M06-2X do, yet predicts BEs of SEFB complexes more reliably. Bader's intermolecular delocalization indices (DIs) confirm that M08-HX is more reliable than M05-2X and M06-2X for SEFB complexes. Larger DIs predicted by M05-2X and M06-2X for SEFB complexes can explain their worse performance after GD3 dispersion correction.

This work does not report performance of various DFAs beyond Minnesota family, such as BHandHLYP, LC-wPBE and wB97X recommended for closed-shell halogen bonding.\cite{Kozuch2013,OteroDeLaRoza2014,Anderson2018} We have assessed several such DFAs with and without various dispersion correction schemes (GD3,\cite{Grimme2010} GD3BJ,\cite{Grimme2011} and XDM(see references in \cite{Johnson2017}), including the role of Hartree-Fock exchange in hybrid functional. This extension will be reported in future.

\section*{Supporting Information:}

Supporting resulting used in the present work can be found in \mbox{\textbf{Figures \ref{fig:S1}-\ref{fig:S13}}} and \mbox{\textbf{Tables \ref{tab:S1}-\ref{tab:S13}}} in Supporting Information.

\section*{ORCID}

Govinda Prasad Khanal: \url{https://orcid.org/0000-0002-0991-7005}

Elangannan Arunan: \url{https://orcid.org/0000-0001-9669-4307}

\section*{Conflicts of interest}

The authors do not have any conflicts to disclose.

\section*{Acknowledgements}

G.P.K. and E.A. thank Prof. Sai G. Ramesh for maintaining the computational facilities at the Department of Inorganic and Physical Chemistry. Prof. Debashree Ghosh is acknowledged for fruitful suggestions. The authors also acknowledge the Supercomputer Education and Research Centre, Indian Institute of Science, for providing computational resources. We further thank Mrs. Sara Taniguchi and the Q-Chem team for generously extending the Q-Chem license throughout this work.

\section*{References}
{\setstretch{1}\fontsize{8}{9.5}\selectfont\setlength{\parskip}{1pt}\justifying\setlength{\parindent}{0pt}
\bibliographystyle{gpk}\bibliography{refs}
}
\clearpage
\captionsetup{format=gpkS}\setcounter{figure}{0}\setcounter{table}{0}\renewcommand{\thefigure}{S\arabic{figure}}\renewcommand{\thetable}{S\arabic{table}}\renewcommand{\theHfigure}{S\arabic{figure}}\renewcommand{\theHtable}{S\arabic{table}}\renewcommand{\figurename}{Figure}\renewcommand{\tablename}{Table}

\section*{Supporting Information for:}

{\fontsize{15}{18}\selectfont\bfseries Single-electron fluorine bonding. II. Benchmarking Minnesota density functional approximations and popular \emph{ab initio} methods for binding energies and F-bond distances\par}\vspace{6pt}

Govinda Prasad Khanal\textsuperscript{a}, and Elangannan Arunan\textsuperscript{a}*

\textsuperscript{a}Inorganic and Physical Chemistry, Indian Institute of Science, Bangalore, India -- 560 012.

\textbf{*E-mail Id:} \href{mailto:arunan@iisc.ac.in}{\nolinkurl{arunan@iisc.ac.in}}

\begin{table}[!htbp]\captionsetup{type=figure}\centering
\par\vspace{2pt}\noindent\hfil\includegraphics[width=0.70\linewidth]{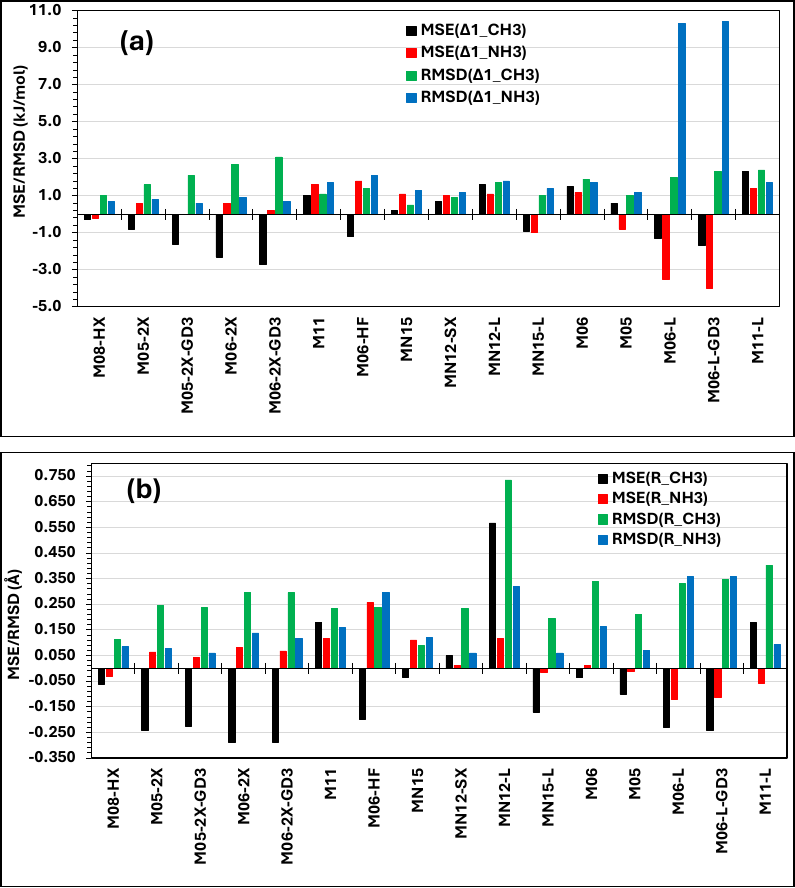}\hfil\null\par\nointerlineskip\vspace{13pt}
\par\vspace{-1pt}\caption{Mean Signed Error (MSE) and Root Mean Square Deviation (RMSD) on binding energies (BEs) \textbf{(a)} and that on the F-bond distances R(F\textperiodcentered{}\textperiodcentered{}\textperiodcentered{}C/N) \textbf{(b)} for DF\textperiodcentered{}\textperiodcentered{}\textperiodcentered{}CH\textsubscript{3}/NH\textsubscript{3} complexes, calculated at various Minnesota DFAs using aug-cc-pVTZ basis set. Reference BEs and R values were calculated at DF-RMP2-F12/CBS{[}V(T,Q)Z-F12{]}//DF-RMP2-F12/VDZ-F12 and DF-RMP2-F12/VDZ-F12 levels of theory, respectively. Only seven DFAs (M08-HX, M05-2X, M05-2X-GD3, M06-2X, M06-2X-GD3, M11 and M06-HF) produced F\textsubscript{2}\textperiodcentered{}\textperiodcentered{}\textperiodcentered{}CH\textsubscript{3} complex. These seven DFAs are used for assessing overall performance of Minnesota DFAs.}\label{fig:S1}
\vspace{10pt}
\end{table}

\begin{table}[!htbp]\captionsetup{type=figure}\centering
\par\vspace{2pt}\noindent\hfil\includegraphics[width=\linewidth]{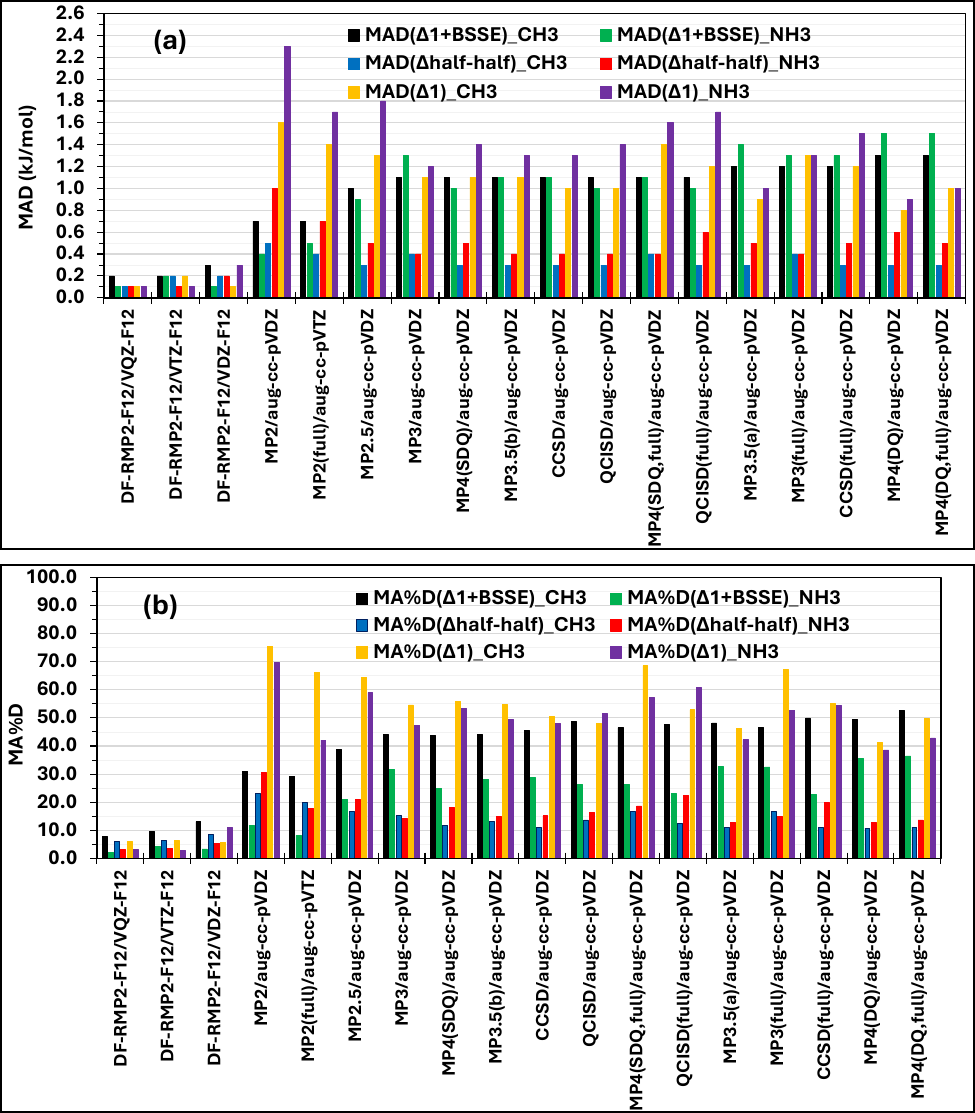}\hfil\null\par\nointerlineskip\vspace{13pt}
\par\vspace{-1pt}\caption{Mean absolute deviations (MADs) on BSSE-uncorrected (\ensuremath{\Delta}\textsubscript{1}), BSSE-corrected (\ensuremath{\Delta}\textsubscript{1+BSSE}) and half-half (\ensuremath{\Delta}\textsubscript{half-half}) binding energies (kJ/mol) \textbf{(a)}, and respective mean absolute percentage deviations (MA\%Ds) on the same \textbf{(b)} for DF\textperiodcentered{}\textperiodcentered{}\textperiodcentered{}CH\textsubscript{3}/NH\textsubscript{3} complexes using various \emph{ab initio} methods. Reference binding energies were calculated at DF-RMP2-F12/CBS{[}V(T,Q)Z-F12{]}//DF-RMP2-F12/VDZ-F12 level.}\label{fig:S2}
\vspace{10pt}
\end{table}

\begin{table}[!htbp]\captionsetup{type=figure}\centering
\par\vspace{2pt}\noindent\hfil\includegraphics[width=\linewidth]{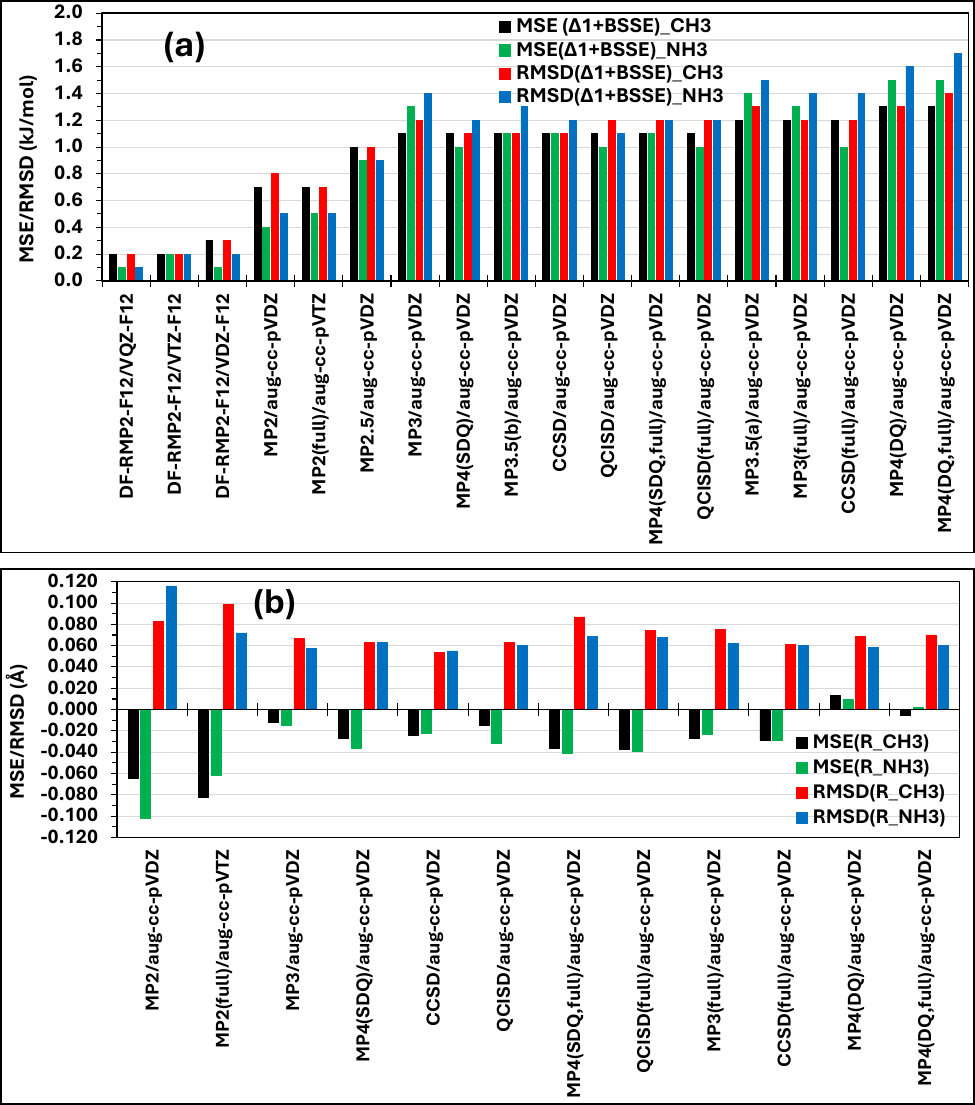}\hfil\null\par\nointerlineskip\vspace{13pt}
\par\vspace{-1pt}\caption{Mean signed errors (MSEs) and root mean square deviations (RMSDs) on BSSE-corrected binding energies (kJ/mol) \textbf{(a)}, and those on the F-bond distances (Å) \textbf{(b)}, for DF\textperiodcentered{}\textperiodcentered{}\textperiodcentered{}CH\textsubscript{3}/NH\textsubscript{3} complexes using several \emph{ab initio} methods. Reference binding energies and F-bond distances were calculated at DF-RMP2-F12/CBS{[}V(T,Q)Z-F12{]}//DF-RMP2-F12/VDZ-F12 and DF-RMP2-F12/VDZ-F12 levels of theory, respectively.}\label{fig:S3}
\vspace{10pt}
\end{table}

\begin{table}[!htbp]\captionsetup{type=figure}\centering
\par\vspace{2pt}\noindent\hfil\includegraphics[width=\linewidth]{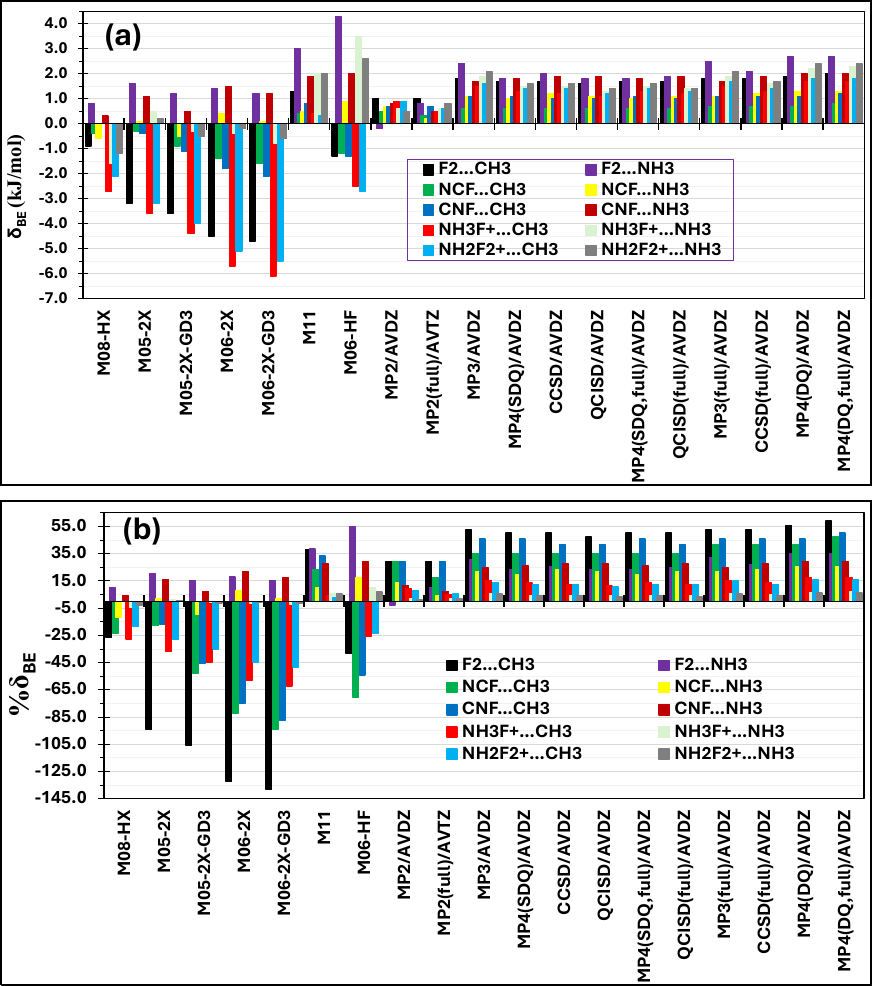}\hfil\null\par\nointerlineskip\vspace{13pt}
\par\vspace{-1pt}\caption{Deviations (\ensuremath{\delta}\textsubscript{BE}) \textbf{(a)} and percentage deviations (\%\ensuremath{\delta}\textsubscript{BE}) \textbf{(b)} on binding energies (kJ/mol) and deviations (\ensuremath{\delta}\textsubscript{R}) \textbf{(c)} and percentage deviations (\%\ensuremath{\delta}\textsubscript{R}) \textbf{(d)} on F-bond distances (Å), for selected complexes DF\textperiodcentered{}\textperiodcentered{}\textperiodcentered{}CH\textsubscript{3}/NH\textsubscript{3} (DF = F\textsubscript{2}, NCF, CNF, NH\textsubscript{3}F\textsuperscript{+} and NH\textsubscript{2}F\textsubscript{2}\textsuperscript{+}) calculated at seven Minnesota DFAs that worked for the F\textsubscript{2}\textperiodcentered{}\textperiodcentered{}\textperiodcentered{}CH\textsubscript{3} (M08-HX, M05-2X, M05-2X-GD3, M06-2X, M06-2X-GD3, M11 and M06-HF) with AVTZ basis set, and several \emph{ab initio} methods. Binding energies calculated at \emph{ab initio} methods are BSSE-corrected.}\label{fig:S4}
\vspace{10pt}
\end{table}

\begin{table}[!htbp]\captionsetup{type=figure}\centering
\par\vspace{2pt}\noindent\hfil\includegraphics[width=\linewidth]{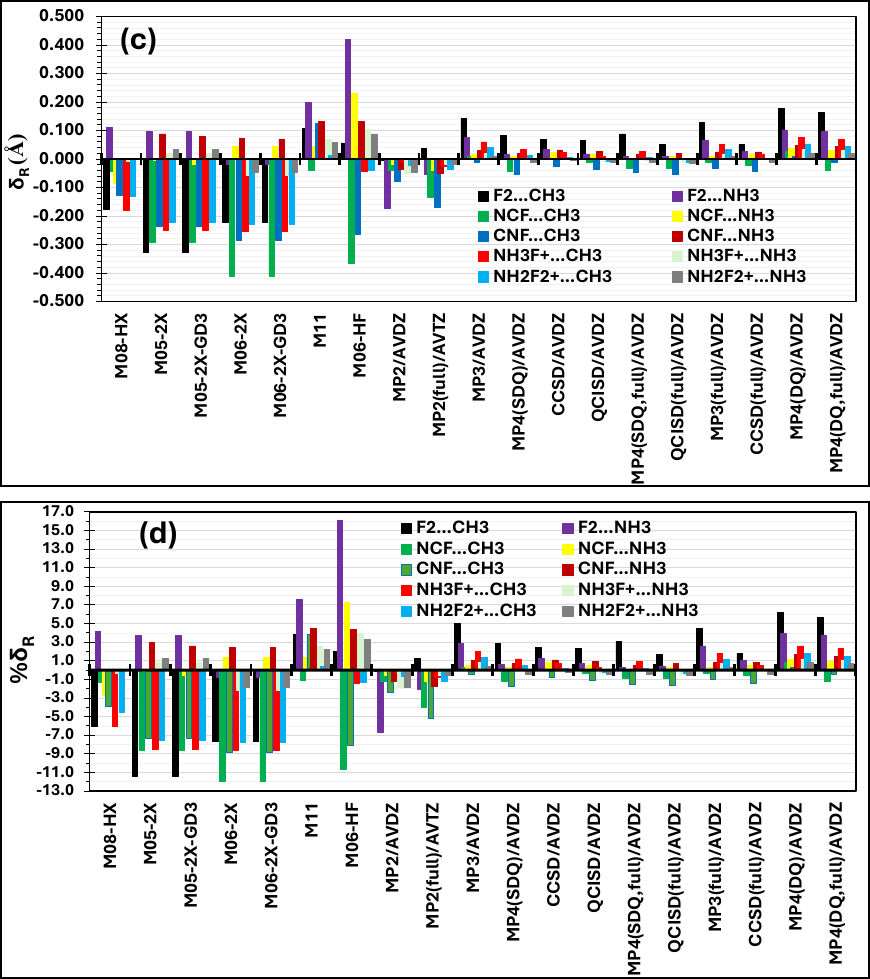}\hfil\null\par\nointerlineskip\vspace{13pt}
\par\vspace{-1pt}\ContinuedFloat\caption[]{(Continued.)}
\vspace{10pt}
\end{table}

\begin{table}[!htbp]\captionsetup{type=figure}\centering
\par\vspace{2pt}\noindent\hfil\includegraphics[width=\linewidth]{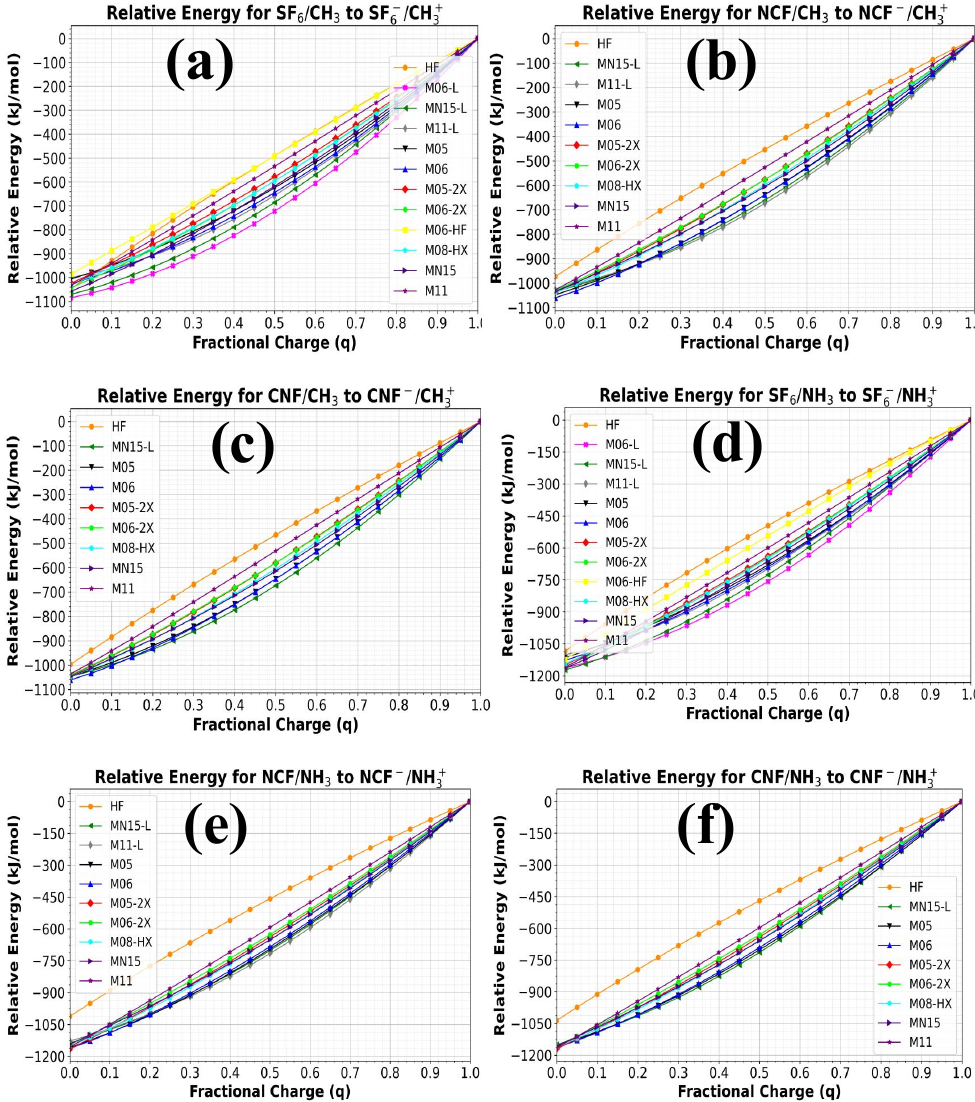}\hfil\null\par\nointerlineskip\vspace{13pt}
\par\vspace{-1pt}\caption{Plots showing relative energy (in kJ/mol) with respect to fractional charge transfer (q) for various F-bonded complexes DF\textperiodcentered{}\textperiodcentered{}\textperiodcentered{}CH\textsubscript{3}/NH\textsubscript{3} (DF = SF\textsubscript{6}, NCF and CNF), considered at infinite separation: \textbf{(a)} SF\textsubscript{6}/CH\textsubscript{3} \ensuremath{\rightarrow} SF\textsubscript{6}\textsuperscript{-}/CH\textsubscript{3}\textsuperscript{+}, \textbf{(b)} NCF/CH\textsubscript{3} \ensuremath{\rightarrow} NCF\textsuperscript{-}/CH\textsubscript{3}\textsuperscript{+}, \textbf{(c)} CNF/CH\textsubscript{3} \ensuremath{\rightarrow} CNF\textsuperscript{-}/CH\textsubscript{3}\textsuperscript{+}, \textbf{(d)} SF\textsubscript{6}/NH\textsubscript{3} \ensuremath{\rightarrow} SF\textsubscript{6}\textsuperscript{-}/NH\textsubscript{3}\textsuperscript{+}, \textbf{(e)} NCF/NH\textsubscript{3} \ensuremath{\rightarrow} NCF\textsuperscript{-}/NH\textsubscript{3}\textsuperscript{+} and \textbf{(f)} CNF/NH\textsubscript{3} \ensuremath{\rightarrow} CNF\textsuperscript{-}/NH\textsubscript{3}\textsuperscript{+}. Energies of right-hand side complexes are considered as zero to calculate relative energies for each fractional charge-transfer value. Calculations were carried out using Q-Chem 6.0.2 program\cite{Epifanovsky2021} in conjunction with AVTZ basis set.}\label{fig:S5}
\vspace{10pt}
\end{table}

\begin{table}[!htbp]\captionsetup{type=figure}\centering
\par\vspace{2pt}\noindent\hfil\includegraphics[width=\linewidth]{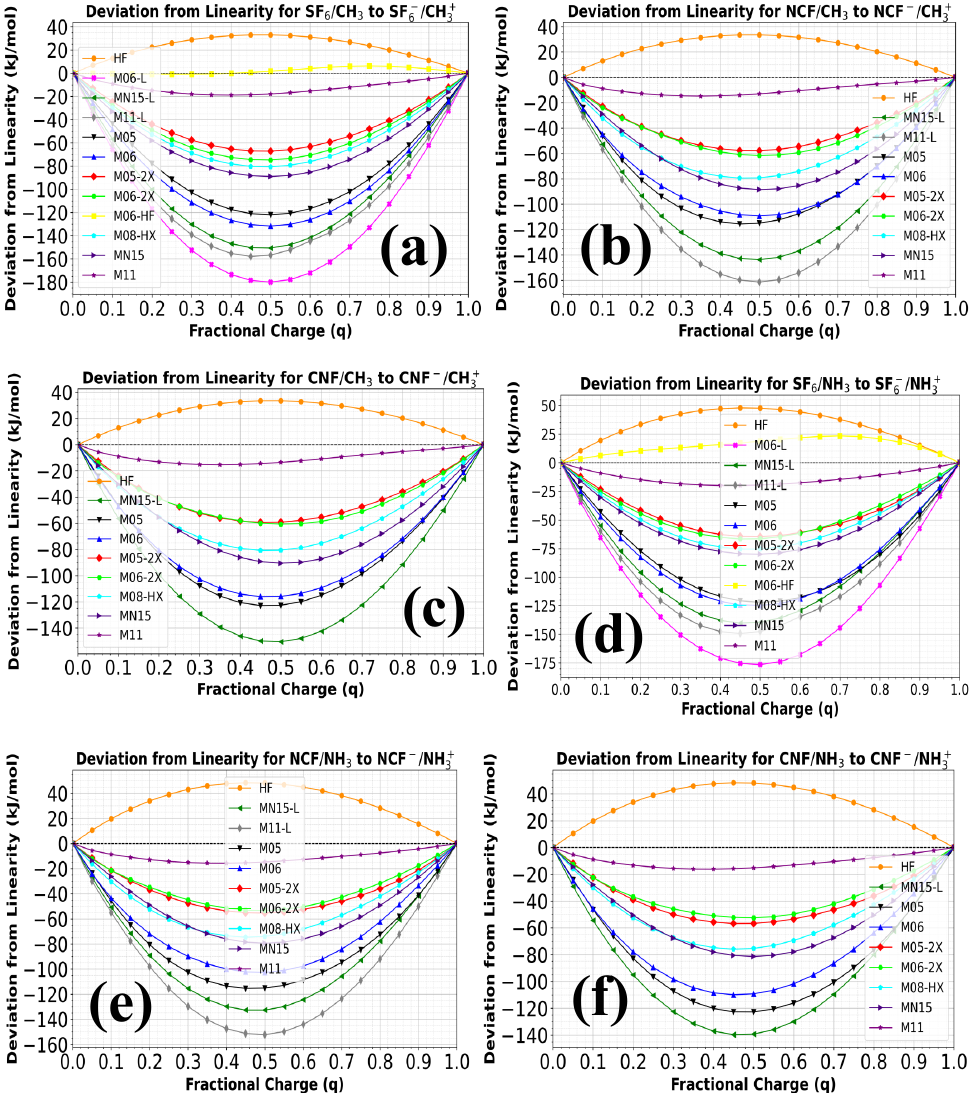}\hfil\null\par\nointerlineskip\vspace{13pt}
\par\vspace{-1pt}\caption{Plots showing deviations from linearity on total energies (in kJ/mol) with respect to fractional charge transfer (q) for various F-bonded complexes DF\textperiodcentered{}\textperiodcentered{}\textperiodcentered{}CH\textsubscript{3}/NH\textsubscript{3} (DF = SF\textsubscript{6}, NCF and CNF), considered at infinite separation: \textbf{(a)} SF\textsubscript{6}/CH\textsubscript{3} \ensuremath{\rightarrow} SF\textsubscript{6}\textsuperscript{-}/CH\textsubscript{3}\textsuperscript{+}, \textbf{(b)} NCF/CH\textsubscript{3} \ensuremath{\rightarrow} NCF\textsuperscript{-}/CH\textsubscript{3}\textsuperscript{+}, \textbf{(c)} CNF/CH\textsubscript{3} \ensuremath{\rightarrow} CNF\textsuperscript{-}/CH\textsubscript{3}\textsuperscript{+}, \textbf{(d)} SF\textsubscript{6}/NH\textsubscript{3} \ensuremath{\rightarrow} SF\textsubscript{6}\textsuperscript{-}/NH\textsubscript{3}\textsuperscript{+}, \textbf{(e)} NCF/NH\textsubscript{3} \ensuremath{\rightarrow} NCF\textsuperscript{-}/NH\textsubscript{3}\textsuperscript{+} and \textbf{(f)} CNF/NH\textsubscript{3} \ensuremath{\rightarrow} CNF\textsuperscript{-}/NH\textsubscript{3}\textsuperscript{+}. Q-Chem 6.0.2 program\cite{Epifanovsky2021} in conjunction with AVTZ basis set has been used for all calculations.}\label{fig:S6}
\vspace{10pt}
\end{table}

\begin{table}[!htbp]\captionsetup{type=figure}\centering
\par\vspace{2pt}\noindent\hfil\includegraphics[width=\linewidth]{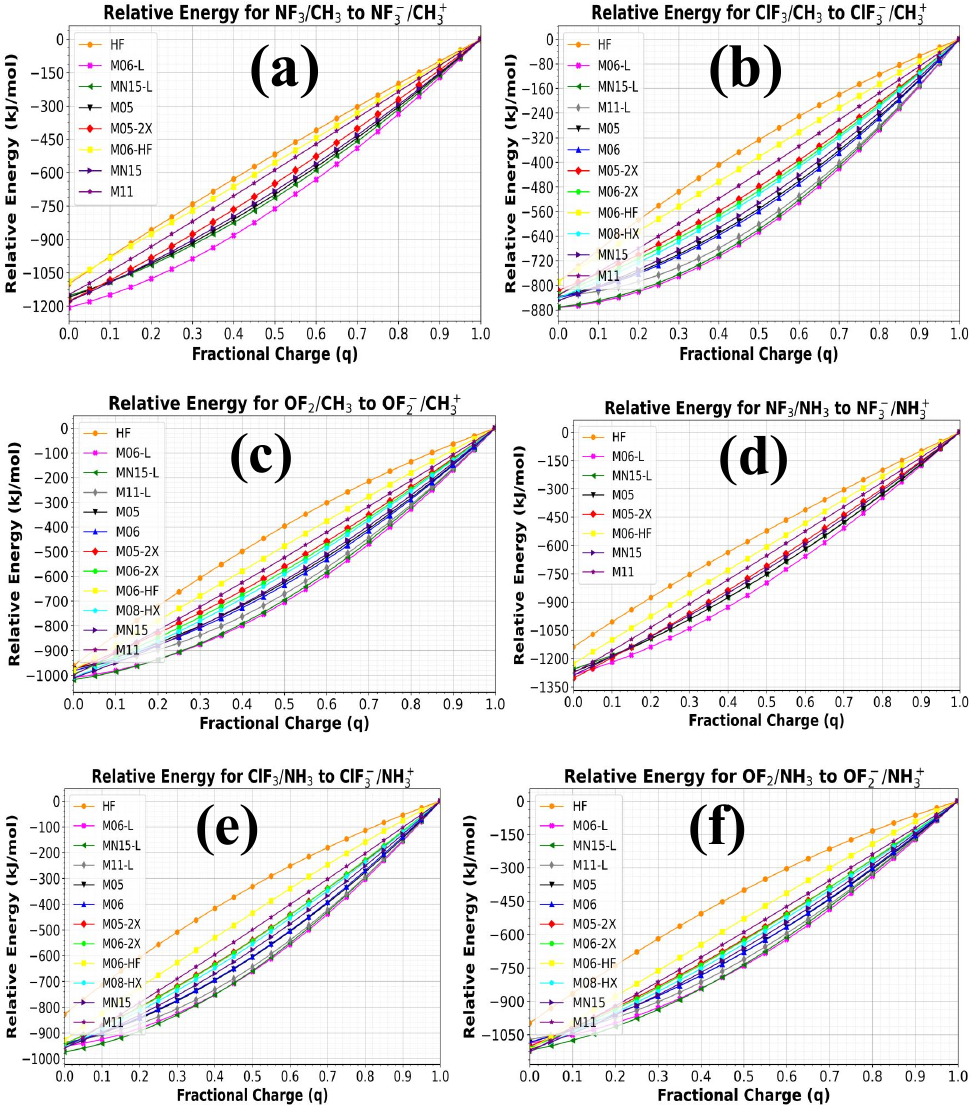}\hfil\null\par\nointerlineskip\vspace{13pt}
\par\vspace{-1pt}\caption{Plots showing relative energy (in kJ/mol) with respect to fractional charge transfer (q) for various F-bonded complexes DF\textperiodcentered{}\textperiodcentered{}\textperiodcentered{}CH\textsubscript{3}/NH\textsubscript{3} (DF = NF\textsubscript{3}, ClF\textsubscript{3} and OF\textsubscript{2}), considered at infinite separation: \textbf{(a)} NF\textsubscript{3}/CH\textsubscript{3} \ensuremath{\rightarrow} NF\textsubscript{3}\textsuperscript{-}/CH\textsubscript{3}\textsuperscript{+}, \textbf{(b)} ClF\textsubscript{3}/CH\textsubscript{3} \ensuremath{\rightarrow} ClF\textsubscript{3}\textsuperscript{-}/CH\textsubscript{3}\textsuperscript{+}, \textbf{(c)} OF\textsubscript{2}/CH\textsubscript{3} \ensuremath{\rightarrow} OF\textsubscript{2}\textsuperscript{-}/CH\textsubscript{3}\textsuperscript{+}, \textbf{(d)} NF\textsubscript{3}/NH\textsubscript{3} \ensuremath{\rightarrow} NF\textsubscript{3}\textsuperscript{-}/NH\textsubscript{3}\textsuperscript{+}, \textbf{(e)} ClF\textsubscript{3}/NH\textsubscript{3} \ensuremath{\rightarrow} ClF\textsubscript{3}\textsuperscript{-}/NH\textsubscript{3}\textsuperscript{+} and \textbf{(f)} OF\textsubscript{2}/NH\textsubscript{3} \ensuremath{\rightarrow} OF\textsubscript{2}\textsuperscript{-}/NH\textsubscript{3}\textsuperscript{+}. Energies of right-hand side complexes are considered as zero to calculate relative energies for each fractional charge-transfer value. Calculations were carried out using Q-Chem 6.0.2 program\cite{Epifanovsky2021} in conjunction with AVTZ basis set.}\label{fig:S7}
\vspace{10pt}
\end{table}

\begin{table}[!htbp]\captionsetup{type=figure}\centering
\par\vspace{2pt}\noindent\hfil\includegraphics[width=\linewidth]{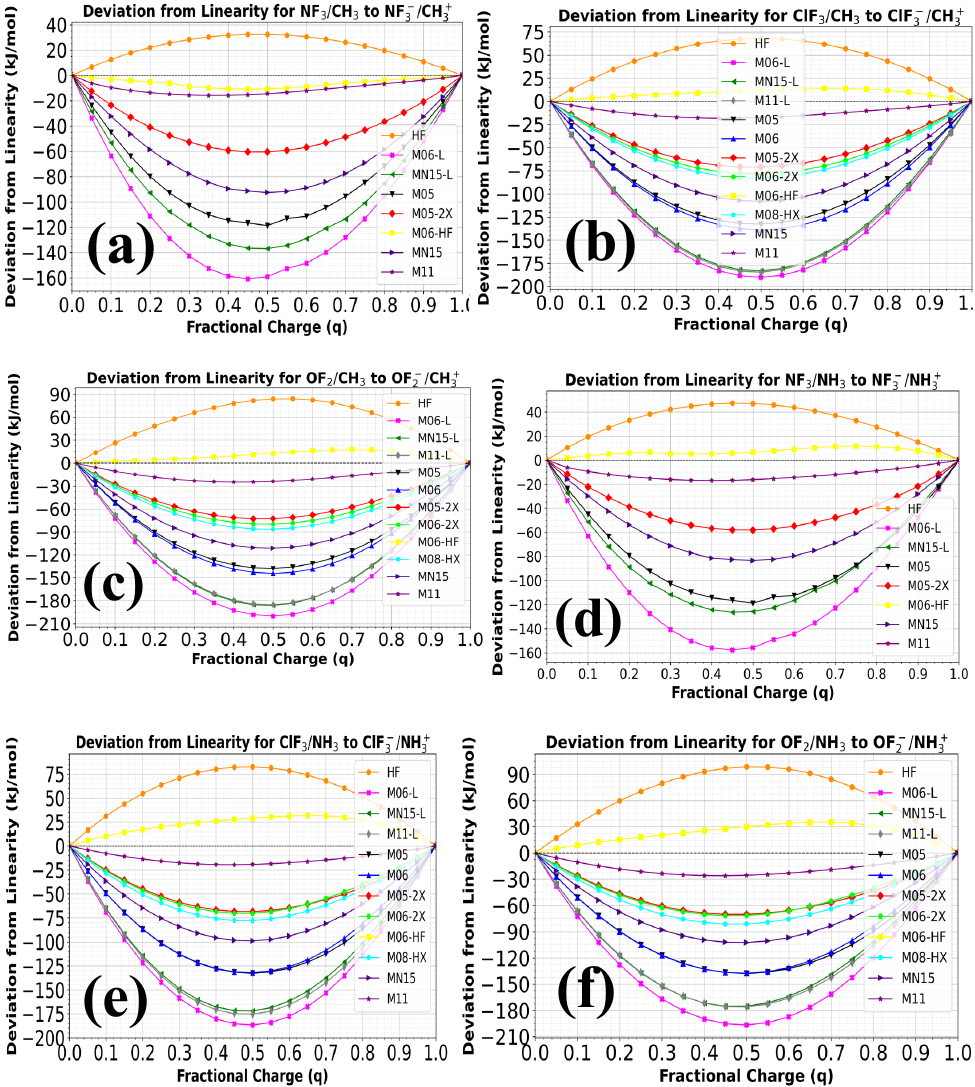}\hfil\null\par\nointerlineskip\vspace{13pt}
\par\vspace{-1pt}\caption{Plots showing deviations from linearity on total energies (in kJ/mol) with respect to fractional charge transfer (q) for various F-bonded complexes DF\textperiodcentered{}\textperiodcentered{}\textperiodcentered{}CH\textsubscript{3}/NH\textsubscript{3} (DF = NF\textsubscript{3}, ClF\textsubscript{3} and OF\textsubscript{2}), considered at infinite separation: \textbf{(a)} NF\textsubscript{3}/CH\textsubscript{3} \ensuremath{\rightarrow} NF\textsubscript{3}\textsuperscript{-}/CH\textsubscript{3}\textsuperscript{+}, \textbf{(b)} ClF\textsubscript{3}/CH\textsubscript{3} \ensuremath{\rightarrow} ClF\textsubscript{3}\textsuperscript{-}/CH\textsubscript{3}\textsuperscript{+}, \textbf{(c)} OF\textsubscript{2}/CH\textsubscript{3} \ensuremath{\rightarrow} OF\textsubscript{2}\textsuperscript{-}/CH\textsubscript{3}\textsuperscript{+}, \textbf{(d)} NF\textsubscript{3}/NH\textsubscript{3} \ensuremath{\rightarrow} NF\textsubscript{3}\textsuperscript{-}/NH\textsubscript{3}\textsuperscript{+}, \textbf{(e)} ClF\textsubscript{3}/NH\textsubscript{3} \ensuremath{\rightarrow} ClF\textsubscript{3}\textsuperscript{-}/NH\textsubscript{3}\textsuperscript{+} and \textbf{(f)} OF\textsubscript{2}/NH\textsubscript{3} \ensuremath{\rightarrow} OF\textsubscript{2}\textsuperscript{-}/NH\textsubscript{3}\textsuperscript{+}. Q-Chem 6.0.2 program\cite{Epifanovsky2021} in conjunction with AVTZ basis set has been used for all calculations.}\label{fig:S8}
\vspace{10pt}
\end{table}

\begin{table}[!htbp]\captionsetup{type=figure}\centering
\par\vspace{2pt}\noindent\hfil\includegraphics[width=\linewidth]{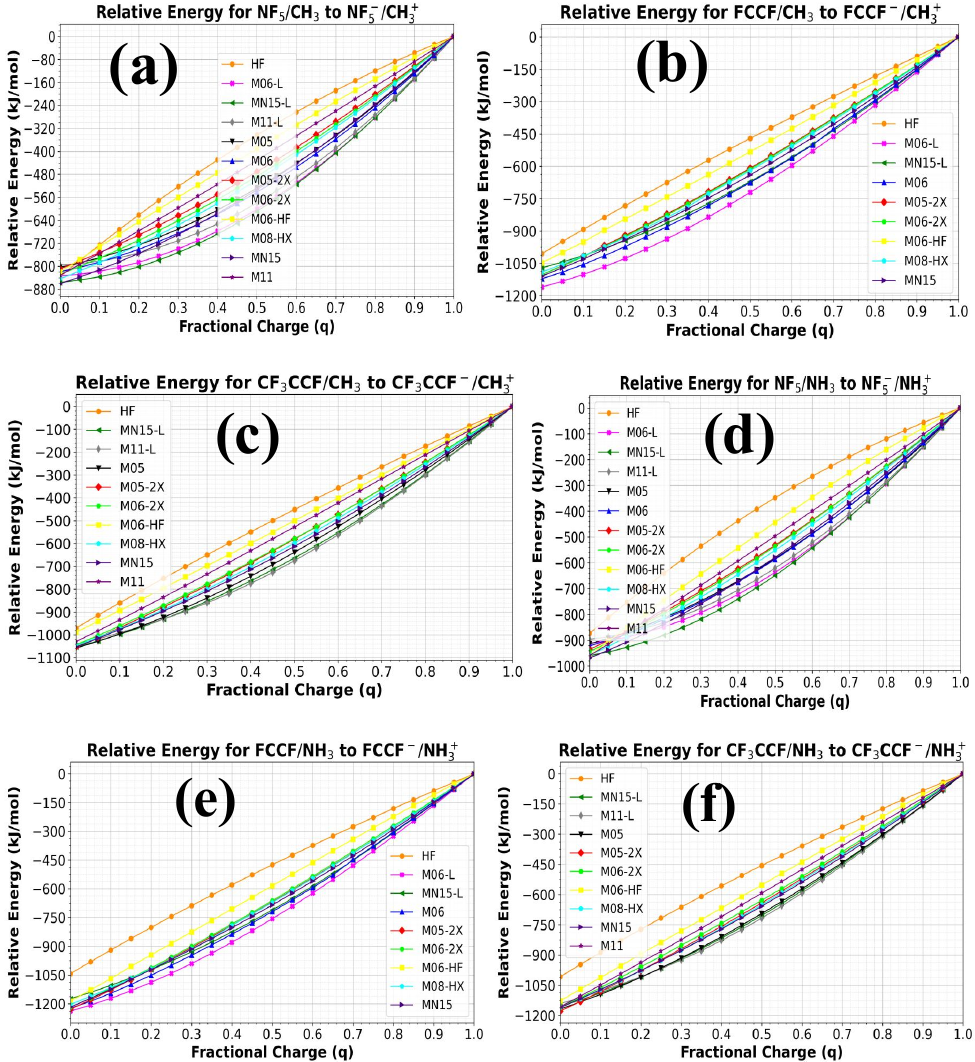}\hfil\null\par\nointerlineskip\vspace{13pt}
\par\vspace{-1pt}\caption{Plots showing relative energy (in kJ/mol) with respect to fractional charge transfer (q) for various F-bonded complexes DF\textperiodcentered{}\textperiodcentered{}\textperiodcentered{}CH\textsubscript{3} /NH\textsubscript{3} (DF = NF\textsubscript{5}, FCCF and CF\textsubscript{3}CCF), considered at infinite separation: \textbf{(a)} NF\textsubscript{5}/CH\textsubscript{3} \ensuremath{\rightarrow} NF\textsubscript{5}\textsuperscript{-}/CH\textsubscript{3}\textsuperscript{+}, \textbf{(b)} FCCF/CH\textsubscript{3} \ensuremath{\rightarrow} FCCF\textsuperscript{-}/CH\textsubscript{3}\textsuperscript{+}, \textbf{(c)} CF\textsubscript{3}CCF/CH\textsubscript{3} \ensuremath{\rightarrow} CF\textsubscript{3}CCF\textsuperscript{-}/CH\textsubscript{3}\textsuperscript{+}, \textbf{(d)} NF\textsubscript{5}/NH\textsubscript{3} \ensuremath{\rightarrow} NF\textsubscript{5}\textsuperscript{-}/NH\textsubscript{3}\textsuperscript{+}, \textbf{(e)} FCCF/NH\textsubscript{3} \ensuremath{\rightarrow} FCCF\textsuperscript{-}/NH\textsubscript{3}\textsuperscript{+} and \textbf{(f)} CF\textsubscript{3}CCF/NH\textsubscript{3} \ensuremath{\rightarrow} CF\textsubscript{3}CCF\textsuperscript{-}/NH\textsubscript{3}\textsuperscript{+}. Energies of right-hand side complexes are considered as zero to calculate relative energies for each fractional charge-transfer value. Calculations were carried out using Q-Chem 6.0.2 program\cite{Epifanovsky2021} in conjunction with AVTZ basis set.}\label{fig:S9}
\vspace{10pt}
\end{table}

\begin{table}[!htbp]\captionsetup{type=figure}\centering
\par\vspace{2pt}\noindent\hfil\includegraphics[width=\linewidth]{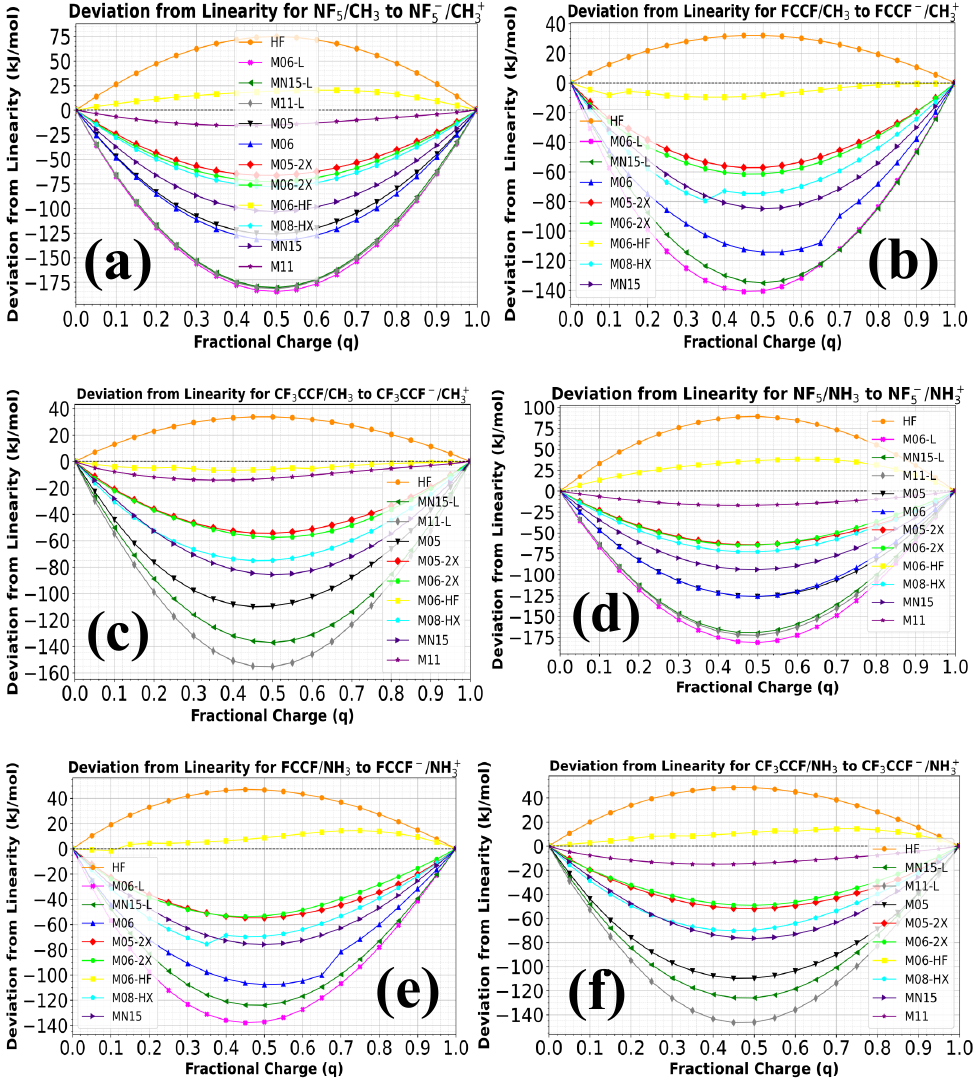}\hfil\null\par\nointerlineskip\vspace{13pt}
\par\vspace{-1pt}\caption{Plots showing deviations from linearity on total energies (in kJ/mol) with respect to fractional charge transfer (q) for various F-bonded complexes DF\textperiodcentered{}\textperiodcentered{}\textperiodcentered{}CH\textsubscript{3}/NH\textsubscript{3}(DF = NF\textsubscript{5}, FCCF and CF\textsubscript{3}CCF) considered at infinite separation: \textbf{(a)} NF\textsubscript{5}/CH\textsubscript{3} \ensuremath{\rightarrow} NF\textsubscript{5}\textsuperscript{-}/CH\textsubscript{3}\textsuperscript{+}, \textbf{(b)} FCCF/CH\textsubscript{3} \ensuremath{\rightarrow} FCCF\textsuperscript{-}/CH\textsubscript{3}\textsuperscript{+}, \textbf{(c)} CF\textsubscript{3}CCF/CH\textsubscript{3} \ensuremath{\rightarrow} CF\textsubscript{3}CCF\textsuperscript{-}/CH\textsubscript{3}\textsuperscript{+}, \textbf{(d)} NF\textsubscript{5}/NH\textsubscript{3} \ensuremath{\rightarrow} NF\textsubscript{5}\textsuperscript{-}/NH\textsubscript{3}\textsuperscript{+}, \textbf{(e)} FCCF/NH\textsubscript{3} \ensuremath{\rightarrow} FCCF\textsuperscript{-}/NH\textsubscript{3}\textsuperscript{+} and \textbf{(f)} CF\textsubscript{3}CCF/NH\textsubscript{3} \ensuremath{\rightarrow} CF\textsubscript{3}CCF\textsuperscript{-}/NH\textsubscript{3}\textsuperscript{+}. Q-Chem 6.0.2 program\cite{Epifanovsky2021} in conjunction with AVTZ basis set has been used for all calculations.}\label{fig:S10}
\vspace{10pt}
\end{table}

\begin{table}[!htbp]\captionsetup{type=figure}\centering
\par\vspace{2pt}\noindent\hfil\includegraphics[width=\linewidth]{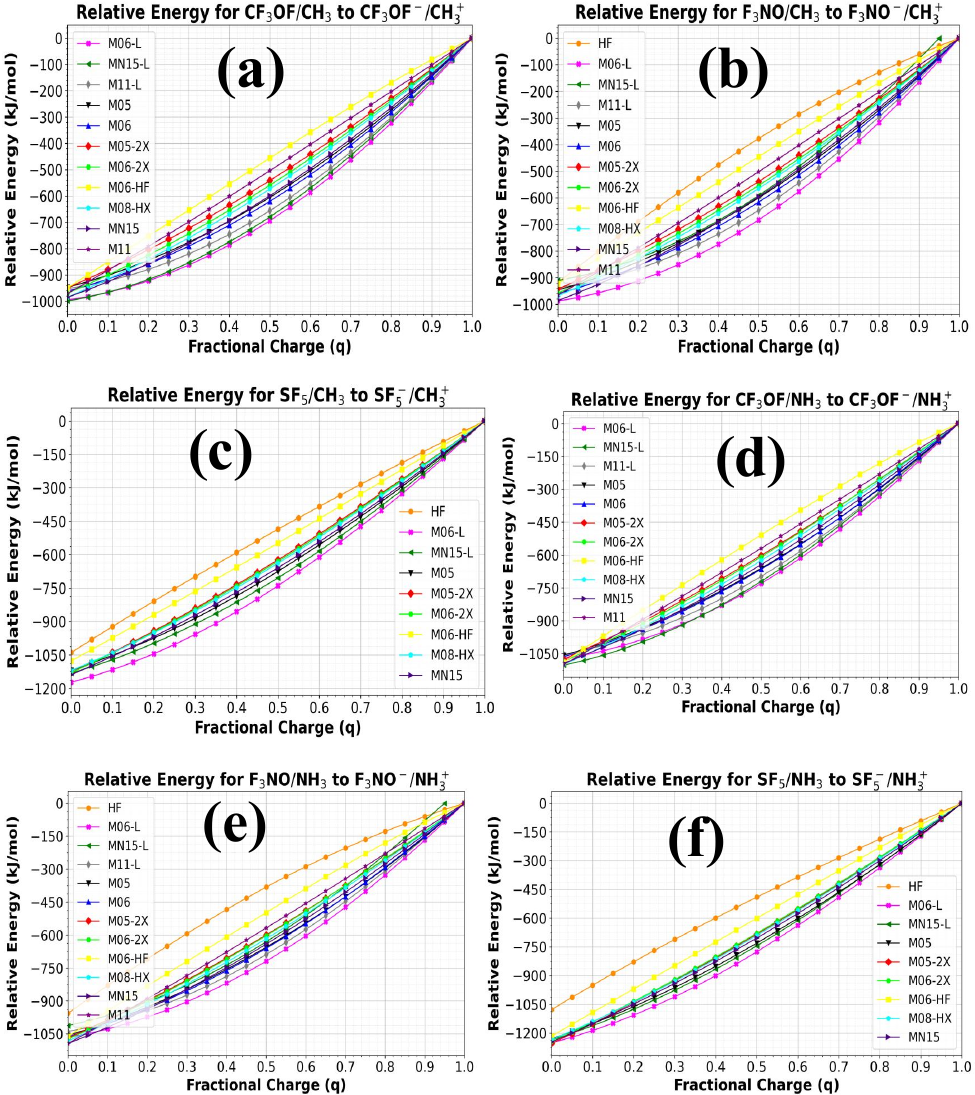}\hfil\null\par\nointerlineskip\vspace{13pt}
\par\vspace{-1pt}\caption{Plots showing relative energy (in kJ/mol) with respect to fractional charge transfer (q) for various F-bonded complexes DF\textperiodcentered{}\textperiodcentered{}\textperiodcentered{}CH\textsubscript{3}/NH\textsubscript{3} (DF = CF\textsubscript{3}OF, F\textsubscript{3}NO and SF\textsubscript{5}), considered at infinite separation: \textbf{(a)} CF\textsubscript{3}OF/CH\textsubscript{3} \ensuremath{\rightarrow} CF\textsubscript{3}OF\textsuperscript{-}/CH\textsubscript{3}\textsuperscript{+}, \textbf{(b)} F\textsubscript{3}NO/CH\textsubscript{3} \ensuremath{\rightarrow} F\textsubscript{3}NO\textsuperscript{-}/CH\textsubscript{3}\textsuperscript{+}, \textbf{(c)} SF\textsubscript{5}/CH\textsubscript{3} \ensuremath{\rightarrow} SF\textsubscript{5}\textsuperscript{-}/CH\textsubscript{3}\textsuperscript{+}, \textbf{(d)} CF\textsubscript{3}OF/NH\textsubscript{3} \ensuremath{\rightarrow} CF\textsubscript{3}OF\textsuperscript{-}/NH\textsubscript{3}\textsuperscript{+}, \textbf{(e)} F\textsubscript{3}NO/NH\textsubscript{3} \ensuremath{\rightarrow} F\textsubscript{3}NO\textsuperscript{-}/NH\textsubscript{3}\textsuperscript{+} and \textbf{(f)} SF\textsubscript{5}/NH\textsubscript{3} \ensuremath{\rightarrow} SF\textsubscript{5}\textsuperscript{-}/NH\textsubscript{3}\textsuperscript{+}. Energies of right-hand side complexes are considered as zero to calculate relative energies for each fractional charge-transfer value. Calculations were carried out using Q-Chem 6.0.2 program\cite{Epifanovsky2021} in conjunction with AVTZ basis set.}\label{fig:S11}
\vspace{10pt}
\end{table}

\begin{table}[!htbp]\captionsetup{type=figure}\centering
\par\vspace{2pt}\noindent\hfil\includegraphics[width=\linewidth]{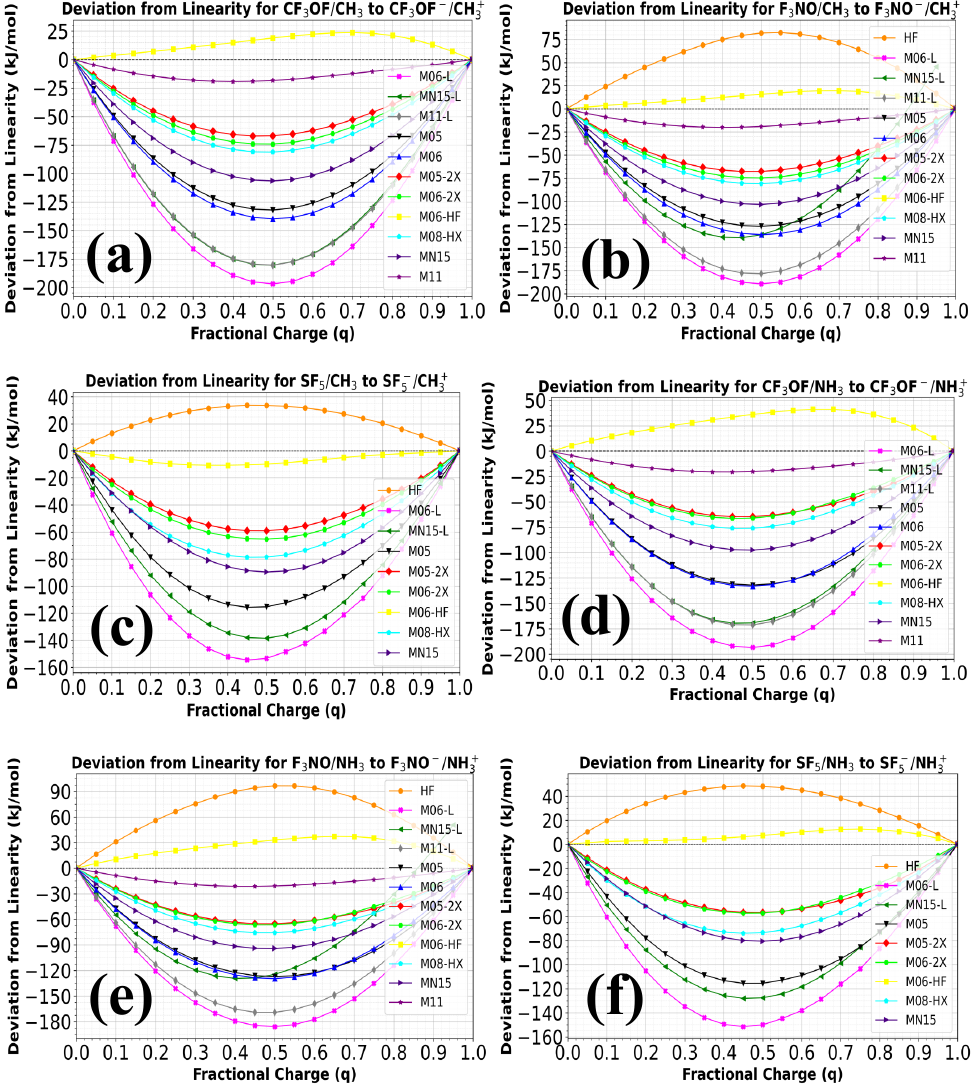}\hfil\null\par\nointerlineskip\vspace{13pt}
\par\vspace{-1pt}\caption{Plots showing deviations from linearity on total energies (in kJ/mol) with respect to fractional charge transfer (q) for various F-bonded complexes DF\textperiodcentered{}\textperiodcentered{}\textperiodcentered{}CH\textsubscript{3}/NH\textsubscript{3} (DF = CF\textsubscript{3}OF, F\textsubscript{3}NO and SF\textsubscript{5}) considered at infinite separation: \textbf{(a)} CF\textsubscript{3}OF/CH\textsubscript{3} \ensuremath{\rightarrow} CF\textsubscript{3}OF\textsuperscript{-}/CH\textsubscript{3}\textsuperscript{+}, \textbf{(b)} F\textsubscript{3}NO/CH\textsubscript{3} \ensuremath{\rightarrow} F\textsubscript{3}NO\textsuperscript{-}/CH\textsubscript{3}\textsuperscript{+}, \textbf{(c)} SF\textsubscript{5}/CH\textsubscript{3} \ensuremath{\rightarrow} SF\textsubscript{5}\textsuperscript{-}/CH\textsubscript{3}\textsuperscript{+}, \textbf{(d)} CF\textsubscript{3}OF/NH\textsubscript{3} \ensuremath{\rightarrow} CF\textsubscript{3}OF\textsuperscript{-}/NH\textsubscript{3}\textsuperscript{+}, \textbf{(e)} F\textsubscript{3}NO/NH\textsubscript{3} \ensuremath{\rightarrow} F\textsubscript{3}NO\textsuperscript{-}/NH\textsubscript{3}\textsuperscript{+} and \textbf{(f)} SF\textsubscript{5}/NH\textsubscript{3} \ensuremath{\rightarrow} SF\textsubscript{5}\textsuperscript{-}/NH\textsubscript{3}\textsuperscript{+}. Q-Chem 6.0.2 program\cite{Epifanovsky2021} in conjunction with AVTZ basis set has been used for all calculations.}\label{fig:S12}
\vspace{10pt}
\end{table}

\begin{table}[!htbp]\captionsetup{type=figure}\centering
\par\vspace{2pt}\noindent\hfil\includegraphics[width=\linewidth]{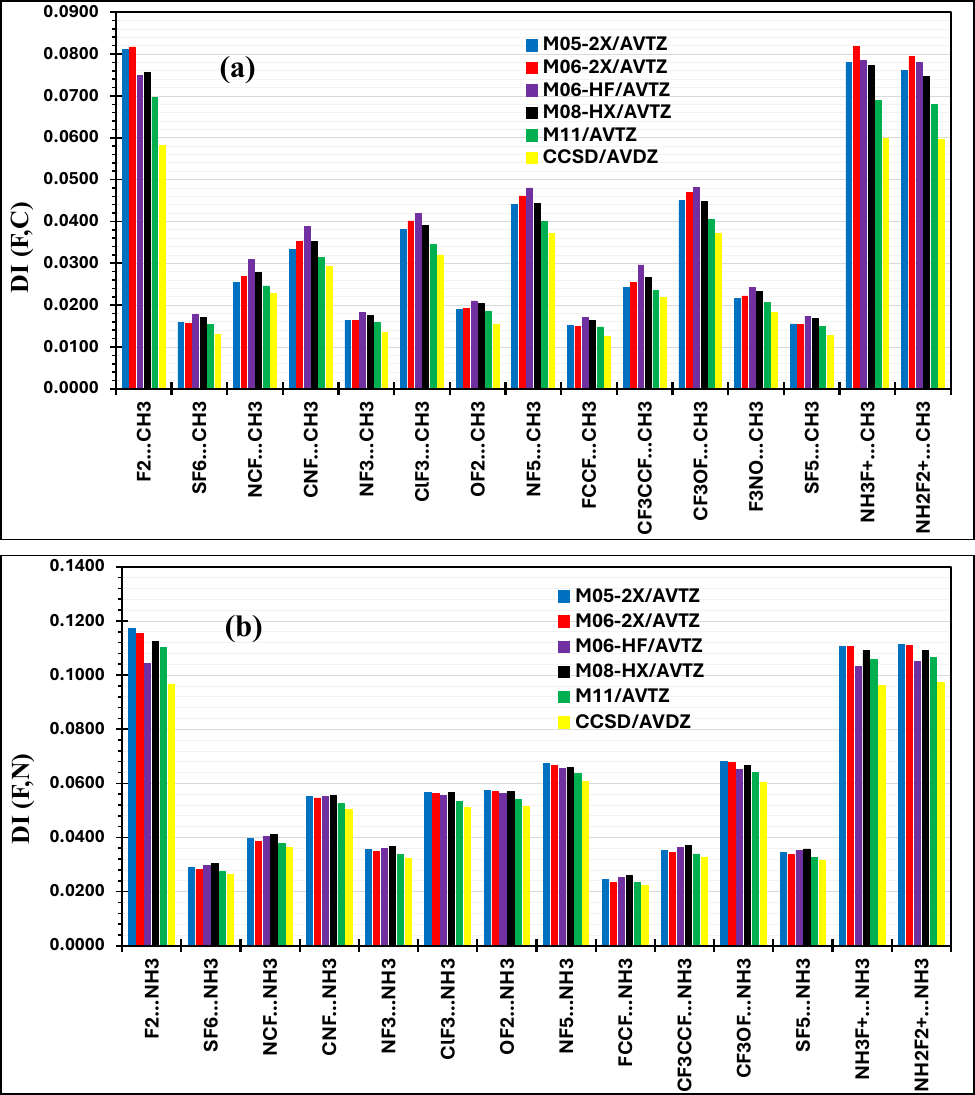}\hfil\null\par\nointerlineskip\vspace{13pt}
\par\vspace{-1pt}\caption{Bader's intermolecular delocalization indices {[}DI(F, C/N){]} for DF\textperiodcentered{}\textperiodcentered{}\textperiodcentered{}CH\textsubscript{3} \textbf{(a)} and DF\textperiodcentered{}\textperiodcentered{}\textperiodcentered{}NH\textsubscript{3} \textbf{(b)} complexes obtained from selected Minnesota DFAs at dimer geometries using geometries optimized at DF-RMP2-F12/VDZ-F12 level of theory. See \mbox{\textbf{Tables \ref{tab:S12}}} and \mbox{\textbf{\ref{tab:S13}}} for respective DIs and BEs, respectively obtained at M05-2X-GD3/AVTZ and M06-2X-GD3/AVTZ level of theories.}\label{fig:S13}
\vspace{10pt}
\end{table}

\begin{table}[!htbp]\centering\setlength{\parskip}{0pt}
\caption{Binding energies (BEs) without BSSE correction \ensuremath{\Delta}\textsubscript{1} (in kJ/\allowbreak{}mol) for F-bonded complexes DF\textperiodcentered{}\textperiodcentered{}\textperiodcentered{}CH\textsubscript{3}/NH\textsubscript{3}\textsubscript{ }using sixteen Minnesota functionals and reference BEs at DF-RMP2-F12/\allowbreak{}CBS[V(T,Q)Z-F12]//DF-RMP2-F12/\allowbreak{}VDZ-F12 level.}\label{tab:S1}\vspace{7pt}\nopagebreak
\adjustbox{width=592pt,max totalheight=0.86\textheight,center}{\fontfamily{ptm}\fontsize{9.2}{10.6}\selectfont\renewcommand{\arraystretch}{1.12}\setlength{\tabcolsep}{1.5pt}
\begin{tabular}{|>{\centering\arraybackslash}m{80.3pt}|>{\centering\arraybackslash}m{27.1pt}|>{\centering\arraybackslash}b{27.1pt}|>{\centering\arraybackslash}m{27.1pt}|>{\centering\arraybackslash}m{27.1pt}|>{\centering\arraybackslash}m{27.1pt}|>{\centering\arraybackslash}b{27.1pt}|>{\centering\arraybackslash}m{27.1pt}|>{\centering\arraybackslash}p{27.1pt}|>{\centering\arraybackslash}m{19.5pt}|>{\centering\arraybackslash}m{27.1pt}|>{\centering\arraybackslash}m{27.1pt}|>{\centering\arraybackslash}m{27.1pt}|>{\centering\arraybackslash}m{27.1pt}|>{\centering\arraybackslash}m{27.1pt}|>{\centering\arraybackslash}m{27.1pt}|>{\centering\arraybackslash}m{19.5pt}|>{\centering\arraybackslash}m{44.8pt}|}
\hline
\multicolumn{1}{|>{\centering\arraybackslash}m{80.3pt}|}{\textbf{DF\textperiodcentered{}\textperiodcentered{}\textperiodcentered{}CH}\textbf{\textsubscript{3}}} & \makebox[\linewidth][c]{\rotatebox{90}{\rbxs{\textbf{M05}}}} & \multicolumn{1}{>{\centering\arraybackslash}m{27.1pt}|}{\makebox[\linewidth][c]{\rotatebox{90}{\rbxs{\textbf{M05-2X}}}}} & \makebox[\linewidth][c]{\rotatebox{90}{\rbxs{\textbf{M05-2X-GD3}}}} & \makebox[\linewidth][c]{\rotatebox{90}{\rbxs{\textbf{M06}}}} & \makebox[\linewidth][c]{\rotatebox{90}{\rbxs{\textbf{M06-2X}}}} & \multicolumn{1}{>{\centering\arraybackslash}m{27.1pt}|}{\makebox[\linewidth][c]{\rotatebox{90}{\rbxs{\textbf{M06-2X-GD3}}}}} & \makebox[\linewidth][c]{\rotatebox{90}{\rbxs{\textbf{M06-HF}}}} & \multicolumn{1}{>{\centering\arraybackslash}m{27.1pt}|}{\makebox[\linewidth][c]{\rotatebox{90}{\rbxs{\textbf{M08-HX}}}}} & \makebox[\linewidth][c]{\rotatebox{90}{\rbxs{\textbf{M11}}}} & \makebox[\linewidth][c]{\rotatebox{90}{\rbxs{\textbf{MN12-SX}}}} & \makebox[\linewidth][c]{\rotatebox{90}{\rbxs{\textbf{MN15}}}} & \makebox[\linewidth][c]{\rotatebox{90}{\rbxs{\textbf{M06-L}}}} & \makebox[\linewidth][c]{\rotatebox{90}{\rbxs{\textbf{M06-L} \textbf{-GD3}}}} & \makebox[\linewidth][c]{\rotatebox{90}{\rbxs{\textbf{M11-L}}}} & \makebox[\linewidth][c]{\rotatebox{90}{\rbxs{\textbf{MN12-L}}}} & \makebox[\linewidth][c]{\rotatebox{90}{\rbxs{\textbf{MN15-L}}}} & \makebox[\linewidth][c]{\rotatebox{90}{\rbxs{\textbf{DF-RMP2-F12}\textbf{/}\textbf{ CBS}\textbf{[}\textbf{V(}\textbf{T,Q}\textbf{)Z-} \textbf{F12]}}}}  \\
\hline
\end{tabular}}
\par\nopagebreak\vspace{3pt}{\setstretch{1}\setlength{\parskip}{1.5pt}\fontsize{9}{10.5}\selectfont\raggedright\noindent \textsuperscript{b}Convergence error during optimization of respective F-bonded complexes.\par}

\end{table}

\begin{table}[!htbp]\centering\setlength{\parskip}{0pt}
\caption{F-bond distance R(F\textperiodcentered{}\textperiodcentered{}\textperiodcentered{}C/\allowbreak{}N) (in Å) for F-bonded complexes DF\textperiodcentered{}\textperiodcentered{}\textperiodcentered{}CH\textsubscript{3}/NH\textsubscript{3} using sixteen Minnesota functionals and reference R values at DF-RMP2-F12/\allowbreak{}VDZ-F12 level.}\label{tab:S2}\vspace{7pt}\nopagebreak
\adjustbox{width=592pt,max totalheight=0.86\textheight,center}{\fontfamily{ptm}\fontsize{9.2}{10.6}\selectfont\renewcommand{\arraystretch}{1.4}\setlength{\tabcolsep}{1.5pt}
\begin{tabular}{|>{\centering\arraybackslash}m{80.1pt}|>{\centering\arraybackslash}m{26.6pt}|>{\centering\arraybackslash}b{26.6pt}|>{\centering\arraybackslash}m{26.6pt}|>{\centering\arraybackslash}m{26.6pt}|>{\centering\arraybackslash}m{33.8pt}|>{\centering\arraybackslash}b{20.5pt}|>{\centering\arraybackslash}m{8.8pt}>{\centering\arraybackslash}m{20.5pt}|>{\centering\arraybackslash}m{26.6pt}|>{\centering\arraybackslash}m{20.5pt}|>{\centering\arraybackslash}m{26.6pt}|>{\centering\arraybackslash}m{26.6pt}|>{\centering\arraybackslash}m{26.6pt}|>{\centering\arraybackslash}m{26.6pt}|>{\centering\arraybackslash}m{26.6pt}|>{\centering\arraybackslash}m{26.6pt}|>{\centering\arraybackslash}m{25.3pt}|>{\centering\arraybackslash}m{37.6pt}|}
\hline
\multicolumn{1}{|>{\centering\arraybackslash}m{80.1pt}|}{\textbf{DF}\textbf{\textperiodcentered{}\textperiodcentered{}\textperiodcentered{}}\textbf{CH}\textbf{\textsubscript{3}}} & \makebox[\linewidth][c]{\rotatebox{90}{\rbxs{\textbf{M05}}}} & \multicolumn{1}{>{\centering\arraybackslash}m{26.6pt}|}{\makebox[\linewidth][c]{\rotatebox{90}{\rbxs{\textbf{M05-2X}}}}} & \makebox[\linewidth][c]{\rotatebox{90}{\rbxs{\textbf{M05-2X-GD3}}}} & \makebox[\linewidth][c]{\rotatebox{90}{\rbxs{\textbf{M06}}}} & \makebox[\linewidth][c]{\rotatebox{90}{\rbxs{\textbf{M06-2X}}}} & \multicolumn{1}{>{\centering\arraybackslash}m{20.5pt}|}{\makebox[\linewidth][c]{\rotatebox{90}{\rbxs{\textbf{M06-2X-GD3}}}}} & \multicolumn{2}{>{\centering\arraybackslash}m{32.6pt}|}{\makebox[\linewidth][c]{\rotatebox{90}{\rbxs{\textbf{M06-HF}}}}} & \makebox[\linewidth][c]{\rotatebox{90}{\rbxs{\textbf{M08-HX}}}} & \makebox[\linewidth][c]{\rotatebox{90}{\rbxs{\textbf{M11}}}} & \makebox[\linewidth][c]{\rotatebox{90}{\rbxs{\textbf{MN12-SX}}}} & \makebox[\linewidth][c]{\rotatebox{90}{\rbxs{\textbf{MN15}}}} & \makebox[\linewidth][c]{\rotatebox{90}{\rbxs{\textbf{M06-L}}}} & \makebox[\linewidth][c]{\rotatebox{90}{\rbxs{\textbf{M06-L} \textbf{-GD3}}}} & \makebox[\linewidth][c]{\rotatebox{90}{\rbxs{\textbf{M11-L}}}} & \makebox[\linewidth][c]{\rotatebox{90}{\rbxs{\textbf{MN12-L}}}} & \makebox[\linewidth][c]{\rotatebox{90}{\rbxs{\textbf{MN15-L}}}} & \makebox[\linewidth][c]{\rotatebox{90}{\rbxs{\textbf{DF-RMP2-F12/\allowbreak{}VDZ-F12}}}}  \\
\hline
\end{tabular}}
\par\nopagebreak\vspace{3pt}{\setstretch{1}\setlength{\parskip}{1.5pt}\fontsize{9}{10.5}\selectfont\raggedright\noindent \textsuperscript{b}Convergence error during optimization of respective F-bonded complexes.\par}

\end{table}

\begin{table}[!htbp]\centering\setlength{\parskip}{0pt}
\caption{F\textbf{-}bond angles (\ensuremath{\angle}DF\textperiodcentered{}\textperiodcentered{}\textperiodcentered{}C/\allowbreak{}N in \textdegree{}) for F-bonded complexes DF\textperiodcentered{}\textperiodcentered{}\textperiodcentered{}CH\textsubscript{3}/NH\textsubscript{3} using sixteen Minnesota functionals and reference F-bond angles at DF-RMP2-F12/\allowbreak{}VDZ-F12 level.}\label{tab:S3}\vspace{7pt}\nopagebreak
\adjustbox{width=592pt,max totalheight=0.86\textheight,center}{\fontfamily{ptm}\fontsize{9.2}{10.6}\selectfont\renewcommand{\arraystretch}{1.4}\setlength{\tabcolsep}{1.5pt}
\begin{tabular}{|>{\centering\arraybackslash}m{78.7pt}|>{\centering\arraybackslash}m{26.5pt}|>{\centering\arraybackslash}b{26.5pt}|>{\centering\arraybackslash}m{26.5pt}|>{\centering\arraybackslash}m{26.5pt}|>{\centering\arraybackslash}m{33.9pt}|>{\centering\arraybackslash}m{26.5pt}|>{\centering\arraybackslash}m{26.5pt}|>{\centering\arraybackslash}m{26.5pt}|>{\centering\arraybackslash}m{26.5pt}|>{\centering\arraybackslash}m{26.5pt}|>{\centering\arraybackslash}m{26.5pt}|>{\centering\arraybackslash}m{26.5pt}|>{\centering\arraybackslash}m{26.5pt}|>{\centering\arraybackslash}m{26.5pt}|>{\centering\arraybackslash}m{26.5pt}|>{\centering\arraybackslash}m{26.5pt}|>{\centering\arraybackslash}m{33.9pt}|}
\hline
\multicolumn{1}{|>{\centering\arraybackslash}m{78.7pt}|}{\textbf{DF}\textbf{\textperiodcentered{}\textperiodcentered{}\textperiodcentered{}}\textbf{CH}\textbf{\textsubscript{3}}} & \makebox[\linewidth][c]{\rotatebox{90}{\rbxs{\textbf{M05}}}} & \multicolumn{1}{>{\centering\arraybackslash}m{26.5pt}|}{\makebox[\linewidth][c]{\rotatebox{90}{\rbxs{\textbf{M05-2X}}}}} & \makebox[\linewidth][c]{\rotatebox{90}{\rbxs{\textbf{M05-2X-GD3}}}} & \makebox[\linewidth][c]{\rotatebox{90}{\rbxs{\textbf{M06}}}} & \makebox[\linewidth][c]{\rotatebox{90}{\rbxs{\textbf{M06-2X}}}} & \makebox[\linewidth][c]{\rotatebox{90}{\rbxs{\textbf{M06-2X-GD3}}}} & \makebox[\linewidth][c]{\rotatebox{90}{\rbxs{\textbf{M06-HF}}}} & \makebox[\linewidth][c]{\rotatebox{90}{\rbxs{\textbf{M08-HX}}}} & \makebox[\linewidth][c]{\rotatebox{90}{\rbxs{\textbf{M11}}}} & \makebox[\linewidth][c]{\rotatebox{90}{\rbxs{\textbf{MN12-SX}}}} & \makebox[\linewidth][c]{\rotatebox{90}{\rbxs{\textbf{MN15}}}} & \makebox[\linewidth][c]{\rotatebox{90}{\rbxs{\textbf{M06-L}}}} & \makebox[\linewidth][c]{\rotatebox{90}{\rbxs{\textbf{M06-L} \textbf{-GD3}}}} & \makebox[\linewidth][c]{\rotatebox{90}{\rbxs{\textbf{M11-L}}}} & \makebox[\linewidth][c]{\rotatebox{90}{\rbxs{\textbf{MN12-L}}}} & \makebox[\linewidth][c]{\rotatebox{90}{\rbxs{\textbf{MN15-L}}}} & \makebox[\linewidth][c]{\rotatebox{90}{\rbxs{\textbf{DF-RMP2-F12}\textbf{/VDZ-F12}}}}  \\
\hline
\end{tabular}}
\par\nopagebreak\vspace{3pt}{\setstretch{1}\setlength{\parskip}{1.5pt}\fontsize{9}{10.5}\selectfont\raggedright\noindent \textsuperscript{b}Convergence error during optimization of respective F-bonded complexes.\par}

\end{table}

\begin{sidewaystable}[!htbp]\centering\setlength{\parskip}{0pt}
\caption{Binding energies (BEs) (in kJ/\allowbreak{}mol) with BSSE correction (\ensuremath{\Delta}\textsubscript{1+BSS}) and without BSSE-correction (\ensuremath{\Delta}\textsubscript{1}) for F-bonded complexes DF\textperiodcentered{}\textperiodcentered{}\textperiodcentered{}CH\textsubscript{3}/NH\textsubscript{3} using various \emph{ab}\emph{ }\emph{initio} methods.}\label{tab:S4}\vspace{7pt}\nopagebreak
\makebox[\textheight][c]{\adjustbox{width=688pt,max totalheight=440pt}{\fontfamily{ptm}\fontsize{8.03}{9.2}\selectfont\renewcommand{\arraystretch}{1.12}\setlength{\tabcolsep}{1.5pt}
\begin{tabular}{|>{\centering\arraybackslash}m{73.2pt}|>{\centering\arraybackslash}m{18.6pt}>{\centering\arraybackslash}m{18.6pt}>{\centering\arraybackslash}m{18.6pt}>{\centering\arraybackslash}m{18.6pt}|>{\centering\arraybackslash}m{18.6pt}>{\centering\arraybackslash}m{18.6pt}|>{\centering\arraybackslash}m{18.6pt}>{\centering\arraybackslash}m{18.6pt}|>{\centering\arraybackslash}m{18.6pt}>{\centering\arraybackslash}m{18.6pt}|>{\centering\arraybackslash}m{18.6pt}>{\centering\arraybackslash}m{18.6pt}|>{\centering\arraybackslash}m{18.6pt}>{\centering\arraybackslash}m{18.9pt}|>{\centering\arraybackslash}m{18.6pt}>{\centering\arraybackslash}m{18.9pt}|>{\centering\arraybackslash}m{18.9pt}>{\centering\arraybackslash}m{18.9pt}|>{\centering\arraybackslash}m{18.9pt}>{\centering\arraybackslash}m{18.9pt}|>{\centering\arraybackslash}m{18.9pt}>{\centering\arraybackslash}m{18.9pt}|>{\centering\arraybackslash}m{18.9pt}>{\centering\arraybackslash}m{18.9pt}|>{\centering\arraybackslash}m{18.9pt}>{\centering\arraybackslash}m{10.4pt}>{\centering\arraybackslash}m{10.4pt}|>{\centering\arraybackslash}m{18.6pt}>{\centering\arraybackslash}m{18.9pt}|}
\hline
\multicolumn{1}{|>{\centering\arraybackslash}m{73.2pt}|}{\multirow{3}{56.3pt}{\centering \textbf{DF}\textbf{\textperiodcentered{}\textperiodcentered{}\textperiodcentered{}}\textbf{CH}\textbf{\textsubscript{3}}\textbf{/} \textbf{NH}\textbf{\textsubscript{3}}}} & \multicolumn{4}{>{\centering\arraybackslash}m{86.2pt}|}{\textbf{DF-RMP2-F12}} & \multicolumn{2}{>{\centering\arraybackslash}m{41.1pt}|}{\multirow{2}{31.6pt}{\centering \makebox[\linewidth][c]{\rotatebox{90}{\rbx{\textbf{MP2/} \textbf{AVDZ}}}}}} & \multicolumn{2}{>{\centering\arraybackslash}m{41.1pt}|}{\multirow{2}{31.6pt}{\centering \makebox[\linewidth][c]{\rotatebox{90}{\rbx{\textbf{MP3/} \textbf{AVDZ}}}}}} & \multicolumn{2}{>{\centering\arraybackslash}m{41.1pt}|}{\multirow{2}{31.6pt}{\centering \makebox[\linewidth][c]{\rotatebox{90}{\rbx{\textbf{MP4(DQ)/} \textbf{AVDZ}}}}}} & \multicolumn{2}{>{\centering\arraybackslash}m{41.1pt}|}{\multirow{2}{31.6pt}{\centering \makebox[\linewidth][c]{\rotatebox{90}{\rbx{\textbf{MP4(SDQ)/} \textbf{AVDZ}}}}}} & \multicolumn{2}{>{\centering\arraybackslash}m{41.3pt}|}{\multirow{2}{31.8pt}{\centering \makebox[\linewidth][c]{\rotatebox{90}{\rbx{\textbf{CCSD/} \textbf{AVDZ}}}}}} & \multicolumn{2}{>{\centering\arraybackslash}m{41.3pt}|}{\multirow{2}{31.8pt}{\centering \makebox[\linewidth][c]{\rotatebox{90}{\rbx{\textbf{QCISD/} \textbf{AVDZ}}}}}} & \multicolumn{2}{>{\centering\arraybackslash}m{41.6pt}|}{\multirow{2}{32.0pt}{\centering \makebox[\linewidth][c]{\rotatebox{90}{\rbx{\textbf{MP2(FULL)/\allowbreak{}AVTZ}}}}}} & \multicolumn{2}{>{\centering\arraybackslash}m{41.6pt}|}{\multirow{2}{32.0pt}{\centering \makebox[\linewidth][c]{\rotatebox{90}{\rbx{\textbf{MP3(FULL)/\allowbreak{}AVDZ}}}}}} & \multicolumn{2}{>{\centering\arraybackslash}m{41.6pt}|}{\multirow{2}{32.0pt}{\centering \makebox[\linewidth][c]{\rotatebox{90}{\rbx{\textbf{MP4(}\textbf{DQ,FULL}\textbf{)/} \textbf{AVDZ}}}}}} & \multicolumn{2}{>{\centering\arraybackslash}m{41.6pt}|}{\multirow{2}{32.0pt}{\centering \makebox[\linewidth][c]{\rotatebox{90}{\rbx{\textbf{MP4(}\textbf{SDQ,FULL}\textbf{)/} \textbf{AVDZ}}}}}} & \multicolumn{3}{>{\centering\arraybackslash}m{43.3pt}|}{\multirow{2}{33.3pt}{\centering \makebox[\linewidth][c]{\rotatebox{90}{\rbx{\textbf{CCSD(FULL)/\allowbreak{}AVDZ}}}}}} & \multicolumn{2}{>{\centering\arraybackslash}m{41.3pt}|}{\multirow{2}{31.8pt}{\centering \makebox[\linewidth][c]{\rotatebox{90}{\rbx{\textbf{QCISD} \textbf{(FULL)/} \textbf{AVDZ}}}}}}  \\
\hline
\end{tabular}}}
\par\nopagebreak\vspace{3pt}{\setstretch{1}\setlength{\parskip}{1.5pt}\fontsize{9}{10.5}\selectfont\raggedright\noindent \textsuperscript{b}Convergence error during optimization of respective F-bonded complexes.\par}
\end{sidewaystable}

\begin{sidewaystable}[!htbp]\centering\setlength{\parskip}{0pt}
\caption*{\textbf{Table S4 (contd.).} Binding energies (BEs) (in kJ/\allowbreak{}mol) with BSSE correction (\ensuremath{\Delta}\textsubscript{1+BSS}) and without BSSE-correction (\ensuremath{\Delta}\textsubscript{1}) for F-bonded complexes DF\textperiodcentered{}\textperiodcentered{}\textperiodcentered{}CH\textsubscript{3}/NH\textsubscript{3} using various \emph{ab}\emph{ }\emph{initio} methods and reference BEs at DF-RMP2-F12/\allowbreak{}CBS[V(T,Q)Z-F12]//DF-RMP2-F12/\allowbreak{}VDZ-F12 level.}\vspace{7pt}\nopagebreak
\makebox[\textheight][c]{\adjustbox{width=688pt,max totalheight=440pt}{\fontfamily{ptm}\fontsize{8.73}{10.0}\selectfont\renewcommand{\arraystretch}{1.12}\setlength{\tabcolsep}{1.5pt}
\begin{tabular}{|>{\centering\arraybackslash}m{81.1pt}|>{\centering\arraybackslash}m{20.9pt}>{\centering\arraybackslash}m{20.9pt}>{\centering\arraybackslash}m{20.9pt}>{\centering\arraybackslash}m{20.9pt}|>{\centering\arraybackslash}m{20.9pt}>{\centering\arraybackslash}m{20.9pt}|>{\centering\arraybackslash}m{20.9pt}>{\centering\arraybackslash}m{16.0pt}|>{\centering\arraybackslash}m{20.9pt}>{\centering\arraybackslash}m{16.0pt}|>{\centering\arraybackslash}m{23.0pt}|>{\centering\arraybackslash}m{81.1pt}|>{\centering\arraybackslash}m{20.9pt}>{\centering\arraybackslash}m{20.9pt}>{\centering\arraybackslash}m{20.9pt}>{\centering\arraybackslash}m{20.9pt}|>{\centering\arraybackslash}m{20.9pt}>{\centering\arraybackslash}m{20.9pt}|>{\centering\arraybackslash}m{10.4pt}>{\centering\arraybackslash}m{10.4pt}>{\centering\arraybackslash}m{20.9pt}|>{\centering\arraybackslash}m{20.9pt}>{\centering\arraybackslash}m{20.9pt}|>{\centering\arraybackslash}m{23.0pt}|}
\hline
\multicolumn{1}{|>{\centering\arraybackslash}m{81.1pt}|}{\multirow{3}{62.4pt}{\centering \textbf{DF}\textbf{\textperiodcentered{}\textperiodcentered{}\textperiodcentered{}}\textbf{CH}\textbf{\textsubscript{3}}}} & \multicolumn{4}{>{\centering\arraybackslash}m{95.4pt}|}{\textbf{DF-RMP2-F12}} & \multicolumn{2}{>{\centering\arraybackslash}m{45.8pt}|}{\multirow{2}{35.2pt}{\centering \makebox[\linewidth][c]{\rotatebox{90}{\rbx{\textbf{MP2.5/} \textbf{AVDZ}}}}}} & \multicolumn{2}{>{\centering\arraybackslash}m{40.8pt}|}{\multirow{2}{31.4pt}{\centering \makebox[\linewidth][c]{\rotatebox{90}{\rbx{\textbf{MP3.5(a)/} \textbf{AVDZ}}}}}} & \multicolumn{2}{>{\centering\arraybackslash}m{40.8pt}|}{\multirow{2}{31.4pt}{\centering \makebox[\linewidth][c]{\rotatebox{90}{\rbx{\textbf{MP3.5(b)/} \textbf{AVDZ}}}}}} & \multirow{3}{17.7pt}{\centering \makebox[\linewidth][c]{\rotatebox{90}{\rbx{\textbf{DF-RMP2-F12}\textbf{/}\textbf{CBS}\textbf{[}\textbf{V(}\textbf{T,Q}\textbf{)Z-F12]}}}}} & \multirow{3}{62.4pt}{\centering \textbf{DF}\textbf{\textperiodcentered{}\textperiodcentered{}\textperiodcentered{}}\textbf{NH}\textbf{\textsubscript{3}}} & \multicolumn{4}{>{\centering\arraybackslash}m{95.4pt}|}{\textbf{DF-RMP2-F12}} & \multicolumn{2}{>{\centering\arraybackslash}m{45.8pt}|}{\multirow{2}{35.2pt}{\centering \makebox[\linewidth][c]{\rotatebox{90}{\rbx{\textbf{MP2.5/} \textbf{AVDZ}}}}}} & \multicolumn{3}{>{\centering\arraybackslash}m{47.8pt}|}{\multirow{2}{36.8pt}{\centering \makebox[\linewidth][c]{\rotatebox{90}{\rbx{\textbf{MP3.5(a)/} \textbf{AVDZ}}}}}} & \multicolumn{2}{>{\centering\arraybackslash}m{45.8pt}|}{\multirow{2}{35.2pt}{\centering \makebox[\linewidth][c]{\rotatebox{90}{\rbx{\textbf{MP3.5(b)/} \textbf{AVDZ}}}}}} & \multirow{3}{17.7pt}{\centering \makebox[\linewidth][c]{\rotatebox{90}{\rbx{\textbf{DF-RMP2-F12}\textbf{/}\textbf{CBS}\textbf{[}\textbf{V(}\textbf{T,Q}\textbf{)Z-F12]}}}}} \\
\hline
\end{tabular}}}
\par\nopagebreak\vspace{3pt}{\setstretch{1}\setlength{\parskip}{1.5pt}\fontsize{9}{10.5}\selectfont\raggedright\noindent \textsuperscript{b}Convergence error during optimization of respective F-bonded complexes.\par}
\end{sidewaystable}

\begin{table}[!htbp]\centering\setlength{\parskip}{0pt}
\caption{Binding energies (BEs) with half-half correction \ensuremath{\Delta}\textsubscript{half-half }(in kJ/\allowbreak{}mol) for F-bonded complexes DF\textperiodcentered{}\textperiodcentered{}\textperiodcentered{}CH\textsubscript{3}/NH\textsubscript{3} using various \emph{ab}\emph{ }\emph{initio }methods and reference BEs at DF-RMP2-F12/\allowbreak{}CBS[V(T,Q)Z-F12]//DF-RMP2-F12/\allowbreak{}VDZ-F12 level.}\label{tab:S5}\vspace{7pt}\nopagebreak
\adjustbox{width=592pt,max totalheight=0.86\textheight,center}{\fontfamily{ptm}\fontsize{9.2}{10.6}\selectfont\renewcommand{\arraystretch}{1.5}\setlength{\tabcolsep}{2.2pt}
\begin{tabular}{|>{\centering\arraybackslash}m{76.8pt}|>{\centering\arraybackslash}m{31.8pt}>{\centering\arraybackslash}m{23.8pt}>{\centering\arraybackslash}m{28.7pt}|>{\centering\arraybackslash}m{23.8pt}|>{\centering\arraybackslash}m{19.0pt}|>{\centering\arraybackslash}m{23.8pt}|>{\centering\arraybackslash}m{23.8pt}|>{\centering\arraybackslash}m{28.6pt}|>{\centering\arraybackslash}m{23.8pt}|>{\centering\arraybackslash}m{23.8pt}|>{\centering\arraybackslash}m{28.6pt}|>{\centering\arraybackslash}m{23.8pt}|>{\centering\arraybackslash}m{28.7pt}|>{\centering\arraybackslash}m{23.8pt}|>{\centering\arraybackslash}m{19.0pt}|>{\centering\arraybackslash}m{19.0pt}|>{\centering\arraybackslash}m{23.8pt}|>{\centering\arraybackslash}m{19.0pt}|>{\centering\arraybackslash}m{24.0pt}|}
\hline
\multicolumn{1}{|>{\centering\arraybackslash}m{76.8pt}|}{\multirow{2}{69.8pt}{\centering \textbf{DF}\textbf{\textperiodcentered{}\textperiodcentered{}\textperiodcentered{}}\textbf{CH}\textbf{\textsubscript{3}}}} & \multicolumn{3}{>{\centering\arraybackslash}m{90.9pt}|}{\textbf{DF-RMP2-F12}} & \multirow{2}{21.6pt}{\centering \makebox[\linewidth][c]{\rotatebox{90}{\rbx{\textbf{MP2/\allowbreak{}AVDZ}}}}} & \multirow{2}{17.3pt}{\centering \makebox[\linewidth][c]{\rotatebox{90}{\rbx{\textbf{MP3/\allowbreak{}AVDZ}}}}} & \multirow{2}{21.6pt}{\centering \makebox[\linewidth][c]{\rotatebox{90}{\rbx{\textbf{MP4(DQ)/\allowbreak{}AVDZ}}}}} & \multirow{2}{21.6pt}{\centering \makebox[\linewidth][c]{\rotatebox{90}{\rbx{\textbf{MP4(SDQ)/\allowbreak{}AVDZ}}}}} & \multirow{2}{26.0pt}{\centering \makebox[\linewidth][c]{\rotatebox{90}{\rbx{\textbf{MP2.5/\allowbreak{}AVDZ}}}}} & \multirow{2}{21.6pt}{\centering \makebox[\linewidth][c]{\rotatebox{90}{\rbx{\textbf{MP3.5(a)/\allowbreak{}AVDZ}}}}} & \multirow{2}{21.6pt}{\centering \makebox[\linewidth][c]{\rotatebox{90}{\rbx{\textbf{MP3.5(b)/\allowbreak{}AVDZ}}}}} & \multirow{2}{26.0pt}{\centering \makebox[\linewidth][c]{\rotatebox{90}{\rbx{\textbf{CCSD/\allowbreak{}AVDZ}}}}} & \multirow{2}{21.6pt}{\centering \makebox[\linewidth][c]{\rotatebox{90}{\rbx{\textbf{QCISD/\allowbreak{}AVDZ}}}}} & \multirow{2}{26.1pt}{\centering \makebox[\linewidth][c]{\rotatebox{90}{\rbx{\textbf{MP2(FULL)/\allowbreak{}AVTZ}}}}} & \multirow{2}{21.6pt}{\centering \makebox[\linewidth][c]{\rotatebox{90}{\rbx{\textbf{MP3(FULL)/\allowbreak{}AVDZ}}}}} & \multirow{2}{17.3pt}{\centering \makebox[\linewidth][c]{\rotatebox{90}{\rbx{\textbf{MP4(}\textbf{DQ,FULL}\textbf{)/\allowbreak{}AVDZ}}}}} & \multirow{2}{17.3pt}{\centering \makebox[\linewidth][c]{\rotatebox{90}{\rbx{\textbf{MP4(}\textbf{SDQ,FULL}\textbf{)/\allowbreak{}AVDZ}}}}} & \multirow{2}{21.6pt}{\centering \makebox[\linewidth][c]{\rotatebox{90}{\rbx{\textbf{CCSD(FULL)/\allowbreak{}AVDZ}}}}} & \multirow{2}{17.3pt}{\centering \makebox[\linewidth][c]{\rotatebox{90}{\rbx{\textbf{QCISD(FULL)/\allowbreak{}AVDZ}}}}} & \multirow{2}{21.8pt}{\centering \makebox[\linewidth][c]{\rotatebox{90}{\rbx{\textbf{DF-RMP2-F12}\textbf{/}\textbf{CBS}\textbf{[}\textbf{V(}\textbf{T,Q}\textbf{)Z-F12]}}}}}  \\
\hline
\end{tabular}}
\par\nopagebreak\vspace{3pt}{\setstretch{1}\setlength{\parskip}{1.5pt}\fontsize{9}{10.5}\selectfont\raggedright\noindent \textsuperscript{b}Convergence error during optimization of respective F-bonded complexes.\par}

\end{table}

\begin{sidewaystable}[!htbp]\centering\setlength{\parskip}{0pt}
\caption{F\textbf{-}bond distances R(F\textperiodcentered{}\textperiodcentered{}\textperiodcentered{}C/\allowbreak{}N) (in Å) and F-bond angles (\ensuremath{\angle}DF\textperiodcentered{}\textperiodcentered{}\textperiodcentered{}C/\allowbreak{}N in \textdegree{}) for F-bonded complexes DF\textperiodcentered{}\textperiodcentered{}\textperiodcentered{}CH\textsubscript{3}/NH\textsubscript{3} using various \emph{ab}\emph{ }\emph{initio} methods and M06-2X/\allowbreak{}AVTZ.}\label{tab:S6}\vspace{7pt}\nopagebreak
\makebox[\textheight][c]{\adjustbox{width=688pt,max totalheight=440pt}{\fontfamily{ptm}\fontsize{7.7}{8.9}\selectfont\renewcommand{\arraystretch}{1.12}\setlength{\tabcolsep}{1.5pt}
\begin{tabular}{|>{\centering\arraybackslash}m{69.4pt}|>{\centering\arraybackslash}m{20.8pt}>{\centering\arraybackslash}m{18.9pt}|>{\centering\arraybackslash}m{18.9pt}>{\centering\arraybackslash}m{18.9pt}|>{\centering\arraybackslash}m{18.9pt}>{\centering\arraybackslash}m{18.9pt}|>{\centering\arraybackslash}m{18.9pt}>{\centering\arraybackslash}m{18.9pt}|>{\centering\arraybackslash}m{18.9pt}>{\centering\arraybackslash}m{18.9pt}|>{\centering\arraybackslash}m{18.9pt}>{\centering\arraybackslash}m{18.9pt}|>{\centering\arraybackslash}m{18.9pt}>{\centering\arraybackslash}m{18.9pt}|>{\centering\arraybackslash}m{18.9pt}>{\centering\arraybackslash}m{18.9pt}|>{\centering\arraybackslash}m{18.9pt}>{\centering\arraybackslash}m{18.9pt}|>{\centering\arraybackslash}m{18.9pt}>{\centering\arraybackslash}m{18.9pt}|>{\centering\arraybackslash}m{18.9pt}>{\centering\arraybackslash}m{18.9pt}|>{\centering\arraybackslash}m{18.9pt}>{\centering\arraybackslash}m{18.9pt}|>{\centering\arraybackslash}m{18.9pt}>{\centering\arraybackslash}m{18.9pt}|>{\centering\arraybackslash}m{18.9pt}>{\centering\arraybackslash}m{18.9pt}|}
\hline
\multicolumn{1}{|>{\centering\arraybackslash}m{69.4pt}|}{\multirow{3}{53.4pt}{\centering \textbf{DF}\textbf{\textperiodcentered{}\textperiodcentered{}\textperiodcentered{}}\textbf{CH}\textbf{\textsubscript{3}}}} & \multicolumn{2}{>{\centering\arraybackslash}m{43.6pt}|}{\makebox[\linewidth][c]{\rotatebox{90}{\rbx{\textbf{DF-RMP2-F12/\allowbreak{}VDZ-F12}}}}} & \multicolumn{2}{>{\centering\arraybackslash}m{41.6pt}|}{\makebox[\linewidth][c]{\rotatebox{90}{\rbx{\textbf{MP2/\allowbreak{}AVDZ}}}}} & \multicolumn{2}{>{\centering\arraybackslash}m{41.6pt}|}{\makebox[\linewidth][c]{\rotatebox{90}{\rbx{\textbf{MP3/\allowbreak{}AVDZ}}}}} & \multicolumn{2}{>{\centering\arraybackslash}m{41.6pt}|}{\makebox[\linewidth][c]{\rotatebox{90}{\rbx{\textbf{MP4(DQ)/} \textbf{AVDZ}}}}} & \multicolumn{2}{>{\centering\arraybackslash}m{41.6pt}|}{\makebox[\linewidth][c]{\rotatebox{90}{\rbx{\textbf{MP4(SDQ)/} \textbf{AVDZ}}}}} & \multicolumn{2}{>{\centering\arraybackslash}m{41.6pt}|}{\makebox[\linewidth][c]{\rotatebox{90}{\rbx{\textbf{CCSD/} \textbf{AVDZ}}}}} & \multicolumn{2}{>{\centering\arraybackslash}m{41.6pt}|}{\makebox[\linewidth][c]{\rotatebox{90}{\rbx{\textbf{QCISD/} \textbf{AVDZ}}}}} & \multicolumn{2}{>{\centering\arraybackslash}m{41.6pt}|}{\makebox[\linewidth][c]{\rotatebox{90}{\rbx{\textbf{MP2(FULL)/\allowbreak{}AVTZ}}}}} & \multicolumn{2}{>{\centering\arraybackslash}m{41.6pt}|}{\makebox[\linewidth][c]{\rotatebox{90}{\rbx{\textbf{MP3(FULL)/\allowbreak{}AVDZ}}}}} & \multicolumn{2}{>{\centering\arraybackslash}m{41.6pt}|}{\makebox[\linewidth][c]{\rotatebox{90}{\rbx{\textbf{MP4(}\textbf{DQ,FULL}\textbf{)/} \textbf{AVDZ}}}}} & \multicolumn{2}{>{\centering\arraybackslash}m{41.6pt}|}{\makebox[\linewidth][c]{\rotatebox{90}{\rbx{\textbf{MP4(}\textbf{SDQ,FULL}\textbf{)/} \textbf{AVDZ}}}}} & \multicolumn{2}{>{\centering\arraybackslash}m{41.6pt}|}{\makebox[\linewidth][c]{\rotatebox{90}{\rbx{\textbf{CCSD(FULL)/} \textbf{AV}\textbf{D}\textbf{Z}}}}} & \multicolumn{2}{>{\centering\arraybackslash}m{41.6pt}|}{\makebox[\linewidth][c]{\rotatebox{90}{\rbx{\textbf{QCISD(FULL)/\allowbreak{}AVDZ}}}}} & \multicolumn{2}{>{\centering\arraybackslash}m{41.6pt}|}{\makebox[\linewidth][c]{\rotatebox{90}{\rbx{\textbf{M06-2X/\allowbreak{}AVTZ}}}}}  \\
\hline
\end{tabular}}}
\par\nopagebreak\vspace{3pt}{\setstretch{1}\setlength{\parskip}{1.5pt}\fontsize{9}{10.5}\selectfont\raggedright\noindent \textsuperscript{b}Convergence error during optimization of respective F-bonded complexes.\par}
\end{sidewaystable}

\begin{sidewaystable}[!htbp]\centering\setlength{\parskip}{0pt}
\caption{Minimum relative energy (kJ/\allowbreak{}mol) and corresponding fractional charge-transfer (q) for infinitely separated F-bonded complexes (DF\textperiodcentered{}\textperiodcentered{}\textperiodcentered{}CH\textsubscript{3}/NH\textsubscript{3}) from SEFB-15 and CSFB-14 data sets predicted by several Minnesota DFA in conjunction with AVTZ basis set.}\label{tab:S7}\vspace{7pt}\nopagebreak
\makebox[\textheight][c]{\adjustbox{width=688pt,max totalheight=440pt}{\fontfamily{ptm}\fontsize{7.9}{9.1}\selectfont\renewcommand{\arraystretch}{1.12}\setlength{\tabcolsep}{1.5pt}
\begin{tabular}{|>{\centering\arraybackslash}p{65.8pt}|>{\centering\arraybackslash}p{15.7pt}>{\centering\arraybackslash}p{27.4pt}|>{\centering\arraybackslash}p{17.3pt}>{\centering\arraybackslash}p{27.4pt}|>{\centering\arraybackslash}p{17.3pt}>{\centering\arraybackslash}p{27.4pt}|>{\centering\arraybackslash}p{17.3pt}>{\centering\arraybackslash}p{27.4pt}|>{\centering\arraybackslash}p{17.3pt}>{\centering\arraybackslash}p{27.4pt}|>{\centering\arraybackslash}p{17.3pt}>{\centering\arraybackslash}p{27.4pt}>{\centering\arraybackslash}m{10.4pt}|>{\centering\arraybackslash}m{10.4pt}>{\centering\arraybackslash}p{27.4pt}>{\centering\arraybackslash}m{10.4pt}|>{\centering\arraybackslash}m{10.4pt}>{\centering\arraybackslash}p{27.4pt}>{\centering\arraybackslash}m{10.4pt}|>{\centering\arraybackslash}m{10.4pt}>{\centering\arraybackslash}p{27.4pt}|>{\centering\arraybackslash}p{15.7pt}>{\centering\arraybackslash}p{27.4pt}|>{\centering\arraybackslash}p{15.7pt}>{\centering\arraybackslash}p{27.4pt}|>{\centering\arraybackslash}p{15.7pt}>{\centering\arraybackslash}p{27.4pt}|}
\hline
\multicolumn{28}{|>{\centering\arraybackslash}p{708.5pt}|}{\textbf{SEFB-15 data set: }\textbf{DF}\textbf{\textperiodcentered{}\textperiodcentered{}\textperiodcentered{}}\textbf{CH}\textbf{\textsubscript{3}}\textbf{ complexes}}  \\
\hline
\end{tabular}}}
\par\nopagebreak\vspace{3pt}{\setstretch{1}\setlength{\parskip}{1.5pt}\fontsize{9}{10.5}\selectfont\raggedright\noindent \textsuperscript{b}Convergence error during optimization of respective F-bonded complexes.\par}
\end{sidewaystable}

\begin{sidewaystable}[!htbp]\centering\setlength{\parskip}{0pt}
\caption{Maximum deviations on electronic energies from linearity (kJ/\allowbreak{}mol) and corresponding fractional charge-transfer (q) for infinitely separated DF\textperiodcentered{}\textperiodcentered{}\textperiodcentered{}CH\textsubscript{3}/NH\textsubscript{3} complexes from SEFB-15 and CSFB-14 data sets predicted by several Minnesota DFAs in conjunction with AVTZ basis set. Positive values on deviations of electronic energies from linearity represent inherent localization error in the functionals under consideration, whereas negative values signify delocalization error present in the corresponding functionals.}\label{tab:S8}\vspace{7pt}\nopagebreak
\makebox[\textheight][c]{\adjustbox{width=688pt,max totalheight=440pt}{\fontfamily{ptm}\fontsize{9.18}{10.6}\selectfont\renewcommand{\arraystretch}{1.12}\setlength{\tabcolsep}{1.5pt}
\begin{tabular}{|>{\centering\arraybackslash}m{78.5pt}|>{\centering\arraybackslash}m{20.7pt}>{\centering\arraybackslash}m{26.4pt}|>{\centering\arraybackslash}m{18.9pt}>{\centering\arraybackslash}m{27.9pt}|>{\centering\arraybackslash}m{18.9pt}>{\centering\arraybackslash}m{27.6pt}|>{\centering\arraybackslash}m{18.9pt}>{\centering\arraybackslash}m{27.6pt}|>{\centering\arraybackslash}m{18.9pt}>{\centering\arraybackslash}m{27.6pt}|>{\centering\arraybackslash}m{18.9pt}>{\centering\arraybackslash}m{27.6pt}|>{\centering\arraybackslash}m{18.9pt}>{\centering\arraybackslash}m{22.4pt}|>{\centering\arraybackslash}m{18.9pt}>{\centering\arraybackslash}m{22.4pt}|>{\centering\arraybackslash}m{20.7pt}>{\centering\arraybackslash}m{24.6pt}|>{\centering\arraybackslash}m{18.9pt}>{\centering\arraybackslash}m{22.4pt}|>{\centering\arraybackslash}m{18.9pt}>{\centering\arraybackslash}m{27.6pt}|>{\centering\arraybackslash}m{18.9pt}>{\centering\arraybackslash}m{22.4pt}|}
\hline
\multicolumn{25}{|>{\centering\arraybackslash}m{708.5pt}|}{\textbf{SEFB-15 data set: }\textbf{DF}\textbf{\textperiodcentered{}\textperiodcentered{}\textperiodcentered{}}\textbf{CH}\textbf{\textsubscript{3}}\textbf{ complexes}}  \\
\hline
\end{tabular}}}
\par\nopagebreak\vspace{3pt}{\setstretch{1}\setlength{\parskip}{1.5pt}\fontsize{9}{10.5}\selectfont\raggedright\noindent \textsuperscript{b}Convergence error during optimization of respective F-bonded complexes.\par}
\end{sidewaystable}

\begin{table}[!htbp]\centering\setlength{\parskip}{0pt}
\caption{R\textsuperscript{2} values for correlation from DI versus BEs, DI versus ref. BEs, as well as DI(DFAs) versus DI(CCSD) for F-bonded complexes (DF\textperiodcentered{}\textperiodcentered{}\textperiodcentered{}CH\textsubscript{3}/NH\textsubscript{3}) predicted by selected Minnesota DFAs with AVTZ basis set. For comparison, results with CCSD/\allowbreak{}AVDZ level of theory are also included. All BEs and DIs are calculated at fully optimized geometries at corresponding level of theories. BSSE corrections are not considered for BE calculations. For necessary data used to obtain results included here, see \mbox{\textbf{Table}}\textbf{s}\textbf{ S}\textbf{10}\textbf{ }and \textbf{S}\textbf{11}.}\label{tab:S9}\vspace{7pt}\nopagebreak
\adjustbox{width=592pt,max totalheight=0.86\textheight,center}{\fontfamily{ptm}\fontsize{9.2}{10.6}\selectfont\renewcommand{\arraystretch}{1.12}\setlength{\tabcolsep}{1.5pt}
\begin{tabular}{|>{\centering\arraybackslash}m{169.2pt}|>{\centering\arraybackslash}m{50.2pt}>{\centering\arraybackslash}m{50.8pt}|>{\centering\arraybackslash}m{50.2pt}>{\centering\arraybackslash}m{50.8pt}|>{\centering\arraybackslash}m{50.2pt}>{\centering\arraybackslash}m{50.8pt}|>{\centering\arraybackslash}m{50.2pt}>{\centering\arraybackslash}m{50.8pt}|}
\hline
\multicolumn{9}{|>{\centering\arraybackslash}m{599.5pt}|}{\textbf{R}\textbf{\textsuperscript{2 }}\textbf{values for correlation of DI with BEs, ref. BEs and DI(CCSD) for}\textbf{ DF}\textbf{\textperiodcentered{}\textperiodcentered{}\textperiodcentered{}}\textbf{CH}\textbf{\textsubscript{3}}\textbf{/NH}\textbf{\textsubscript{3}}\textbf{ }\textbf{complexes using corresponding optimized geometries}}  \\
\hline
\multicolumn{1}{|>{\centering\arraybackslash}m{169.2pt}|}{\multirow{2}{153.8pt}{\centering \textbf{Methods}}} & \multicolumn{2}{>{\centering\arraybackslash}m{104.3pt}|}{\textbf{DI(DFAs) vs DI(CCSD)}} & \multicolumn{2}{>{\centering\arraybackslash}m{104.3pt}|}{\textbf{BE(DFAs) vs ref. BEs}} & \multicolumn{2}{>{\centering\arraybackslash}m{104.3pt}|}{\textbf{MAD (kJ/\allowbreak{}mol)}} & \multicolumn{2}{>{\centering\arraybackslash}m{104.3pt}|}{\textbf{MA\%}\textbf{D(}\textbf{\%)}}  \\
\hline
\end{tabular}}

\end{table}

\begin{table}[!htbp]\centering\setlength{\parskip}{0pt}
\caption{Binding energies (BEs) (kJ/\allowbreak{}mol) and intermolecular delocalization indices [DI(F,C/\allowbreak{}N)] for F-bonded complexes (DF\textperiodcentered{}\textperiodcentered{}\textperiodcentered{}CH\textsubscript{3}/NH\textsubscript{3}) predicted by several Minnesota DFAs with AVTZ basis set. For comparison results with CCSD/\allowbreak{}AVDZ level of theory are also included. All BEs and DIs are calculated at fully optimized geometries at corresponding level of theories. BSSE corrections are not considered for BE calculations. Also, reference BEs obtained at DF-RMP2-F12/\allowbreak{}CBS//DF-RMP2-F12/\allowbreak{}VDZ-F12 level (denoted as “Ref. BE”) are also included. Largest DI values for each complex are highlighted in bold.}\label{tab:S10}\vspace{7pt}\nopagebreak
\adjustbox{width=592pt,max totalheight=0.86\textheight,center}{\fontfamily{ptm}\fontsize{7.22}{8.3}\selectfont\renewcommand{\arraystretch}{1.12}\setlength{\tabcolsep}{1.5pt}
\begin{tabular}{|>{\centering\arraybackslash}m{49.4pt}|>{\centering\arraybackslash}b{29.8pt}>{\centering\arraybackslash}m{29.8pt}|>{\centering\arraybackslash}m{29.8pt}>{\centering\arraybackslash}m{29.8pt}|>{\centering\arraybackslash}m{29.8pt}>{\centering\arraybackslash}m{29.8pt}|>{\centering\arraybackslash}b{29.8pt}>{\centering\arraybackslash}m{29.8pt}|>{\centering\arraybackslash}m{29.8pt}>{\centering\arraybackslash}m{29.8pt}|>{\centering\arraybackslash}p{29.8pt}>{\centering\arraybackslash}m{29.8pt}|>{\centering\arraybackslash}m{29.8pt}>{\centering\arraybackslash}m{29.8pt}|>{\centering\arraybackslash}m{29.8pt}>{\centering\arraybackslash}b{29.8pt}|>{\centering\arraybackslash}m{16.6pt}|}
\hline
\multicolumn{18}{|>{\centering\arraybackslash}m{599.5pt}|}{\textbf{SEFB-15 data set: }\textbf{DF}\textbf{\textperiodcentered{}\textperiodcentered{}\textperiodcentered{}}\textbf{CH}\textbf{\textsubscript{3}}\textbf{ complexes}}  \\
\cline{2-18}
\multicolumn{1}{|>{\centering\arraybackslash}m{49.4pt}|}{} & \multicolumn{2}{>{\centering\arraybackslash}m{63.0pt}|}{\textbf{M05-2X(opt)}} & \multicolumn{2}{>{\centering\arraybackslash}m{63.0pt}|}{\textbf{M05-2X-GD3(opt)}} & \multicolumn{2}{>{\centering\arraybackslash}m{63.0pt}|}{\textbf{M06-2X(opt)}} & \multicolumn{2}{>{\centering\arraybackslash}m{63.0pt}|}{\textbf{M06-2X-GD3(opt)}} & \multicolumn{2}{>{\centering\arraybackslash}m{63.0pt}|}{\textbf{M06-HF} \textbf{(opt)}} & \multicolumn{2}{>{\centering\arraybackslash}m{63.0pt}|}{\textbf{M08-HX} \textbf{(opt)}} & \multicolumn{2}{>{\centering\arraybackslash}m{63.0pt}|}{\textbf{M11} \textbf{(opt)}} & \multicolumn{2}{>{\centering\arraybackslash}m{63.0pt}|}{\textbf{CCSD} \textbf{(opt)}} & \multirow{2}{15.1pt}{\centering \textbf{Ref. BE}}  \\
\hline
\end{tabular}}
\par\nopagebreak\vspace{3pt}{\setstretch{1}\setlength{\parskip}{1.5pt}\fontsize{9}{10.5}\selectfont\raggedright\noindent \textsuperscript{b}Convergence error during optimization of respective F-bonded complexes.\par}

\end{table}

\begin{table}[!htbp]\centering\setlength{\parskip}{0pt}
\caption{Binding energies (BEs) (kJ/\allowbreak{}mol) and intermolecular delocalization indices [DI(F,C/\allowbreak{}N)] for F-bonded complexes (DF\textperiodcentered{}\textperiodcentered{}\textperiodcentered{}CH\textsubscript{3}/NH\textsubscript{3}) predicted by several Minnesota DFAs with AVTZ basis set considering geometries optimized at DF-RMP2-F12/\allowbreak{}VDZ-F12 level of theory (denoted as “DI(sp)”) . All BEs were without.BSSE correction. Reference BEs obtained at DF-RMP2-F12/\allowbreak{}CBS//DF-RMP2-F12/\allowbreak{}VDZ-F12 level (denoted as “Ref. BE”) are also included. For comparison results with CCSD/\allowbreak{}AVDZ level of theory are also included.}\label{tab:S11}\vspace{7pt}\nopagebreak
\adjustbox{width=592pt,max totalheight=0.86\textheight,center}{\fontfamily{ptm}\fontsize{7.94}{9.1}\selectfont\renewcommand{\arraystretch}{1.12}\setlength{\tabcolsep}{1.5pt}
\begin{tabular}{|>{\centering\arraybackslash}m{55.7pt}|>{\centering\arraybackslash}m{29.3pt}>{\centering\arraybackslash}m{29.3pt}|>{\centering\arraybackslash}m{29.3pt}>{\centering\arraybackslash}m{29.3pt}|>{\centering\arraybackslash}m{29.3pt}>{\centering\arraybackslash}m{29.3pt}|>{\centering\arraybackslash}m{29.3pt}>{\centering\arraybackslash}m{29.3pt}|>{\centering\arraybackslash}m{29.3pt}>{\centering\arraybackslash}m{29.3pt}|>{\centering\arraybackslash}m{29.3pt}>{\centering\arraybackslash}m{29.3pt}|>{\centering\arraybackslash}m{29.3pt}>{\centering\arraybackslash}m{29.3pt}|>{\centering\arraybackslash}m{29.3pt}>{\centering\arraybackslash}m{29.3pt}|>{\centering\arraybackslash}m{19.0pt}|}
\hline
\multicolumn{18}{|>{\centering\arraybackslash}m{599.5pt}|}{\textbf{SEFB-15 data set: }\textbf{DF}\textbf{\textperiodcentered{}\textperiodcentered{}\textperiodcentered{}}\textbf{CH}\textbf{\textsubscript{3}}\textbf{ complexes}}  \\
\cline{2-18}
\multicolumn{1}{|>{\centering\arraybackslash}m{55.7pt}|}{} & \multicolumn{2}{>{\centering\arraybackslash}m{61.9pt}|}{\textbf{M05-2X(}\textbf{sp}\textbf{)}} & \multicolumn{2}{>{\centering\arraybackslash}m{61.9pt}|}{\textbf{M05-2X-GD3(}\textbf{sp}\textbf{)}} & \multicolumn{2}{>{\centering\arraybackslash}m{61.9pt}|}{\textbf{M06-2X(}\textbf{sp}\textbf{)}} & \multicolumn{2}{>{\centering\arraybackslash}m{61.9pt}|}{\textbf{M06-2X-GD3(}\textbf{sp}\textbf{)}} & \multicolumn{2}{>{\centering\arraybackslash}m{61.9pt}|}{\textbf{M06-}\textbf{HF(}\textbf{sp}\textbf{)}} & \multicolumn{2}{>{\centering\arraybackslash}m{61.9pt}|}{\textbf{M08-}\textbf{HX(}\textbf{sp}\textbf{)}} & \multicolumn{2}{>{\centering\arraybackslash}m{61.9pt}|}{\textbf{M11(}\textbf{sp}\textbf{)}} & \multicolumn{2}{>{\centering\arraybackslash}m{61.9pt}|}{\textbf{CCSD(}\textbf{sp}\textbf{)}} & \multirow{2}{17.3pt}{\centering \textbf{Ref. BE}}  \\
\hline
\end{tabular}}
\par\nopagebreak\vspace{3pt}{\setstretch{1}\setlength{\parskip}{1.5pt}\fontsize{9}{10.5}\selectfont\raggedright\noindent \textsuperscript{b}Convergence error during optimization of respective F-bonded complexes.\par}

\end{table}

\begin{table}[!htbp]\centering\setlength{\parskip}{0pt}
\caption{Intermolecular delocalization indices [DI(F,C/\allowbreak{}N)], denoted as DI(sp), for F-bonded complexes (DF\textperiodcentered{}\textperiodcentered{}\textperiodcentered{}CH\textsubscript{3}/NH\textsubscript{3}) predicted by several Minnesota DFAs with AVTZ basis set using geometries optimized at DF-RMP2-F12/\allowbreak{}VDZ-F12 level. For comparison results with CCSD/\allowbreak{}AVDZ and HF/\allowbreak{}AVTZ levels are also included. Largest DI values for each complex are highlighted in bold.}\label{tab:S12}\vspace{7pt}\nopagebreak
\adjustbox{width=592pt,max totalheight=0.86\textheight,center}{\fontfamily{ptm}\fontsize{8.55}{9.8}\selectfont\renewcommand{\arraystretch}{1.12}\setlength{\tabcolsep}{1.5pt}
\begin{tabular}{|>{\centering\arraybackslash}m{61.1pt}|>{\centering\arraybackslash}m{32.3pt}|>{\centering\arraybackslash}m{32.3pt}|>{\centering\arraybackslash}m{32.3pt}|>{\centering\arraybackslash}m{32.3pt}|>{\centering\arraybackslash}m{32.3pt}|>{\centering\arraybackslash}m{32.3pt}|>{\centering\arraybackslash}m{33.4pt}|>{\centering\arraybackslash}m{33.4pt}|>{\centering\arraybackslash}m{33.4pt}|>{\centering\arraybackslash}m{33.4pt}|>{\centering\arraybackslash}m{32.3pt}|>{\centering\arraybackslash}m{32.3pt}|>{\centering\arraybackslash}m{32.3pt}|>{\centering\arraybackslash}m{32.3pt}|>{\centering\arraybackslash}m{32.3pt}|}
\hline
\multicolumn{16}{|>{\centering\arraybackslash}m{599.5pt}|}{\textbf{SEFB-15 data set: DF}\textbf{\textperiodcentered{}\textperiodcentered{}\textperiodcentered{}}\textbf{CH}\textbf{\textsubscript{3}}\textbf{ complexes using geometries optimized at DF-RMP2-F12/\allowbreak{}VDZ-F12 level}}  \\
\cline{2-16}
\multicolumn{1}{|>{\centering\arraybackslash}m{61.1pt}|}{} & \textbf{HF} & \textbf{M06-L} & \textbf{MN15-L} & \textbf{M11-L } & \textbf{M05} & \textbf{M06} & \textbf{M05-2X} & \textbf{M05-2X-GD3} & \textbf{M06-2X} & \textbf{M06-2X-GD3} & \textbf{M06-HF} & \textbf{M08-HX} & \textbf{MN15} & \textbf{M11} & \textbf{CCSD}  \\
\hline
\multicolumn{1}{|>{\centering\arraybackslash}m{61.1pt}|}{\textbf{DF}\textbf{\textperiodcentered{}\textperiodcentered{}\textperiodcentered{}}\textbf{CH}\textbf{\textsubscript{3}}} & \textbf{DI(}\textbf{sp}\textbf{)} & \textbf{DI(}\textbf{sp}\textbf{)} & \textbf{DI(}\textbf{sp}\textbf{)} & \textbf{DI(}\textbf{sp}\textbf{)} & \textbf{DI(}\textbf{sp}\textbf{)} & \textbf{DI(}\textbf{sp}\textbf{)} & \textbf{DI(}\textbf{sp}\textbf{)} & \textbf{DI(}\textbf{sp}\textbf{)} & \textbf{DI(}\textbf{sp}\textbf{)} & \textbf{DI(}\textbf{sp}\textbf{)} & \textbf{DI(}\textbf{sp}\textbf{)} & \textbf{DI(}\textbf{sp}\textbf{)} & \textbf{DI(}\textbf{sp}\textbf{)} & \textbf{DI(}\textbf{sp}\textbf{)} & \textbf{DI(}\textbf{sp}\textbf{)}  \\
\hline
\end{tabular}}
\par\nopagebreak\vspace{3pt}{\setstretch{1}\setlength{\parskip}{1.5pt}\fontsize{9}{10.5}\selectfont\raggedright\noindent \textsuperscript{b}Convergence error during optimization of respective F-bonded complexes.\par}

\end{table}

\begin{table}[!htbp]\centering\setlength{\parskip}{0pt}
\caption{Binding energies (BEs) in kJ/\allowbreak{}mol for F-bonded complexes (DF\textperiodcentered{}\textperiodcentered{}\textperiodcentered{}CH\textsubscript{3}/NH\textsubscript{3}) predicted by several Minnesota DFAs with AVTZ basis set using geometries optimized at DF-RMP2-F12/\allowbreak{}VDZ-F12 level. For comparison, results with CCSD/\allowbreak{}AVDZ level are also included. Reference BEs obtained at DF-RMP2-F12/\allowbreak{}CBS//DF-RMP2-F12/\allowbreak{}VDZ-F12 level (denoted as “Ref. BE”) are also included.\textbf{ }}\label{tab:S13}\vspace{7pt}\nopagebreak
\adjustbox{width=592pt,max totalheight=0.86\textheight,center}{\fontfamily{ptm}\fontsize{9.2}{10.6}\selectfont\renewcommand{\arraystretch}{1.12}\setlength{\tabcolsep}{1.5pt}
}
\par\nopagebreak\vspace{3pt}{\setstretch{1}\setlength{\parskip}{1.5pt}\fontsize{9}{10.5}\selectfont\raggedright\noindent \textsuperscript{b}Convergence error during optimization of respective F-bonded complexe\par}

\end{table}

\end{document}